\documentclass[12pt]{article}
\usepackage{amssymb}
\usepackage{amsfonts}
\usepackage{amsmath}
\usepackage{eurosym}
\usepackage{makeidx}
\usepackage{amssymb}
\usepackage{times}
\usepackage{anysize}
\usepackage{cite}
\usepackage{lineno}

\marginsize{2cm}{2cm}{1cm}{1cm}
\newtheorem{theorem}{Theorem}[section]

\newtheorem{corollary}[theorem]{Corollary}

\newtheorem{definition}[theorem]{Definition}

\newtheorem{lemma}[theorem]{Lemma}

\newtheorem{proposition}[theorem]{Proposition}
\newtheorem{remark}[theorem]{Remark}

\newenvironment{proof}[1][Proof]{\noindent\textbf{#1.} }{\ \rule{0.5em}{0.5em}}
\input{tcilatex}
\begin{document}

\title{On the Equivalence of the KMS Condition and the Variational Principle
for Quantum Lattice Systems with Mean-Field Interactions}
\author{J.-B. Bru \and W. de Siqueira Pedra \and R. S. Yamaguti Miada}
\date{\today }
\maketitle

\begin{abstract}
We extend Araki's well-known results on the equivalence of the KMS condition
and the variational principle for equilibrium states of quantum lattice
systems with short-range interactions, to a large class of models possibly
containing mean-field interactions (representing an extreme form of
long-range interactions). This result is reminiscent of van Hemmen's work on
equilibrium states for mean-field models. The extension was made possible by
our recent outcomes on states minimizing the free energy density of
mean-field models on the lattice, as well as on the infinite volume dynamics
for such models. \bigskip

\noindent \textbf{Keywords:} Mean-field, self-consistency, quantum lattice
systems, KMS, equilibrium states.\bigskip

\noindent \textbf{AMS Subject Classification:} 82B03, 82B10, 82B20, 46N55
\end{abstract}

\setcounter{tocdepth}{2}
\tableofcontents%

\section{Introduction}

As is widely accepted in theoretical and mathematical physics, the
equilibrium state of a given \emph{finite} quantum system is the Gibbs state
associated with its Hamiltonian. Among other things, this state is
stationary with respect to the dynamics generated by the Hamiltonian. In
general, Gibbs states are ill-defined for infinite systems and, thus,
several notions of thermodynamic equilibrium were proposed for these
systems, based on properties of finite volume Gibbs states, like free energy
minimization, complete passivity, Kubo-Martin-Schwinger (KMS) and Gibbs
property, to mention the most commonly used. All the listed properties
uniquely define Gibbs states of any finite quantum system. In particular,
they are all equivalent\footnote{%
More precisely, in contrast to the other conditions, the complete passivity
does not fixes the temperature of the system, i.e., this condition only
implies that the state is Gibbs for some temperature.} to each other for
finite systems, but that is not the case for infinite systems. It is
therefore important, from both the mathematical and the physical viewpoints,
to understand the relation between these mathematical definitions of
equilibrium states of infinite quantum systems. In this scope, here we focus
on fermions and quantum spins on infinite lattices and analyze the relation
between the minimization of the free energy density and the KMS property of
space-homogeneous states, in presence of mean-field, or long-range\footnote{%
We frequently prefer the term \textquotedblleft
long-range\textquotedblright\ instead of \textquotedblleft
mean-field\textquotedblright , because the latter can refer to different
scalings and has thus some ambiguity, whereas the former is precisely
defined later on.}, interactions.

Because of their elegant definition and good mathematical properties (see,
e.g., \cite[Sections 5.3-5.4]{BrattelliRobinson}), KMS states certainly
constitute the most popular notion of equilibrium states for infinite
systems. For instance, they provide the Kelvin-Planck statement of the 2nd
law of thermodynamics with a precise mathematical meaning, the complete
passivity of these states \cite{PW}. The KMS property refers to a \emph{%
dynamical} notion of equilibrium. By contrast, a \emph{static} notion of
thermodynamic equilibrium can also be given by defining equilibrium states
via the minimization of the free energy density of the quantum system under
consideration. In opposition to the dynamical viewpoint, this approach
requires some space homogeneity of states. On the other hand, the static
approach has the big advantage of getting around the problem of the
existence of quantum dynamics in the infinite volume limit. Moreover, the
existence of equilibrium states in the static sense can, in general, be
easily ensured (by standard compactness and lower semi-continuity
arguments), whereas, in many cases, the mere existence of KMS states is an
issue, even if the infinite volume dynamics exists and is continuous in the
appropriate sense.

Giving a precise and convenient mathematical meaning to the infinite volume
dynamics of quantum systems can be highly non-trivial: On the one hand, the
Green-function method (see, e.g., \cite[Section 6.3.4]{BrattelliRobinson})
allows one to implement, under very general conditions, the infinite volume
limit of quantum dynamics as a strongly continuous group of unitaries acting
in some abstract Hilbert space. Note however that this approach needs
(initial) states to be fixed. On the other hand, the KMS approach requires a
strongly continuous group of $\ast $-auto%
\-%
morphisms of the (original) $C^{\ast }$-algebra of the given quantum system.
Such a group of $\ast $-auto%
\-%
morphisms should, in principle, not depend on the particular choices of
states for the system. In fact, for lattice fermion and quantum spin systems
with\emph{\ short-range} interactions, such a continuous group of $\ast $%
-auto%
\-%
morphisms is well-defined in the infinite volume limit, independently of
some choice of states. See \cite{Araki-Moriya,brupedraLR} for the fermionic
case and \cite{BrattelliRobinson,Israel} for the quantum spin one. For
short-range interactions, the notion of KMS states naturally applies and the
equivalence between the KMS condition and the minimization of the infinite
volume free energy density (at the same inverse temperature) can be proven
for space-homogeneous states: Inspired by works of Dobrushin and
Lanford-Ruelle on classical spins, Araki contributed in 1974 \cite{Araki}
such an equivalence proof for quantum spins with short-range interactions on
a (infinite square) lattice of arbitrary dimension. The case of lattice
fermions (with short-range interactions) was only studied in detail much
later, in 2003, by Araki and Moriya \cite{Araki-Moriya}.

Quantum many-body systems in the continuum are expected to have, in general,
no such dynamics in the thermodynamic limit, but only a \emph{state-dependent%
} Heisenberg dynamics (see, e.g., \cite[Section 6.3]{BrattelliRobinson}). To
get around this issue in the continuum, in presence of interactions, one may
use a UV cut-off of some kind, as it is recently discussed and proven in
\cite{LRB-continuum}. The same issue is also expected for interactions
slowly decaying in space, even in the lattice case. For instance, as already
remarked in the seminal work \cite{haag62} on the mathematics of the BCS
theory, in presence of mean-field interactions, the finite volume quantum
dynamics does \emph{not} generally converge within the (original) $C^{\ast }$%
-algebra of the given system, when the volume diverges. See also \cite[%
Section 4.3]{BruPedra-MFII}.

From the sixties to the nineties, there were a lot of studies on infinite
volume dynamics of lattice fermion and quantum spin systems with mean-field
interactions, by B\'{o}na, Duffield, Duffner, Haag, Hepp, Lieb, van Hemmen,
Rieckers, Thirring, Unnerstall, Wehrl, Werner, and others. Mathematically
rigorous results were obtained for models that are permutation-invariant,
almost exclusively for quantum spins (except, e.g., in \cite{Hemmen,haag62}%
). In this case, the infinite volume dynamics refers to an effective
time-evolution on a finite-dimensional Hilbert space\footnote{%
The one-site Hilbert space for quantum spins and the fermionic Fock space
associated with the one-site Hilbert space for lattice fermions.}, whereas
the KMS\ theory were proposed to deal with infinite systems, the finite case
being trivial from the point of view of this theory. During the last two
decades, many studies on the dynamics of fermion systems in the continuum
with mean-field interactions were also performed, by Benedikter, Elgart, Erd%
\"{o}s, Jak\v{s}i\'{c}, Petrat, Pickl, Porta, Schlein, Yau, and others, but
(positive temperature) thermal states were not considered. In this context,
the KMS\ condition is irrelevant. For a detailed discussion on this subject,
see \cite[Section 1]{BruPedra-MFII} and references therein.

In 2013, \cite{BruPedra2} contributed the complete characterization of
space-homogeneous free-energy%
\-%
-density-minimizing states of quantum lattice systems with mean-field
interactions, in terms of free energy density minimizers of their so-called
Bogoliubov (short-range) approximations. In fact, in contrast to the
short-range case \cite{Israel,BrattelliRobinson,Araki-Moriya}, note that
strict minimizers of the free energy density may not exist, but the (weak$%
^{\ast }$) limits of minimizing sequences of states offer a good, more
general, notion of equilibrium states, in the long-range case.

In 2021, \cite{BruPedra-MFII,BruPedra-MFIII} derived the infinite volume
dynamics of space homogeneous quantum lattice systems with mean-field, or
long-range, interactions in great generality. In fact, it was shown that the
infinite volume dynamics of such systems is always equivalent to an
intricate combination of classical and short-range quantum dynamics. A
simple illustration of this general result is given in \cite%
{Bru-pedra-proceeding,Bru-pedra-MF-IV}. \cite{BruPedra-MFIII} shows in
particular that, within the cyclic representation associated with the
initial state, the infinite volume limit of dynamics is well-defined in the $%
\sigma $-weak operator topology and corresponds to a \emph{state-dependent}
Heisenberg dynamics.

Combined with \cite{BruPedra2}, the outcomes of the recent papers \cite%
{BruPedra-MFII,BruPedra-MFIII} pave the way to a study of the relations
between static and dynamical notions of equilibrium, respectively
characterized by the minimization of the free energy density and the KMS
conditions, for very general infinite quantum lattice systems with
mean-field interactions. This is the main objective of this paper. In
particular, the precise status of the KMS condition in presence of
mean-field, or long-range, interactions will be examined.

Our main results are summarized as follows: To simplify discussions, we take
a translation-invariant\footnote{%
See Remark \ref{Periodic quantum lattice systems}.} model $\mathfrak{m}$ for
fermions on the lattice with mean-field interactions, within a very general
class of models. Fix once and for all the (non-zero) temperature of the
system. By \cite[Theorem 2.12]{BruPedra2}, the infinite volume
(grand-canonical) pressure $\mathrm{P}$ of the model $\mathfrak{m}$ is given
by the following variational problem
\begin{equation}
\mathrm{P}=-\inf f\left( E_{1}\right) \in \mathbb{R}\ ,  \label{P}
\end{equation}%
where $f:E_{1}\rightarrow \mathbb{R}$ is the free energy density functional
associated with $\mathfrak{m}$, defined on the (metrizable weak$^{\ast }$%
-compact) convex set $E_{1}$ of all translation-invariant states of the
system. Define the (metrizable weak$^{\ast }$-compact) convex set $\mathit{%
\Omega }\subseteq E_{1}$ of so-called generalized equilibrium states as
being the set of all weak$^{\ast }$-limit points of minimizing sequences of
this variational problem. We then have two observations at this point: \
\medskip

\noindent \textbf{Observation 1:} As explained above, there is no
state-independent Heisenberg dynamics in infinite volume, but a
state-dependent one actually exists, in relation with the ergodic
decomposition of translation-invariant states. See \cite[Theorem 4.3]%
{BruPedra-MFIII} or Equation (\ref{long-range dyn}) below.\medskip

\noindent \textbf{Observation 2:} By \cite[Theorem 2.39]{BruPedra2}, the set
$\mathcal{E}(\mathit{\Omega })$ of extreme points of the weak$^{\ast }$%
-compact convex set $\mathit{\Omega }$ belongs to the set
\begin{equation*}
\mathbf{M}\doteq \underset{d\in \mathcal{C}}{\cup }\mathit{M}_{\Phi (d)}
\end{equation*}%
of strict minimizers of the free energy densities associated with effective
translation-invariant \emph{short-range} interactions $\Phi (d)$, $d\in
\mathcal{C}$, which refer to Bogoliubov approximations of the full
interaction of the model $\mathfrak{m}$ \cite[Section 2.10.1]{BruPedra2}. In
other words, $\mathcal{E}(\mathit{\Omega })\subseteq \mathbf{M}\subseteq
E_{1}$. The set $\mathcal{C}$ (of parameters $d$) and the corresponding
Bogoliubov approximations strongly depend on the particular choice of the
model $\mathfrak{m}$ and will be precisely defined in the sequel.\medskip

By \cite[Theorem 1]{Araki} (quantum spins) and \cite[Corollary 6.7 and
Theorem 12.11]{Araki-Moriya} (lattice fermions), the elements of $\mathit{M}%
_{\Phi (d)}$ are KMS states with respect to a well-defined dynamics on the
underlying $C^{\ast }$-algebra $\mathcal{U}$, since the interactions $\Phi
(d)$, $d\in \mathcal{C}$, are short-range. By Observations 1 and 2, this
suggests that one should decompose arbitrary generalized equilibrium states
of $\mathfrak{m}$ in terms of extreme ones, in order to understand the
status of the KMS condition in presence of mean-field interactions. In fact,
one can apply the Choquet theorem (see, for instance, \cite[Theorem 10.18]%
{BruPedra2}) to the metrizable and weak$^{\ast }$-compact convex set $%
\mathit{\Omega }\subseteq E_{1}$: Any generalized equilibrium state $\omega
\in \mathit{\Omega }$ is the barycenter of a unique probability measure $\nu
_{\omega }$ which is supported on the set $\mathcal{E}(\mathit{\Omega })$,
i.e.,
\begin{equation}
\omega \left( \cdot \right) =\int_{\mathcal{E}(\mathit{\Omega })}\hat{\omega}%
\left( \cdot \right) \nu _{\omega }\left( \mathrm{d}\hat{\omega}\right) \ .
\label{decomp1}
\end{equation}%
There are two caveats in relation to this strategy:\medskip

\noindent \textbf{Problem (a):} By \cite[Lemma 9.8]{BruPedra2}, there are
uncountably many translation-invariant lattice fermion or quantum spin
systems with mean-field interactions such that $\mathcal{E}(\mathit{\Omega }%
)\nsubseteq \mathcal{E}(E_{1})$, where $\mathcal{E}(E_{1})$ is the set of
extreme points of $E_{1}$. This is problematic because it prevents us from
using the ergodicity of extreme states of $E_{1}$. See \cite[Theorem 1.16]%
{BruPedra2}. \medskip

\noindent \textbf{Problem (b):} We prove in Theorem \ref{lemma explosion l
du mec copacabana2} the existence of uncountably many models with mean-field
interactions having generalized equilibrium states $\omega \in \mathit{%
\Omega }$, whose Choquet measure $\nu _{\omega }$ is non-orthogonal. This is
problematic because it prevents us from using the Effros Theorem \cite[%
Theorem 4.4.9]{BrattelliRobinsonI}, in the scope of the theory of direct
integrals of measurable families of Hilbert spaces, operators, von Neumann
algebras, and $C^{\ast }$-algebra representations. See \cite[Sections 5-6]%
{BruPedra-MFIII} for more details. \medskip

However, one can also apply the Choquet theorem \cite[Theorem 10.18]%
{BruPedra2} to the metrizable and weak$^{\ast }$-compact convex set $%
E_{1}\supseteq \mathit{\Omega }$ of all translation-invriant states of the
system: Each generalized equilibrium state $\omega \in \mathit{\Omega }$ is
again the barycenter of a unique probability measure $\mu _{\omega }$ which
is supported on the set $\mathcal{E}(E_{1})$, i.e.,
\begin{equation}
\omega \left( \cdot \right) =\int_{\mathcal{E}(E_{1})}\hat{\omega}\left(
\cdot \right) \mu _{\omega }\left( \mathrm{d}\hat{\omega}\right) \ .
\label{decomp2}
\end{equation}%
This time, $\mu _{\omega }$ is always an orthogonal measure, as stated, for
instance, in \cite[Theorem 5.1]{BruPedra-MFIII}. Compare with Problem (b).
There is however an accessory problem in decomposing generalized equilibrium
states in the set $\mathcal{E}(E_{1})$ of extreme points of $E_{1}$, instead
of decomposing them in $\mathcal{E}(\mathit{\Omega })$, i.e., in terms of
extreme generalized equilibrium states of $\mathfrak{m}$: \medskip

\noindent \textbf{Problem (c):} Being solution to the variational problem (%
\ref{P}), extreme generalized equilibrium states satisfy a sort of
Euler-Lagrange equation in relation with the approximating interactions $%
\Phi (d)$, $d\in \mathcal{C}$, also called gap equations in the Physics
literature. See \cite[Theorem 2.39]{BruPedra2}. If $\mathcal{E}(\mathit{%
\Omega })\nsubseteq \mathcal{E}(E_{1})$ (cf. \cite[Lemma 9.8]{BruPedra2})
then the above probability measure $\mu _{\omega }$ is supported on a set of
states that do not a priori satisfy these equations, which turn out to be
essential in the present study.\medskip

Problems (a)--(c) are direct consequences of the presence of mean-field
interactions. To properly deal with all of them, Theorem \ref{Thm cool} is a
key technical result showing that the unique orthogonal probability measure $%
\mu _{\omega }$ representing a generalized equilibrium state $\omega \in
\mathit{\Omega }\subseteq E_{1}$ in terms of ergodic states, via Equation (%
\ref{decomp2}), is supported on the set $\mathbf{M}$ (of minimizers of the
free energy density functional of short-range Bogoliubov approximations),
i.e.,
\begin{equation}
\mu _{\omega }\left( \mathcal{E}(E_{1})\cap \mathbf{M}\right) =1.
\label{dfdf}
\end{equation}%
See Observation 2.

Recall now Observation 1, i.e., the fact that, in presence of mean-field
interactions, there is generally no well-defined infinite volume dynamics on
the corresponding $C^{\ast }$-algebra, but rather a state-dependent one. A
first very natural way to get around this problem is to weaken the KMS
property by imposing it only \textquotedblleft fiberwise\textquotedblright ,
in the ergodic decomposition (\ref{decomp2}). By doing so, Problems (a)--(b)
are solved. In order to solve Problem (c), we consider the set $\mathrm{B}$
of all Bogoliubov states of the given long-range model, which are defined
here as being states $\rho \in E_{1}$ satisfying the above mentioned
Euler-Lagrange equations \textquotedblleft fiberwise\textquotedblright ,
i.e., for any $\hat{\rho}\in E_{1}$ these equations are $\mu _{\rho }$%
-almost surely satisfied for some $d_{\hat{\rho}}\in \mathcal{C}$. In this
way we get an extension of \cite[Theorem 1]{Araki} (quantum spins) and \cite[%
Corollary 6.7, Proposition 12.1, Theorems 7.5 and 12.11]{Araki-Moriya}
(lattice fermions) to a very general class of quantum lattice systems with
mean-field, or long-range, interactions:

\begin{theorem}
$\mathit{\Omega }\cap \mathrm{B}=\mathrm{K}\cap \mathrm{B}$, where $\mathrm{K%
}\subseteq E_{1}$ is the set of states $\rho \in E_{1}$ such that $\hat{\rho}%
\in E_{1}$ is $\mu _{\rho }$-almost surely a KMS state for the strongly
continuous group of $\ast $-automorphisms generated by the approximating
interaction $\Phi (d_{\hat{\rho}})$ for some $d_{\hat{\rho}}\in \mathcal{C}$.
\end{theorem}

\noindent This assertion corresponds to Theorem \ref{lr - equilibrium and
kms states (1)}. Meanwhile, Theorem \ref{Thm cool}, i.e., Equation (\ref%
{dfdf}), allows us to identify a very large subclass of quantum lattice
systems with mean-field interactions for which the Choquet decompositions of
generalized equilibrium states $\omega $ in $\mathit{\Omega }$ and $E_{1}$
are identical, i.e., $\nu _{\omega }=\mu _{\omega }$ and $\mathcal{E}(%
\mathit{\Omega })\subseteq \mathcal{E}(E_{1})$ in Equations (\ref{decomp1})
and (\ref{decomp2}):%
\begin{equation}
\omega \left( \cdot \right) =\int_{\mathcal{E}(E_{1})}\hat{\omega}\left(
\cdot \right) \mu _{\omega }\left( \mathrm{d}\hat{\omega}\right) =\int_{%
\mathcal{E}(\mathit{\Omega })}\hat{\omega}\left( \cdot \right) \mu _{\omega
}\left( \mathrm{d}\hat{\omega}\right) \ .  \label{ergodic decomposition}
\end{equation}%
In this case, Problems (a)--(c) \emph{disappear} and one then obtains the
following:

\begin{corollary}
$\mathit{\Omega }=\mathrm{K}\cap \mathrm{B}$.
\end{corollary}

\noindent This assertion refers to Corollary \ref{lr - equilibrium and kms
states (2)} and shows that generalized equilibrium states are \emph{%
\textquotedblleft fiberwise\textquotedblright } KMS\ states. The subclass of
quantum lattice systems used in this corollary basically includes all
lattice fermion and quantum spin systems of condensed matter physics with
mean-field interactions -- at least to our knowledge.

In order to get the usual (global) KMS property for generalized equilibrium
states one shall first use the cyclic representations associated with these
states, to make sense of the infinite volume dynamics, by \cite[Theorem 4.3]%
{BruPedra-MFIII}. We apply this procedure to the above subclass of quantum
lattice systems. In this case, for any generalized equilibrium state $\omega
\in \mathit{\Omega }$, whose associated cyclic representation of the $%
C^{\ast }$-algebra $\mathcal{U}$ is denoted by $(\mathcal{H}_{\omega },\pi
_{\omega },\Omega _{\omega })$, we show in Proposition \ref{long-range dyn
on gen equilibrium states} the existence of a $\sigma $-weakly continuous
group $\mathbf{\Lambda }^{\omega }\equiv (\mathbf{\Lambda }_{t}^{\omega
})_{t\in \mathbb{R}}$ of $\ast $-automorphisms of the von Neumann algebra $%
\pi _{\omega }(\mathcal{U})^{\prime \prime }$, which results from the
infinite volume limit of the full (long-range) dynamics in this
representation. Then, we get the following statement:

\begin{theorem}
The normal extension $\tilde{\omega}$ of the generalized equilibrium state $%
\omega \in \mathit{\Omega }$ to the von Neumann algebra $\pi _{\omega }(%
\mathcal{U})^{\prime \prime }$ is a KMS state for the $\sigma $-weakly
continuous group $\mathbf{\Lambda }^{\omega }$.
\end{theorem}

\noindent This assertion refers to Theorem \ref{gen eq states as KMS states}%
. It represents another natural extension of \cite[Theorem 1]{Araki}
(quantum spins) and \cite[Corollary 6.7 and Theorem 12.11]{Araki-Moriya}
(lattice fermions) to quantum lattice systems with mean-field, or
long-range, interactions.

In our opinion, these results improve the mathematical status of mean-field
quantum models, like the BCS model of (conventional) superconductivity
theory, as they allow to use well-known powerful machineries, like the
Tomita-Takesaki modular theory \cite[Section 2.5]{BrattelliRobinsonI} and
the KMS theory \cite[Sections 5.3-5.4]{BrattelliRobinson}. For instance, by\
Corollary \ref{modular group}, in the cyclic representation $(\mathcal{H}%
_{\omega },\pi _{\omega },\Omega _{\omega })$ of any generalized equilibrium
state $\omega \in \mathit{\Omega }$, the infinite volume limit of the full
dynamics is nothing else than the (time-rescaled) modular $\ast $%
-automorphism group associated with (the cyclic and separating vector) $%
\Omega _{\omega }$ and the von Neumann algebra $\pi _{\omega }(\mathcal{U}%
)^{\prime \prime }$, just as in the short-range case.

To conclude, the paper is organized as follows: Section \ref{Section FERMI0}
presents the general mathematical framework. Observe that we focus on
lattice fermion systems. See Remark \ref{Quantum spin systems}. In Section %
\ref{Section Banach space interaction}, we give a brief account on the
theory of fermion systems on the lattice with short-range interactions,
which is used as a springboard to introduce the theory of mean-field, or
long-range, models in Section \ref{Long-Range Models}. Note that the
mathematical setting used in the current paper is -- up to minor
modifications -- the one of \cite{BruPedra-MFII,BruPedra2}, including the
notation. We thus provide it in a concise way. The main results, i.e.,
Corollary \ref{lr - equilibrium and kms states (2)}, Theorems \ref{lr -
equilibrium and kms states (1)} and \ref{gen eq states as KMS states}, are
finally given in Section \ref{mainresult}.

\begin{remark}[Quantum spin systems]
\label{Quantum spin systems}\mbox{ }\newline
Our study focuses on lattice fermion systems, which are, from a technical
point of view, slightly more difficult than quantum spin systems, because of
a non-commutativity issue at different lattice sites. However, all the
results presented here hold true for quantum spin systems, via obvious
modifications.
\end{remark}

\begin{remark}[Periodic quantum lattice systems]
\label{Periodic quantum lattice systems}\mbox{ }\newline
Our study focuses on lattice fermion systems that are translation invariant
(in space). However, all the results presented here hold true for
(space-)periodic lattice fermion systems, by appropriately redefining the
spin set\footnote{%
In fact, one can see the lattice points in a (space) period as a single
point in an equivalent lattice on which particles have an enlarged spin set.}%
. A similar argument holds true for quantum spin systems.
\end{remark}

\begin{remark}[Inverse temperature of quantum lattice systems]
\label{remark temp}\mbox{ }\newline
In all the paper, we fix once and for all the inverse temperature of quantum
systems via the strictly positive parameter $\beta \in \mathbb{R}^{+}$. This
parameter is usually not referred to in the notation, unless it is important
for the reader's comprehension.
\end{remark}

\section{Algebraic Formulation of Lattice Fermion Systems\label{Section
FERMI0}}

\subsection{CAR Algebra for Lattice Fermions\label{Algebra of Lattices}}

\subsubsection{Background Lattice}

For some dimension $d\in \mathbb{N}$, which is fixed once and for all in the
sequel, let $\mathfrak{L}\doteq \mathbb{Z}^{d}$ and $\mathcal{P}%
_{f}\subseteq 2^{\mathfrak{L}}$ be the set of all non-empty finite subsets
of $\mathfrak{L}$. In order to define the thermodynamic limit, we use the
cubic boxes%
\begin{equation}
\Lambda _{L}\doteq \{(x_{1},\ldots ,x_{d})\in \mathfrak{L}:|x_{1}|,\ldots
,|x_{d}|\leq L\}\in \mathcal{P}_{f}\ ,\qquad L\in \mathbb{N},
\label{eq:def lambda n}
\end{equation}%
as a so-called van Hove sequence. We also fix, once and for all, a
positive-valued symmetric function $\mathbf{F}:\mathfrak{L}^{2}\rightarrow
(0,1]$ satisfying $\mathbf{F}\left( x,x\right) =1$ for all $x\in \mathfrak{L}
$,
\begin{equation}
\left\Vert \mathbf{F}\right\Vert _{1,\mathfrak{L}}\doteq \underset{y\in
\mathfrak{L}}{\sup }\sum_{x\in \mathfrak{L}}\mathbf{F}\left( x,y\right) \in %
\left[ 1,\infty \right) \ ,  \label{(3.1) NS}
\end{equation}%
as well as%
\begin{equation}
\mathbf{D}\doteq \underset{x,y\in \mathfrak{L}}{\sup }\sum_{z\in \mathfrak{L}%
}\frac{\mathbf{F}\left( x,z\right) \mathbf{F}\left( z,y\right) }{\mathbf{F}%
\left( x,y\right) }<\infty \ .  \label{(3.2) NS}
\end{equation}%
Examples of functions satisfying all these conditions, are given by
\begin{equation*}
\mathbf{F}\left( x,y\right) =\mathrm{e}^{-\varsigma \left\vert
x-y\right\vert }(1+\left\vert x-y\right\vert )^{-(d+\epsilon )}\ ,\qquad
x,y\in \mathfrak{L}^{2},
\end{equation*}%
for every positive parameter $\varsigma \in \mathbb{R}_{0}^{+}$ and $%
\epsilon \in \mathbb{R}^{+}$.

\subsubsection{The CAR $C^{\ast }$-Algebra}

For any subset $\Lambda \subseteq \mathfrak{L}$, $\mathcal{U}_{\Lambda }$
denotes the universal unital $C^{\ast }$-algebra generated by elements $%
\{a_{x,\mathrm{s}}\}_{x\in \Lambda ,\mathrm{s}\in \mathrm{S}}$ satisfying
the canonical anti-commutation relations (CAR):
\begin{equation*}
\left\{
\begin{array}{ccc}
a_{x,\mathrm{s}}a_{y,\mathrm{t}}+a_{y,\mathrm{t}}a_{x,\mathrm{s}} & = & 0 \\
a_{x,\mathrm{s}}^{\ast }a_{y,\mathrm{t}}+a_{y,\mathrm{t}}a_{x,\mathrm{s}%
}^{\ast } & = & \delta _{x,y}\delta _{\mathrm{s},\mathrm{t}}\mathfrak{1}%
\end{array}%
\right. ,\text{\qquad }x,y\in \Lambda ,\text{ }\mathrm{s},\mathrm{t}\in
\mathrm{S},
\end{equation*}%
where $\mathfrak{1}$ stands for the unit of the algebra, $\delta _{\cdot
,\cdot }$ is the Kronecker delta and $\mathrm{S}$\ is some finite set
(representing an orthonormal basis of spin modes), which is fixed once and
for all. We use the notation
\begin{equation}
\left\vert A\right\vert ^{2}\doteq A^{\ast }A,\qquad A\in \mathcal{U}%
_{\Lambda },\ \Lambda \subseteq \mathfrak{L},  \label{carre de A}
\end{equation}%
to shorten the equations, in particular in the context of long-range models.

By identifying the generators $\{a_{x,\mathrm{s}}\}_{x\in \Lambda \cap
\Lambda ^{\prime },\mathrm{s}\in \mathrm{S}}$ of both $C^{\ast }$-algebras $%
\mathcal{U}_{\Lambda }$ and $\mathcal{U}_{\Lambda ^{\prime }}$, $\{\mathcal{U%
}_{\Lambda }\}_{\Lambda \in 2^{\mathfrak{L}}}$ is a net of $C^{\ast }$%
-algebras with respect to inclusion: For all subsets $\Lambda ,\Lambda
^{\prime }\subseteq \mathfrak{L}$ so that $\Lambda \subseteq \Lambda
^{\prime }$, one has $\mathcal{U}_{\Lambda }\subseteq \mathcal{U}_{\Lambda
^{\prime }}$. For $\Lambda =\mathfrak{L}$ we use the notation $\mathcal{U}%
\equiv \mathcal{U}_{\mathfrak{L}}$. Observe additionally that the subspace
\begin{equation}
\mathcal{U}_{0}\doteq \bigcup_{\Lambda \in \mathcal{P}_{f}}\mathcal{U}%
_{\Lambda }\subseteq \mathcal{U}\equiv \mathcal{U}_{\mathfrak{L}}
\label{simple}
\end{equation}%
is a dense $\ast $-algebra of the CAR $C^{\ast }$-algebra $\mathcal{U}$ of
the infinite lattice. In particular, $\mathcal{U}$\ is separable, because $%
\mathcal{U}_{\Lambda }$ has finite dimension for all (finite subsets) $%
\Lambda \in \mathcal{P}_{f}$ and the set $\mathcal{P}_{f}$\ is countable.
Elements of $\mathcal{U}_{0}$ are called local elements of $\mathcal{U}$.
The (real) Banach subspace of all self-adjoint elements of $\mathcal{U}$ is
denoted by $\mathcal{U}^{\mathbb{R}}\varsubsetneq \mathcal{U}$.

The local causality of quantum field theory is broken in CAR algebras and
physical quantities are therefore defined from the subalgebra of even
elements, which are defined as follows: Given a fixed parameter $\theta \in
\mathbb{R}/(2\pi \mathbb{Z)}$, the condition
\begin{equation}
\mathrm{g}_{\theta }(a_{x,\mathrm{s}})=\mathrm{e}^{-i\theta }a_{x,\mathrm{s}%
}\ ,\qquad x\in \mathbb{Z}^{d},\ \mathrm{s}\in \mathrm{S},
\label{automorphism gauge invariance}
\end{equation}%
defines a unique $\ast $-automorphism $\mathrm{g}_{\theta }$ of the $C^{\ast
}$-algebra $\mathcal{U}$. Note that, for any $\Lambda \subseteq \mathfrak{L}$%
, $\mathrm{g}_{\theta }(\mathcal{U}_{\Lambda })\subseteq \mathcal{U}%
_{\Lambda }$ and thus $\mathrm{g}_{\theta }$ canonically defines a $\ast $%
-automorphism of the subalgebra $\mathcal{U}_{\Lambda }$. A special role is
played by $\mathrm{g}_{\pi }$. Elements $A,B\in \mathcal{U}_{\Lambda }$, $%
\Lambda \subseteq \mathfrak{L}$, satisfying $\mathrm{g}_{\pi }(A)=A$ and $%
\mathrm{g}_{\pi }(B)=-B$ are respectively called even and odd. (Elements $%
A\in \mathcal{U}_{\Lambda }$ satisfying $\mathrm{g}_{\theta }(A)=A$ for any $%
\theta \in \mathbb{R}/(2\pi \mathbb{Z)}$ are called gauge invariant.) The
space of even elements of $\mathcal{U}$ is denoted by
\begin{equation}
\mathcal{U}^{+}\doteq \{A\in \mathcal{U}:A=\mathrm{g}_{\pi }(A)\}\subseteq
\mathcal{U}\text{ }.  \label{definition of even operators}
\end{equation}%
In fact, it is a $C^{\ast }$-subalgebra of the $C^{\ast }$-algebra $\mathcal{%
U}$. In Physics, $\mathcal{U}^{+}$ is seen as more fundamental than $%
\mathcal{U}$, because of the local causality in quantum field theory, which
holds in the first $C^{\ast }$-algebra, but not in the second one. See,
e.g., discussions in \cite[Section 2.3]{BruPedra-MFII}.

\subsection{States of Lattice Fermion Systems}

\subsubsection{Even States}

States on the $C^{\ast }$-algebra $\mathcal{U}$ are, by definition, linear
functionals $\rho :\mathcal{U}\rightarrow \mathbb{C}$ which are positive,
i.e., for all elements $A\in \mathcal{U}$, $\rho (\left\vert A\right\vert
^{2})\geq 0$, and normalized, i.e., $\rho (\mathfrak{1})=1$. Equivalently,
the linear functional $\rho $ is a state iff $\rho (\mathfrak{1})=1$ and $%
\Vert \rho \Vert _{\mathcal{U}^{\ast }}=1$. The set of all states on $%
\mathcal{U}$ is denoted
\begin{equation}
E\doteq \bigcap\limits_{A\in \mathcal{U}}\{\rho \in \mathcal{U}^{\ast }:\rho
(\mathfrak{1})=1,\;\rho (|A|^{2})\geq 0\}\ .  \label{set of states}
\end{equation}%
This convex set is metrizable and compact with respect to the weak$^{\ast }$
topology. Mutatis mutandis, for every $\Lambda \subseteq \mathfrak{L}$, we
define the set $E_{\Lambda }$ of all states on the sub-algebra $\mathcal{U}%
_{\Lambda }\subseteq \mathcal{U}$. For any $\Lambda \subseteq \mathfrak{L}$,
we use the symbol $\rho _{\Lambda }$ to denote the restriction of any $\rho
\in E$ to the sub-algebra $\mathcal{U}_{\Lambda }$. This restriction is
clearly a state on $\mathcal{U}_{\Lambda }$.

Even states on $\mathcal{U}_{\Lambda }$, $\Lambda \subseteq \mathfrak{L}$,
are, by definition, the states $\rho \in E_{\Lambda }$ satisfying $\rho
\circ \mathrm{g}_{\pi }=\rho $. In other words, the even states\ on $%
\mathcal{U}_{\Lambda }$ are exactly those vanishing on all odd elements of $%
\mathcal{U}_{\Lambda }$. The set of even states on $\mathcal{U}$ can be
canonically identified with the set of states on the $C^{\ast }$-subalgebra $%
\mathcal{U}^{+}$ of even elements, by \cite[Proof of Proposition 2.1]%
{BruPedra-MFII}. As a consequence, physically relevant states on $\mathcal{U}
$ are even.

\subsubsection{Translation-Invariant States\label{Sect Periodic-State Space}}

Lattice translations refer to the group homomorphism $x\mapsto \alpha _{x}$
from $(\mathbb{Z}^{d},+)$ to the group of $\ast $-automorphisms of the CAR $%
C^{\ast }$-algebra $\mathcal{U}$ of the (infinite) lattice $\mathfrak{L}$,
which is uniquely defined by the condition%
\begin{equation}
\alpha _{x}(a_{y,\mathrm{s}})=a_{y+x,\mathrm{s}}\ ,\quad y\in \mathfrak{L},\;%
\mathrm{s}\in \mathrm{S}\text{ }.  \label{transl}
\end{equation}%
This group homomorphism is used to define the translation invariance of
states and interactions of lattice fermion systems.

The state $\rho \in E$ is said to be translation-invariant iff it satisfies $%
\rho \circ \alpha _{x}=\rho $ for all $x\in \mathbb{Z}^{d}$. The space of
translation-invariant states on $\mathcal{U}$ is the convex set%
\begin{equation}
E_{1}\doteq \bigcap\limits_{x\in \mathbb{Z}^{d},\text{ }A\in \mathcal{U}%
}\{\rho \in \mathcal{U}^{\ast }:\rho (\mathfrak{1})=1,\;\rho (|A|^{2})\geq
0,\;\rho =\rho \circ \alpha _{x}\}\ ,  \label{periodic invariant states}
\end{equation}%
which is again metrizable and compact with respect to the weak$^{\ast }$
topology. Any translation-invariant state is even, by \cite[Lemma 1.8]%
{BruPedra2}. Thanks to the Krein-Milman theorem \cite[Theorem 3.23]{Rudin}, $%
E_{1}$ is the weak$^{\ast }$-closure of the convex hull of the (non-empty)
set $\mathcal{E}\left( E_{1}\right) $ of its extreme points, which turns out
to be a weak$^{\ast }$-dense ($G_{\delta }$) subset \cite%
{Bru-pedra-MF-I,BruPedra-MFII}:
\begin{equation}
E_{1}=\overline{\mathrm{co}}\mathcal{E}\left( E_{1}\right) =\overline{%
\mathcal{E}\left( E_{1}\right) }\ ,  \label{cov heull l perio}
\end{equation}%
where $\overline{\mathrm{co}}(K)$ denotes the weak$^{\ast }$-closed convex
hull of a set $K$. This fact is well-known and is also true for quantum spin
systems on lattices \cite[Example 4.3.26 and discussions p. 464]%
{BrattelliRobinsonI}.

Since $E_{1}$ is metrizable (by separability of $\mathcal{U}$), the Choquet
theorem applies: We denote by $\Sigma _{E}$ the (Borel) $\sigma $-algebra
generated by weak$^{\ast }$-open subsets of the set $E$ of all states. $%
\mathcal{E}\left( E_{1}\right) $ is of course a Borel set for it is a $%
G_{\delta }$. Recall meanwhile that any positive functional $\rho \in
\mathcal{U}_{+}^{\ast }$ is associated with a unique (up to unitary
equivalence) cyclic representation $(\mathcal{H}_{\rho },\pi _{\rho },\Omega
_{\rho })$ of $\mathcal{U}$. Two positive linear functionals $\rho _{1},\rho
_{2}\in \mathcal{U}^{\ast }$ are said to be orthogonal whenever
\begin{equation*}
(\mathcal{H}_{\rho _{1}}\oplus \mathcal{H}_{\rho _{2}},\pi _{\rho
_{1}}\oplus \pi _{\rho _{2}},\Omega _{\rho _{1}}\oplus \Omega _{\rho _{2}})
\end{equation*}%
is the cyclic representation of $\mathcal{U}$ associated with the positive
functional $\rho _{1}+\rho _{2}\in \mathcal{U}^{\ast }$. A probability
measure $\mu $ on $E$ is called orthogonal whenever $\rho _{\mu _{\mathfrak{B%
}}}$ and $\rho _{\mu _{E\backslash \mathfrak{B}}}$ are orthogonal for any $%
\mathfrak{B}\in \Sigma _{E}$, where
\begin{equation*}
\rho _{\mu _{\mathfrak{B}_{0}}}\left( A\right) \doteq \int_{\mathfrak{B}%
_{0}}\rho \left( A\right) \mu \left( \mathrm{d}\rho \right) \text{ },\qquad
A\in \mathcal{U},\ \mathfrak{B}_{0}\in \Sigma _{E}\text{ }.
\end{equation*}%
See \cite[Lemma 4.1.19 and Definition 4.1.20]{BrattelliRobinsonI}. With
these definitions we are in a position to state \cite[Theorem 5.1]%
{BruPedra-MFIII} for translation-invariant states, which provides a
(stronger) version of the Choquet theorem (in this particular case):

\begin{theorem}[Ergodic orthogonal decomposition of translation-invariant
states]
\label{theorem choquet}\mbox{ }\newline
For any $\rho \in E_{1}$, there is a unique probability measure\footnote{%
For $E$ is a metrizable compact space, any finite measure is regular and
tight. Thus, here, probabilities measures are just the same as normalized
Borel measures.} $\mu _{\rho }$ on $E$ such that $\mu _{\rho }\left(
\mathcal{E}\left( E_{1}\right) \right) =1$ and%
\begin{equation*}
\rho \left( A\right) =\int_{E_{1}}\hat{\rho}\left( A\right) \mu _{\rho
}\left( \mathrm{d}\hat{\rho}\right) \text{ },\qquad A\in \mathcal{U}\text{ }.
\end{equation*}%
Moreover, $\mu _{\rho }\ $is an orthogonal measure on $(E,\Sigma _{E})$.
\end{theorem}

\noindent In particular, $E_{1}$ is a Choquet simplex. In fact, up to an
affine homeomorphism, $E_{1}$ is the so-called Poulsen simplex \cite[Theorem
1.12]{BruPedra2}.

The unique decomposition of a translation-invariant state $\rho \in E_{1}$
in terms of extreme translation-invariant states $\hat{\rho}\in \mathcal{E}%
(E_{1})$, given in Theorem \ref{theorem choquet}, is also called the \emph{%
ergodic} decomposition of $\rho $ because of the following fact: Define the
space-averages of any element $A\in \mathcal{U}$ by%
\begin{equation}
A_{L}\doteq \frac{1}{|\Lambda _{L}|}\sum\limits_{x\in \Lambda _{L}}\alpha
_{x}\left( A\right) \ ,\mathrm{\qquad }L\in \mathbb{N}\text{ }.
\label{Limit of Space-Averages}
\end{equation}%
Then, by definition, a translation-invariant state $\hat{\rho}\in E_{1}$ is
\emph{ergodic} iff
\begin{equation}
\lim\limits_{L\rightarrow \infty }\hat{\rho}(\left\vert A_{L}\right\vert
^{2})=|\hat{\rho}(A)|^{2}\ ,\mathrm{\qquad }A\in \mathcal{U}\text{ }.
\label{Ergodicity}
\end{equation}%
(Recall Equation (\ref{carre de A}).) By \cite[Theorem 1.16]{BruPedra2}, any
extreme translation-invariant state is ergodic and conversely. In other
words, the set of extreme translation-invariant states is equal to%
\begin{equation}
\mathcal{E}(E_{1})=\left\{ \hat{\rho}\in E_{1}:\hat{\rho}\text{ is ergodic}%
\right\} =\bigcap\limits_{A\in \mathcal{U}}\left\{ \hat{\rho}\in
E_{1}:\lim\limits_{L\rightarrow \infty }\hat{\rho}(\left\vert
A_{L}\right\vert ^{2})=|\hat{\rho}(A)|^{2}\right\} \ .  \label{Ergodicity2}
\end{equation}

\begin{remark}[Periodic states]
\label{periodic states}\mbox{ }\newline
For any given $\vec{\ell}\doteq (\ell _{1},\ldots ,\ell _{d})\in \mathbb{N}%
^{d}$, let $(\mathbb{Z}_{\vec{\ell}}^{d},+)\subseteq (\mathbb{Z}^{d},+)$ be
the subgroup defined by $\mathbb{Z}_{\vec{\ell}}^{d}\doteq \ell _{1}\mathbb{Z%
}\times \cdots \times \ell _{d}\mathbb{Z}$. Any state $\rho \in E$
satisfying $\rho \circ \alpha _{x}=\rho $ for all $x\in \mathbb{Z}_{\vec{\ell%
}}^{d}$ is called $\vec{\ell}$\emph{-}periodic. All properties given above
on translation-invariant states hold true for $\vec{\ell}$\emph{-}periodic
states, via obvious modifications. In particular, the weak$^{\ast }$-compact
convex set of $\vec{\ell}$-periodic states is again the so-called Poulsen
simplex \cite[Theorem 1.12]{BruPedra2}. By \cite[Proposition 2.3]%
{BruPedra-MFII}, the union of all sets of $\vec{\ell}$-periodic states, $%
\vec{\ell}\in \mathbb{N}^{d}$, is a weak$^{\ast }$-dense subset of the set
of all even states, the physically relevant ones.
\end{remark}

\section{Infinite Volume Short-Range Models\label{Section Banach space
interaction}}

\subsection{Short-Range Interactions}

A (complex) \emph{interaction} is, by definition, any mapping $\Phi :%
\mathcal{P}_{f}\rightarrow \mathcal{U}^{+}$ from the set $\mathcal{P}%
_{f}\subseteq 2^{\mathfrak{L}}$ of all non-empty finite subsets of $%
\mathfrak{L}$ to the $C^{\ast }$-subalgebra $\mathcal{U}^{+}$ (\ref%
{definition of even operators}) of even elements such that $\Phi _{\Lambda
}\in \mathcal{U}_{\Lambda }$ for all $\Lambda \in \mathcal{P}_{f}$. The set $%
\mathcal{V}$ of all interactions can be naturally endowed with the structure
of a complex vector space (via the usual point-wise vector space
operations), as well as with the antilinear involution
\begin{equation}
\Phi \mapsto \Phi ^{\ast }\doteq (\Phi _{\Lambda }^{\ast })_{\Lambda \in
\mathcal{P}_{f}}\ .  \label{involution}
\end{equation}%
An interaction $\Phi $ is said to be self-adjoint iff $\Phi =\Phi ^{\ast }$.
The set $\mathcal{V}^{\mathbb{R}}$ of all self-adjoint interactions forms a
real subspace of the space of all (complex) interactions.

The (normed) space of short-range interactions is defined by
\begin{equation}
\mathcal{W}\doteq \left\{ \Phi \in \mathcal{V}:\left\Vert \Phi \right\Vert _{%
\mathcal{W}}<\infty \right\} \text{ },  \label{banach space short range}
\end{equation}%
its norm being defined by%
\begin{equation}
\left\Vert \Phi \right\Vert _{\mathcal{W}}\doteq \underset{x,y\in \mathfrak{L%
}}{\sup }\sum\limits_{\Lambda \in \mathcal{P}_{f},\;\Lambda \supseteq
\{x,y\}}\frac{\Vert \Phi _{\Lambda }\Vert _{\mathcal{U}}}{\mathbf{F}\left(
x,y\right) }\ ,\qquad \Phi \in \mathcal{V}\text{ },  \label{iteration0}
\end{equation}%
where $\mathbf{F}$ is the positive-valued symmetric function introduced in
Section \ref{Algebra of Lattices}. $(\mathcal{W},\left\Vert \cdot
\right\Vert _{\mathcal{W}})$ is a separable Banach space. The (real) Banach
subspace of self-adjoint short-range interactions is denoted
\begin{equation*}
\mathcal{W}^{\mathbb{R}}\doteq \mathcal{V}^{\mathbb{R}}\cap \mathcal{W}\ ,
\end{equation*}%
similar to $\mathcal{U}^{\mathbb{R}}\varsubsetneq \mathcal{U}$ and $\mathcal{%
V}^{\mathbb{R}}\varsubsetneq \mathcal{V}$.

By definition, the interaction $\Phi \in \mathcal{V}$ is
translation-invariant\ iff%
\begin{equation}
\Phi _{\Lambda +x}=\alpha _{x}\left( \Phi _{\Lambda }\right) \ ,\qquad x\in
\mathbb{Z}^{d},\ \Lambda \in \mathcal{P}_{f}\text{ },  \label{ti interaction}
\end{equation}%
where $\{\alpha _{x}\}_{x\in \mathbb{Z}^{d}}$ is the family of (translation)
$\ast $-automorphisms of $\mathcal{U}$ defined by (\ref{transl}), while%
\begin{equation}
\Lambda +x\doteq \left\{ y+x\in \mathfrak{L}:y\in \Lambda \right\} \ ,\qquad
x\in \mathbb{Z}^{d},\ \Lambda \in \mathcal{P}_{f}\text{ }.
\label{translation box}
\end{equation}%
The (separable) Banach subspace of translation-invariant and short-range
interactions of $\mathcal{V}$ is denoted
\begin{equation}
\mathcal{W}_{1}\doteq \bigcap\limits_{x\in \mathbb{Z}^{d},\ \Lambda \in
\mathcal{P}_{f}}\left\{ \Phi \in \mathcal{W}:\Phi _{\Lambda +x}=\alpha
_{x}\left( \Phi _{\Lambda }\right) \right\} \varsubsetneq \mathcal{W}%
\varsubsetneq \mathcal{V}\ .  \label{W1}
\end{equation}%
Finally, the (real) Banach subspace of interactions that are simultaneously
self-adjoint, translation-invariant and short-range, is denoted
\begin{equation}
\mathcal{W}_{1}^{\mathbb{R}}\doteq \mathcal{V}^{\mathbb{R}}\cap \mathcal{W}%
_{1}\subseteq \mathcal{W}^{\mathbb{R}}\ ,  \label{W1Re}
\end{equation}%
similar to $\mathcal{U}^{\mathbb{R}}\varsubsetneq \mathcal{U}$, $\mathcal{V}%
^{\mathbb{R}}\varsubsetneq \mathcal{V}$ and $\mathcal{W}^{\mathbb{R}%
}\varsubsetneq \mathcal{W}$.

\subsection{Dynamics Generated by Short-Range Interactions\label{sect
Lieb--Robinson}}

\subsubsection{Limit Derivations}

Local energy elements associated with a given complex interaction $\Phi \in
\mathcal{V}$ correspond to the following sequence within the $C^{\ast }$%
-subalgebra $\mathcal{U}^{+}$ (\ref{definition of even operators}) of even
elements:%
\begin{equation}
U_{L}^{\Phi }\doteq \sum\limits_{\Lambda \subseteq \Lambda _{L}}\Phi
_{\Lambda }\in \mathcal{U}_{\Lambda _{L}}\cap \mathcal{U}^{+}\ ,\qquad L\in
\mathbb{N}\text{ },  \label{equation fininte vol dynam0}
\end{equation}%
where we recall that $\Lambda _{L}$, $L\in \mathbb{N}$, are the cubic boxes (%
\ref{eq:def lambda n}) used to define the thermodynamic limit. If $\Phi \in
\mathcal{V}^{\mathbb{R}}$ is self-adjoint then $(U_{L}^{\Phi })_{L\in
\mathbb{N}}$ is a sequence (of local Hamiltonians) in $\mathcal{U}^{\mathbb{R%
}}$. By straightforward estimates using Equations (\ref{(3.1) NS}) and (\ref%
{iteration0}), note that, for arbitrary short-range interactions $\Phi ,\Psi
\in \mathcal{W}$,
\begin{equation}
\left\Vert U_{L}^{\Phi }-U_{L}^{\Psi }\right\Vert _{\mathcal{U}}=\left\Vert
U_{L}^{\Phi -\Psi }\right\Vert _{\mathcal{U}}\leq \left\vert \Lambda
_{L}\right\vert \left\Vert \mathbf{F}\right\Vert _{1,\mathfrak{L}}\left\Vert
\Phi -\Psi \right\Vert _{\mathcal{W}}\ ,\qquad L\in \mathbb{N}\text{ }.
\label{norm Uphi}
\end{equation}

The sequence $(\delta _{L}^{\Phi })_{L\in \mathbb{N}}\subseteq \mathcal{B}(%
\mathcal{U})$ of local derivations associated with any given interaction $%
\Phi \in \mathcal{V}$ is then defined by%
\begin{equation*}
\delta _{L}^{\Phi }(A)\doteq i\left[ U_{L}^{\Phi },A\right] \doteq i\left(
U_{L}^{\Phi }A-AU_{L}^{\Phi }\right) \ ,\qquad A\in \mathcal{U},\ L\in
\mathbb{N}\text{ }.
\end{equation*}%
If $\Phi \in \mathcal{V}^{\mathbb{R}}$ is self-adjoint then $(\delta
_{L}^{\Phi })_{L\in \mathbb{N}}$ is a sequence of symmetric derivations (or $%
\ast $-derivations). By \cite[Corollary 3.5]{BruPedra-MFII}, for any
short-range interaction $\Phi \in \mathcal{W}$ and every local element $A\in
\mathcal{U}_{0}$, the limit%
\begin{equation}
\delta ^{\Phi }\left( A\right) \doteq \lim_{L\rightarrow \infty }\delta
_{L}^{\Phi }\left( A\right) \doteq \sum\limits_{\Lambda \in \mathcal{P}_{f}}%
\left[ \Phi _{\Lambda },A\right]  \label{****}
\end{equation}%
exists and defines a (densely defined) derivation $\delta ^{\Phi }$ on the $%
C^{\ast }$ algebra $\mathcal{U}$, whose domain is the dense $\ast $-algebra $%
\mathcal{U}_{0}\subseteq \mathcal{U}$ of local elements defined by (\ref%
{simple}). Additionally, $\delta ^{\Phi }$ is symmetric (or a $\ast $%
-derivation) whenever $\Phi $ is self-adjoint, i.e., $\Phi \in \mathcal{W}^{%
\mathbb{R}}$.

\subsubsection{Limit Short-Range Dynamics in the Heisenberg Picture\label%
{sect Lieb--Robinson copy(2)}}

We now consider time-dependent interactions. Let $\Psi \in C(\mathbb{R};%
\mathcal{W})$ be a continuous function from $\mathbb{R}$ to the Banach space
$\mathcal{W}$ of short-range interactions. Then, for any $L\in \mathbb{N}$,
there is a unique (fundamental) solution $(\tau _{t,s}^{(L,\Psi
)})_{_{s,t\in \mathbb{R}}}$ in $\mathcal{B}(\mathcal{U})$ to the (finite
volume) non-auto%
\-%
nomous evolution equations%
\begin{equation}
\forall s,t\in {\mathbb{R}}:\qquad \partial _{s}\tau _{t,s}^{(L,\Psi
)}=-\delta _{L}^{\Psi \left( s\right) }\circ \tau _{t,s}^{(L,\Psi )}\
,\qquad \tau _{t,t}^{(L,\Psi )}=\mathbf{1}_{\mathcal{U}}\ ,  \label{cauchy1}
\end{equation}%
and%
\begin{equation}
\forall s,t\in {\mathbb{R}}:\qquad \partial _{t}\tau _{t,s}^{(L,\Psi )}=\tau
_{t,s}^{(L,\Psi )}\circ \delta _{L}^{\Psi \left( t\right) }\ ,\qquad \tau
_{s,s}^{(L,\Psi )}=\mathbf{1}_{\mathcal{U}}\ .  \label{cauchy2}
\end{equation}%
In these equations, $\mathbf{1}_{\mathcal{U}}$ refers to the identity
mapping from $\mathcal{U}$ to itself. Note also that, for any $L\in \mathbb{N%
}$ and $\Psi \in C\left( \mathbb{R};\mathcal{W}\right) $, $(\tau
_{t,s}^{(L,\Psi )})_{_{s,t\in \mathbb{R}}}$ is a continuous two-para%
\-%
meter family of bounded operators that satisfies the (reverse) cocycle
property%
\begin{equation*}
\tau _{t,s}^{(L,\Psi )}=\tau _{r,s}^{(L,\Psi )}\tau _{t,r}^{(L,\Psi )}\
,\qquad s,r,t\in \mathbb{R}\text{ }.
\end{equation*}%
If $\Psi \in C(\mathbb{R};\mathcal{W}^{\mathbb{R}})$ then $\delta _{L}^{\Psi
\left( t\right) }$ is always a symmetric derivation and thus, in this case, $%
\tau _{t,s}^{(L,\Psi )}$ is a $\ast $-auto%
\-%
morphism of $\mathcal{U}$ for all lengths $L\in \mathbb{N}$ and times $%
s,t\in \mathbb{R}$.

By \cite[Proposition 3.7]{BruPedra-MFII}, in the thermodynamic limit $%
L\rightarrow \infty $, for any fixed $\Psi \in C(\mathbb{R};\mathcal{W}^{%
\mathbb{R}})$, $(\tau _{t,s}^{(L,\Psi )})_{s,t\in {\mathbb{R}}}$ converges
strongly, uniformly for $s,t$ on compacta, to a strongly continuous two-para%
\-%
meter family $(\tau _{t,s}^{\Psi })_{s,t\in {\mathbb{R}}}$ of $\ast $-auto%
\-%
morphisms of $\mathcal{U}$, which is the unique solution in $\mathcal{B}(%
\mathcal{U})$ to the non-auto%
\-%
nomous evolutions equation
\begin{equation}
\forall s,t\in {\mathbb{R}}:\qquad \partial _{t}\tau _{t,s}^{\Psi }=\tau
_{t,s}^{\Psi }\circ \delta ^{\Psi \left( t\right) }\ ,\qquad \tau
_{s,s}^{\Psi }=\mathbf{1}_{\mathcal{U}}\ ,  \label{cauchy trivial1}
\end{equation}%
in the strong sense on the dense $\ast $-algebra $\mathcal{U}_{0}\subseteq
\mathcal{U}$ of local elements defined by (\ref{simple}). In particular, it
satisfies the reverse cocycle property:%
\begin{equation}
\tau _{t,s}^{\Psi }=\tau _{r,s}^{\Psi }\tau _{t,r}^{\Psi }\ ,\qquad s,r,t\in
\mathbb{R}\text{ }.  \label{reverse cocycle}
\end{equation}%
This refers to a non-autonomous limit dynamics in the Heisenberg picture of
quantum mechanics. Taking a constant self-adjoint short-range interaction
\begin{equation*}
\Phi \in \mathcal{W}^{\mathbb{R}}\subseteq C(\mathbb{R};\mathcal{W}^{\mathbb{%
R}})
\end{equation*}%
and fixing $s=0$, we obtain a $C_{0}$-group $\tau ^{\Phi }\equiv (\tau
_{t}^{\Phi })_{t\in {\mathbb{R}}}$ of $\ast $-auto%
\-%
morphisms of $\mathcal{U}$. This refers now to an autonomous limit dynamics,
again in the Heisenberg picture of quantum mechanics.

\subsection{Equilibrium States of Short-Range Interactions\label{sect
Lieb--Robinson copy(1)}}

\subsubsection{Energy Density Functional on Translation-Invariant States
\label{energy density1}}

The energy density of a state $\rho \in E$ with respect to a given
interaction $\Phi \in \mathcal{V}$ is defined by
\begin{equation*}
e_{\Phi }\left( \rho \right) \doteq \underset{L\rightarrow \infty }{\lim
\sup }\frac{\mathrm{Re}\{\rho \left( U_{L}^{\Phi }\right) \}}{\left\vert
\Lambda _{L}\right\vert }+i\ \underset{L\rightarrow \infty }{\lim \sup }%
\frac{\mathrm{Im}\{\rho \left( U_{L}^{\Phi }\right) \}}{\left\vert \Lambda
_{L}\right\vert }\in (-\infty ,\infty ]+i(-\infty ,\infty ]\ ,
\end{equation*}%
$\Lambda _{L}$, $L\in \mathbb{N}$, being the cubic boxes (\ref{eq:def lambda
n}). Observe from Inequality (\ref{norm Uphi}) that
\begin{equation}
\left\vert e_{\Phi }\left( \rho \right) -e_{\Psi }\left( \rho \right)
\right\vert \leq \left\Vert \mathbf{F}\right\Vert _{1,\mathfrak{L}%
}\left\Vert \Phi -\Psi \right\Vert _{\mathcal{W}}\ ,\qquad \Phi ,\Psi \in
\mathcal{W}\ .  \label{inequality a la con}
\end{equation}%
By \cite[Proposition 3.2]{BruPedra-MFII}, for any translation-invariant
state $\rho \in E_{1}$ (\ref{periodic invariant states}) and each
translation-invariant short-range interaction$\ \Phi \in \mathcal{W}_{1}$,
\begin{equation}
e_{\Phi }\left( \rho \right) =\lim\limits_{L\rightarrow \infty }\frac{\rho
\left( U_{L}^{\Phi }\right) }{\left\vert \Lambda _{L}\right\vert }=\rho (%
\mathfrak{e}_{\Phi })\ ,  \label{ssssssssss}
\end{equation}%
where $\mathfrak{e}_{(\cdot )}:\mathcal{W}\rightarrow \mathcal{U}$ is the
continuous mapping from the Banach space $\mathcal{W}$ to the CAR algebra $%
\mathcal{U}$, defined by
\begin{equation}
\mathfrak{e}_{\Phi }\doteq \sum\limits_{\mathcal{Z}\in \mathcal{P}_{f},\;%
\mathcal{Z}\ni 0}\frac{\Phi _{\mathcal{Z}}}{\left\vert \mathcal{Z}%
\right\vert }\in \mathcal{U}\ ,\text{\qquad }\Phi \in \mathcal{W}\text{ }.
\label{eq:enpersite}
\end{equation}%
In particular, for any fixed $\Phi \in \mathcal{W}_{1}$, the mapping $\rho
\mapsto e_{\Phi }(\rho )$ from $E_{1}$ to $\mathbb{C}$ is a weak$^{\ast }$%
-continuous affine functional. Given any fixed translation-invariant state $%
\rho \in E_{1}$, the linear mapping $\Phi \mapsto e_{\Phi }\left( \rho
\right) $ from $\mathcal{W}_{1}$ to $\mathbb{C}$ is continuous, by (\ref%
{inequality a la con}). Note also that, for all $\rho \in E_{1}$, $e_{\Phi
}(\rho )$ is a real number, whenever the interaction $\Phi \in \mathcal{V}$
is short-range, translation-invariant and self-adjoint, i.e., $\Phi \in
\mathcal{W}_{1}^{\mathbb{R}}$.

\subsubsection{Entropy Density Functional on Translation-Invariant States
\label{energy density2}}

The entropy density functional $s:E_{1}\rightarrow \mathbb{R}_{0}^{+}$ is
the von Neumann entropy per unit volume in the thermodynamic limit, that is,%
\begin{equation*}
s(\rho )\doteq -\lim\limits_{L\rightarrow \infty }\left\{ \frac{1}{|\Lambda
_{L}|}\mathrm{Trace}\,\left( \mathrm{d}_{\rho _{\Lambda _{L}}}\ln \mathrm{d}%
_{\rho _{\Lambda _{L}}}\right) \right\} \ ,\qquad \rho \in E_{1}\text{ },
\end{equation*}%
where we recall that $\rho _{\Lambda _{L}}$ is the restriction of the
translation-invariant state $\rho \in E_{1}$ to the finite-dimensional CAR
algebra $\mathcal{U}_{\Lambda _{L}}$ of the cubic box $\Lambda _{L}$ defined
by (\ref{eq:def lambda n}). Here, $\mathrm{d}_{\rho _{\Lambda _{L}}}\in
\mathcal{U}_{\Lambda _{L}}$ is the (uniquely defined) density matrix
representing the state $\rho _{\Lambda _{L}}$ via a trace\footnote{%
For $\Lambda \in \mathcal{P}_{f}$, the trace on the finite-dimensional $%
C^{\ast }$-algebra $\mathcal{U}_{\Lambda }$ refers to the usual trace on the
fermionic Fock space representation.}:
\begin{equation*}
\rho _{\Lambda _{L}}(\cdot )=\mathrm{Trace}\,\left( \;\cdot \;\mathrm{d}%
_{\rho _{\Lambda _{L}}}\right) \ .
\end{equation*}%
By \cite[Lemma 4.15]{BruPedra2}, the functional $s$ is well-defined on the
set $E_{1}$ of translation-invariant states. See also \cite[Section 10.2]%
{Araki-Moriya}. By \cite[Lemma 1.29]{BruPedra2}, the entropy density
functional $s$ is a weak$^{\ast }$-upper semi-continuous affine functional.

\subsubsection{Equilibrium States as Minimizers of the Free Energy Density
\label{Translation Invariant Equilibrium States}}

Equilibrium states of lattice fermion systems are always defined in relation
to a fixed self-adjoint interaction, which determines the energy density of
states as well as the microscopic dynamics. Here, we define equilibrium
states as minimizers of the free energy density functional, in direct
relation with the notion of (grand-canonical) pressure: For a fixed $\beta
\in \mathbb{R}^{+}$, the infinite volume pressure $\mathrm{P}$ is the
real-valued function on the real Banach subspace $\mathcal{W}_{1}^{\mathbb{R}%
}$ (\ref{W1Re}) of interactions that are self-adjoint, translation-invariant
and short-range, defined by%
\begin{equation*}
\Phi \mapsto \mathrm{P}_{\Phi }\doteq \underset{L\rightarrow \infty }{\lim }%
\frac{1}{\beta |\Lambda _{L}|}\ln \mathrm{Trace}(\mathrm{e}^{-\beta
U_{L}^{\Phi }})\ .
\end{equation*}%
Recall that the parameter $\beta \in \mathbb{R}^{+}$ is the inverse
temperature of the system. It is fixed once and for all and, therefore, it
is often omitted in our discussions or notation, see Remark \ref{remark temp}%
. By \cite[Theorem 2.12]{BruPedra2}, the above pressure is well-defined and,
for any $\Phi \in \mathcal{W}_{1}^{\mathbb{R}}$,%
\begin{equation}
\mathrm{P}_{\Phi }=-\inf f_{\Phi }(E_{1})<\infty \text{ },
\label{pressure free energy}
\end{equation}%
the mapping $f_{\Phi }:E_{1}\rightarrow \mathbb{R}$ being the free energy
density functional defined on the set $E_{1}$ of translation-invariant
states by%
\begin{equation}
f_{\Phi }\doteq e_{\Phi }-\beta ^{-1}s\ .  \label{map free energy}
\end{equation}%
Recall that $e_{\Phi }:E_{1}\rightarrow \mathbb{R}$ is the energy density
functional defined in Section \ref{energy density1} for any $\Phi \in
\mathcal{W}_{1}$, while $s:E_{1}\rightarrow \mathbb{R}_{0}^{+}$ is the
entropy density functional presented in Section \ref{energy density2}.

As explained in Sections \ref{energy density1}--\ref{energy density2}, the
functionals $e_{\Phi }$, $\Phi \in \mathcal{W}_{1}^{\mathbb{R}}$, and $%
-\beta ^{-1}s$, $\beta \in \mathbb{R}^{+}$, are weak$^{\ast }$-lower
semi-continuous and affine. In particular, the functional $f_{\Phi }$ (\ref%
{map free energy}) is weak$^{\ast }$-lower semi-continuous and affine.
Therefore, for any $\Phi \in \mathcal{W}_{1}^{\mathbb{R}}$, this functional
has minimizers in the weak$^{\ast }$-compact set $E_{1}$ of
translation-invariant states. Similarly to what is done for
translation-invariant quantum spin systems (see, e.g., \cite%
{BrattelliRobinson,Sewell}), for any $\Phi \in \mathcal{W}_{1}^{\mathbb{R}}$%
, the set $\mathit{M}_{\Phi }$ of translation-invariant equilibrium states\
of fermions on the lattice is, by definition, the (non-empty) set
\begin{equation}
\mathit{M}_{\Phi }\doteq \left\{ \omega \in E_{1}:f_{\Phi }\left( \omega
\right) =\inf \,f_{\Phi }(E_{1})=-\mathrm{P}_{\Phi }\right\}
\label{minimizer short range}
\end{equation}%
of all minimizers of the free energy density functional $f_{\Phi }$ over the
set $E_{1}$. By affineness and weak$^{\ast }$-lower semi-continuity of $%
f_{\Phi }$, $\mathit{M}_{\Phi }$ is a (non-empty) weak$^{\ast }$-closed face
of $E_{1}$ for any $\Phi \in \mathcal{W}_{1}^{\mathbb{R}}$.

Recall that this is not the only reasonable way of defining equilibrium
states. For fixed interactions, they can also be defined as tangent
functionals to the corresponding pressure functional or via other conditions
like the local stability, the Gibbs condition or the Kubo-Martin-Schwinger
(KMS) condition. All these definitions are generally not completely
equivalent to each other. For instance, the free energy density minimizing
property and the tangent property assume translation invariance (or, at
least, periodicity), whereas other conditions like the KMS\ one are
independent of the space invariance of states. We discuss a well-known
partial result on this question in the following section. For more details,
we recommend the paper \cite{Araki-Moriya}.

\subsubsection{Equilibrium States as Kubo-Martin-Schwinger (KMS) States\label%
{Section KMS States}}

The KMS condition was introduced in 1957 by Kubo and Martin, and by
Schwinger in 1959, in the context of thermodynamic Green's functions. It has
led to the notion of \emph{KMS states}. There are various equivalent
definitions of KMS states, see, e.g., \cite[Sections 5.3-5.4]%
{BrattelliRobinson}. Here we use the following one: Given a\ $C_{0}$-group $%
\tau \equiv (\tau _{t})_{t\in {\mathbb{R}}}$ of $\ast $-auto%
\-%
morphisms of $\mathcal{U}$\ and an inverse temperature $\beta \in \mathbb{R}%
^{+}$, a state $\rho \in E$ is a $(\tau ,\beta )$-KMS\ state iff%
\begin{equation}
\int_{\mathbb{R}}f(t-i\beta )\rho \left( A\tau _{t}(B)\right) \mathrm{d}%
t=\int_{\mathbb{R}}f(t)\rho \left( \tau _{t}(B)A\right) \mathrm{d}t
\label{KMS condition}
\end{equation}%
for all $A,B\in \mathcal{U}$ and any function $f$ being the (holomorphic)
Fourier transform of a smooth function with compact support. See, for
instance, \cite[Equation (13) and Lemma III.3.1 in Chapter III]{Israel}.

Given any fixed $\beta \in \mathbb{R}^{+}$ and self-adjoint short-range
interaction $\Phi \in \mathcal{W}^{\mathbb{R}}$, the set of
translation-invariant $(\tau ^{\Phi },\beta )$-KMS states associated with
the $C_{0}$-group $\tau ^{\Phi }\equiv (\tau _{t}^{\Phi })_{t\in {\mathbb{R}}%
}$ of $\ast $-auto%
\-%
morphisms of $\mathcal{U}$ defined in Section \ref{sect Lieb--Robinson
copy(2)} is denoted by
\begin{equation}
\mathit{K}_{\Phi }\doteq \left\{ \omega \in E_{1}:\omega \text{ is a }(\tau
^{\Phi },\beta )\text{-KMS state}\right\} \ .  \label{set of KMS}
\end{equation}%
One big advantage of the KMS property, as compared to other notions of
equilibrium states like the one presented in Section \ref{Translation
Invariant Equilibrium States}, is that the translation-invariance of states
and interactions is not needed. However, as we are dealing with
translation-invariant equilibrium states only, this generalization is not
further considered here.

The set $\mathit{K}_{\Phi }$ is non-empty, weak$^{\ast }$-compact and
convex, for any $\Phi \in \mathcal{W}^{\mathbb{R}}$. In fact, following \cite%
{Araki-Moriya}, we show below that equilibrium states, as minimizers of the
free energy density functional $f_{\Phi }$, are exactly the
translation-invariant $(\tau ^{\Phi },\beta )$-KMS states.

\begin{theorem}[Equilibrium states as KMS states]
\label{eq.tang.bcs.type0 copy(1)}\mbox{ }\newline
For any $\Phi \in \mathcal{W}_{1}^{\mathbb{R}}$, $\mathit{M}_{\Phi }=\mathit{%
K}_{\Phi }$, see Equations (\ref{minimizer short range}) and (\ref{set of
KMS}).
\end{theorem}

\begin{proof}
Fix $\Phi \in \mathcal{W}_{1}^{\mathbb{R}}$. From Equation (\ref{iteration0}%
),
\begin{equation*}
\sum_{\Lambda \in \mathcal{P}_{f},\;\Lambda \supseteq \{0\}}\Vert \Phi
_{\Lambda }\Vert _{\mathcal{U}}\leq \Vert \Phi \Vert _{\mathcal{W}}<\infty
\end{equation*}%
and it follows from the translation invariance of the interaction $\Phi $
that
\begin{equation*}
H_{L}^{\Phi }\doteq \sum_{\Lambda \in \mathcal{P}_{f},\;\Lambda _{L}\cap
\Lambda \neq \emptyset }\Phi _{\Lambda \text{ }},\qquad L\in \mathbb{N}\text{
},
\end{equation*}%
is well-defined and
\begin{equation*}
\delta ^{\Phi }(A)=i[H_{L}^{\Phi },A]\text{ },\qquad A\in \mathcal{U}%
_{\Lambda _{L}}\text{ },
\end{equation*}%
thanks to Equation (\ref{****}), keeping in mind that $\Phi _{\Lambda }\in
\mathcal{U}_{\Lambda }\cap \mathcal{U}^{+}$ for $\Lambda \in \mathcal{P}_{f}$%
. Then, $\Phi $ is a (translation covariant) general potential in the sense
of \cite[Section 5.5]{Araki-Moriya}, which comes from the dynamics (i.e., $%
C_{0}$-group) $\tau ^{\Phi }$. From \cite[Theorem 5.13]{Araki-Moriya}, there
is a unique standard potential $\tilde{\Phi}$, in the sense of \cite[%
Definition 5.10]{Araki-Moriya}, which is associated with the same dynamics $%
\tau ^{\Phi }$. Additionally, since
\begin{equation*}
\frac{1}{|\Lambda _{L}|}\left\Vert U_{L}^{\Phi }-H_{L}^{\Phi }\right\Vert _{%
\mathcal{U}}\leq \frac{1}{|\Lambda _{L}|}\sum_{\Lambda \in \mathcal{P}%
_{f}:\Lambda \nsubseteq \Lambda _{L},\Lambda \cap \Lambda _{L}\neq \emptyset
}\left\Vert \Phi _{\Lambda }\right\Vert _{\mathcal{U}}\ ,
\end{equation*}%
with the right-hand side of this inequality going to zero as $L\rightarrow
\infty $, thanks to Equation (\ref{iteration0}), we use now \cite[Theorem 9.5%
]{Araki-Moriya} to deduce that this standard potential $\tilde{\Phi}$
defines the same free energy density functional (\ref{map free energy}) and,
consequently, the same equilibrium states (defined as minimizers of the free
energy density functional). We can then directly apply, on the one hand,
\cite[Corollary 6.7 and Theorem 12.11]{Araki-Moriya} to conclude that $%
\mathit{M}_{\Phi }\subseteq \mathit{K}_{\Phi }$, and, on the other hand,
\cite[Theorem 7.5 and Proposition 12.1]{Araki-Moriya} to conclude that $%
\mathit{K}_{\Phi }\subseteq \mathit{M}_{\Phi }$.
\end{proof}

Recall that a state $\rho \in E$ on $\mathcal{U}$ is called \emph{faithful}
iff $\rho \left( |A|^{2}\right) >0$ for all nonzero elements $A\in \mathcal{U%
}$. It is, by definition, \emph{modular} iff $\Omega _{\rho }$ is separating
for the von Neumann algebra $\pi _{\rho }(\mathcal{U})^{\prime \prime
}\subseteq \mathcal{B}(\mathcal{H}_{\rho })$, where $(\mathcal{H}_{\rho
},\pi _{\rho },\Omega _{\rho })$ is the cyclic representation of $\mathcal{U}
$ associated with $\rho $. KMS states on simple $C^{\ast }$-algebras like $%
\mathcal{U}$ are modular and faithful. We thus deduce the following from
Theorem \ref{eq.tang.bcs.type0 copy(1)}:

\begin{corollary}[Equilibrium states as faithful states]
\label{eq.tang.bcs.type0 copy(2)}\mbox{ }\newline
For any $\Phi \in \mathcal{W}_{1}$, translation-invariant equilibrium states
$\omega \in \mathit{M}_{\Phi }$ are faithful and modular.
\end{corollary}

\begin{proof}
The modular property of translation-invariant equilibrium states is a direct
consequence of Theorem \ref{eq.tang.bcs.type0 copy(1)} and \cite[Corollary
5.3.9]{BrattelliRobinson}. Given $\Phi \in \mathcal{W}_{1}$, it follows that
a translation-invariant equilibrium state $\omega \in \mathit{M}_{\Phi }$
with associated cyclic representation $(\mathcal{H}_{\omega },\pi _{\omega
},\Omega _{\omega })$ is faithful whenever $\pi _{\omega }$ is a $\ast $%
-isomorphism between $\mathcal{U}$ and $\pi _{\omega }(\mathcal{U})$. The $%
C^{\ast }$-algebra $\mathcal{U}$ is a UHF (uniformly hyperfinite) algebra
and is thus simple, i.e., the only closed two-sided ideals of $\mathcal{U}$
are the trivial one, $\{0\}$, and $\mathcal{U}$ itself. See, e.g., \cite[%
Section 8]{Bru-pedra-MF-I} or \cite[Corollary 2.6.19]{BrattelliRobinsonI}.
As a consequence, $\pi _{\omega }$ must be a $\ast $-isomorphism between $%
\mathcal{U}$ and $\pi _{\omega }(\mathcal{U})$, for, clearly, $\pi _{\omega
}(\mathfrak{1})\neq 0$ (being the identity operator on $\mathcal{H}_{\omega
} $).
\end{proof}

\section{Infinite Volume Long-Range Models\label{Long-Range Models}}

\subsection{Long-Range Models}

Let
\begin{equation*}
\mathbb{S}\doteq \left\{ \Phi \in \mathcal{W}_{1}:\left\Vert \Phi
\right\Vert _{\mathcal{U}}=1\right\}
\end{equation*}%
be the unit sphere of the Banach space $\mathcal{W}_{1}$ (\ref{W1}) of
translation-invariant short-range interactions. For any measure $\mathfrak{a}
$ on $\mathbb{S}$, we define a new measure $\mathfrak{a}^{\ast }$ on $%
\mathbb{S}$ as being the pushforward of $\mathfrak{a}$ through the
(continuous mapping) $(\cdot )^{\ast }:\mathbb{S}\rightarrow \mathbb{S}$. We
say that a measure $\mathfrak{a}$ on $\mathbb{S}$ is self-adjoint iff $%
\mathfrak{a}^{\ast }=\mathfrak{a}$. Denote by $\mathcal{S}_{1}$ the space of
self-adjoint signed Borel measures of bounded variation on $\mathbb{S}$,
which is a real Banach space whose norm is the total variation of measures
\begin{equation*}
\left\Vert \mathfrak{a}\right\Vert _{\mathcal{S}_{1}}\doteq \left\vert
\mathfrak{a}\right\vert \left( \mathbb{S}\right) \text{ },\qquad \mathfrak{a}%
\in \mathcal{S}_{1}\text{ }.
\end{equation*}

We are now in a position to define the space of long-range models. It is the
separable (real) Banach space%
\begin{equation}
\mathcal{M}\doteq \left\{ \mathfrak{m}\in \mathcal{W}^{\mathbb{R}}\times
\mathcal{S}_{1}:\left\Vert \mathfrak{m}\right\Vert _{\mathcal{M}}<\infty
\right\} \ ,  \label{def long range1}
\end{equation}%
whose norm is
\begin{equation}
\left\Vert \mathfrak{m}\right\Vert _{\mathcal{M}}\doteq \left\Vert \Phi
\right\Vert _{\mathcal{W}}+\left\Vert \mathfrak{a}\right\Vert _{\mathcal{S}%
_{1}}\,,\qquad \mathfrak{m}\doteq \left( \Phi ,\mathfrak{a}\right) \in
\mathcal{M}\text{ }.  \label{def long range2}
\end{equation}%
The spaces $\mathcal{W}^{\mathbb{R}}$ and $\mathcal{S}_{1}$ are canonically
seen as subspaces of $\mathcal{M}$, i.e.,
\begin{equation}
\mathcal{W}^{\mathbb{R}}\subseteq \mathcal{M}\qquad \text{and}\qquad
\mathcal{S}_{1}\subseteq \mathcal{M}\ .  \label{identification model}
\end{equation}%
In particular, $\Phi \equiv \left( \Phi ,0\right) \in \mathcal{M}$ for $\Phi
\in \mathcal{W}^{\mathbb{R}}$ and $\mathfrak{a}\equiv \left( 0,\mathfrak{a}%
\right) \in \mathcal{M}$ for $\mathfrak{a}\in \mathcal{S}_{1}$.

We define the dense subspace
\begin{equation}
\mathcal{M}_{0}\doteq \bigcup_{L\in \mathbb{N}}\mathcal{M}_{\Lambda _{L}}
\label{M0}
\end{equation}%
of the space $\mathcal{M}$ of long-range models, where, for any finite
subset $\Lambda \in \mathcal{P}_{f}$,
\begin{equation}
\mathcal{M}_{\Lambda }\doteq \left\{ \left( \Phi ,\mathfrak{a}\right) \in
\mathcal{M}:\left\Vert \mathfrak{a}\right\Vert _{\mathcal{S}_{1}}\doteq
\left\vert \mathfrak{a}\right\vert \left( \mathbb{S}\right) =\left\vert
\mathfrak{a}\right\vert \left( \mathbb{S}\cap \mathcal{W}_{\Lambda }\right)
\right\} \ ,  \label{S00bis}
\end{equation}%
$\mathcal{W}_{\Lambda }$ being the closed subspace of finite-range
translation-invariant interactions defined by
\begin{equation}
\mathcal{W}_{\Lambda }\doteq \left\{ \Phi \in \mathcal{W}_{1}:\Phi _{%
\mathcal{Z}}=0\text{ whenever }\mathcal{Z}\nsubseteq \Lambda \text{, }%
\mathcal{Z}\ni 0\right\} \ .  \label{eq:enpersitebis}
\end{equation}%
(If $0\notin \Lambda \in \mathcal{P}_{f}$ then $\mathcal{W}_{\Lambda }=\{0\}$%
, but this is of course not the case of interest here.)

Long-range models $\mathfrak{m}\doteq \left( \Phi ,\mathfrak{a}\right) $ are
not necessarily translation-invariant, because their short-range component $%
\Phi $ is not required to be translation-invariant. We thus define
\begin{equation}
\mathcal{M}_{1}\doteq \mathcal{W}_{1}^{\mathbb{R}}\times \mathcal{S}%
_{1}\varsubsetneq \mathcal{M}
\label{translatino invariatn long range models}
\end{equation}%
as being the (real) Banach space of translation-invariant long-range models.

\begin{remark}[Equivalent definition of translation-invariant long-range
models]
\label{remark temp copy(1)}\mbox{ }\newline
\cite{BruPedra2} and \cite{BruPedra-MFII,BruPedra-MFIII} have different
definitions of long-range models. We use here the formalism introduced in
\cite{BruPedra-MFII,BruPedra-MFIII}. Compare (\ref{translatino invariatn
long range models}) with \cite[Definition 2.1]{BruPedra2}. However, \cite[%
Section 8]{BruPedra-MFII} shows that the results of \cite{BruPedra2} apply
equally to all translation-invariant long-range models $\mathfrak{m}\in
\mathcal{M}_{1}$.
\end{remark}

\subsection{Purely Repulsive and Purely Attractive Long-Range Models \label%
{Section purely attrac}}

By the Hahn decomposition theorem, any signed measure $\mathfrak{a}$ of
bounded variation on the unit sphere $\mathbb{S}$ of the Banach space $%
\mathcal{W}_{1}$ has a unique decomposition%
\begin{equation}
\mathfrak{a=}\underset{\text{long-range repulsion}}{\underbrace{\mathfrak{a}%
_{+}}}-\underset{\text{long-range attraction}}{\underbrace{\mathfrak{a}_{-}}}
\label{Hahn decomposition}
\end{equation}%
$\mathfrak{a}_{\pm }$ being two positive finite measures vanishing on
disjoint Borel sets, respectively denoted by $\mathbb{S}_{\mp }\subseteq
\mathbb{S}$. Recall that such a decomposition is called the Jordan
decomposition of the measure of bounded variation $\mathfrak{a}$ and $|%
\mathfrak{a}|=\mathfrak{a}_{+}+\mathfrak{a}_{-}$. Long-range attractions are
represented by the measure $\mathfrak{a}_{-}$, whereas $\mathfrak{a}_{+}$
refers to long-range repulsions. A long-range model $\mathfrak{m}\doteq
\left( \Phi ,\mathfrak{a}\right) \in \mathcal{M}$ is said to be \emph{purely
attractive} iff $\mathfrak{a}_{+}=0$, while it is \emph{purely repulsive}
iff $\mathfrak{a}_{-}=0$.

Distinguishing between these two special types of models is important
because the effects of long-range attractions and repulsions on the
structure of corresponding sets of (generalized) equilibrium states can be
very different: By \cite[Theorem 2.25]{BruPedra2}, long-range attractions
have no particular effect on the structure of the set of (generalized,
translation-invariant) equilibrium states, which is still a (non-empty) weak$%
^{\ast }$-closed face of the set $E_{1}$ of translation-invariant states,
like for short-range interactions. By contrast, long-range repulsions have
generally a geometrical effect by possibly breaking the face structure of
the set of (generalized) equilibrium states (see \cite[Lemma 9.8]{BruPedra2}%
). This feature of long-range repulsions leads us to introduce the notion of
\emph{simple} long-range models, in Definition \ref{def simple} below.

\subsection{Dynamics Generated by Long-Range Models\label{sect
Lieb--Robinson copy(3)}}

\subsubsection{Local Derivations and Long-Range Dynamics}

The local Hamiltonians associated with any long-range model $\mathfrak{m}%
\doteq \left( \Phi ,\mathfrak{a}\right) \in \mathcal{M}$ are the
(well-defined) self-adjoint elements%
\begin{equation}
U_{L}^{\mathfrak{m}}\doteq U_{L}^{\Phi }+\frac{1}{\left\vert \Lambda
_{L}\right\vert }\int_{\mathbb{S}}\left\vert U_{L}^{\Psi }\right\vert ^{2}%
\mathfrak{a}\left( \mathrm{d}\Psi \right) \,,\qquad L\in \mathbb{N}\text{ },
\label{equation long range energy}
\end{equation}%
where we recall that $|A|^{2}\doteq A^{\ast }A$ for any $A\in \mathcal{U}$,
see (\ref{carre de A}). Note that $U_{L}^{\left( \Phi ,0\right)
}=U_{L}^{\Phi }$ for any self-adjoint short-range interaction $\Phi \in
\mathcal{W}^{\mathbb{R}}$ (cf. (\ref{identification model})) and
straightforward estimates yield the bound
\begin{equation}
\left\Vert U_{L}^{\mathfrak{m}}\right\Vert _{\mathcal{U}}\leq \left\vert
\Lambda _{L}\right\vert \left\Vert \mathbf{F}\right\Vert _{1,\mathfrak{L}%
}\left\Vert \mathfrak{m}\right\Vert _{\mathcal{M}}\ ,\qquad L\in \mathbb{N}%
\text{ },\ \mathfrak{m}\in \mathcal{M}\text{ },
\label{energy bound long range}
\end{equation}%
by Equations (\ref{norm Uphi}) and (\ref{def long range2}).

The sequence $(\delta _{L}^{\mathfrak{m}})_{L\in \mathbb{N}}$ of local
(symmetric) derivations of the $C^{\ast }$-algebra $\mathcal{U}$, associated
with any fixed long-range model $\mathfrak{m}\in \mathcal{M}$, is defined by
\begin{equation}
\delta _{L}^{\mathfrak{m}}(A)\doteq i\left[ U_{L}^{\mathfrak{m}},A\right]
\doteq i\left( U_{L}^{\mathfrak{m}}A-AU_{L}^{\mathfrak{m}}\right) \ ,\qquad
A\in \mathcal{U}\text{ },\ L\in \mathbb{N}\text{ }.
\label{equation derivation long range}
\end{equation}%
Note that $\delta _{L}^{\left( \Phi ,0\right) }=\delta _{L}^{\Phi }$ for
every length $L\in \mathbb{N}$ and any self-adjoint short-range interaction $%
\Phi \in \mathcal{W}^{\mathbb{R}}$ (cf. (\ref{identification model})). For
any $\mathfrak{m}\in \mathcal{M}$ and $L\in \mathbb{N}$, the local
long-range dynamics is defined to be the continuous group $(\tau _{t}^{(L,%
\mathfrak{m})})_{t\in \mathbb{R}}$ of $\ast $-auto%
\-%
morphisms of $\mathcal{U}$ generated by the bounded derivation $\delta _{L}^{%
\mathfrak{m}}$. Equivalently,
\begin{equation}
\tau _{t}^{(L,\mathfrak{m})}(A)\doteq \mathrm{e}^{itU_{L}^{\mathfrak{m}}}A%
\mathrm{e}^{-itU_{L}^{\mathfrak{m}}}\ ,\qquad A\in \mathcal{U}\text{ },\
t\in \mathbb{R}\text{ }.  \label{finite vol long range dyna}
\end{equation}%
Note that $\tau _{t}^{(L,(\Phi ,0))}=\tau _{t}^{(L,\Phi )}$ for any
self-adjoint short-range interaction $\Phi \in \mathcal{W}^{\mathbb{R}}$,\
every length $L\in \mathbb{N}$ and all times $t\in \mathbb{R}$. See Section %
\ref{sect Lieb--Robinson copy(2)}.

\subsubsection{Dynamical Self-Consistency Equations}

Generically, long-range dynamics in infinite volume are equivalent to
intricate combinations of a classical and short-range (infinite volume)
quantum dynamics. This fact results from the existence of a solution to a
(dynamical) \emph{self-consistency equation}. In order to present this
equation, we need some preliminary definitions: For any long-range model $%
\mathfrak{m}=(\Phi ,\mathfrak{\mathfrak{a}})\in \mathcal{M}$ and every
function $c=(c_{\Psi })_{\Psi \in \mathbb{S}}\in L^{2}(\mathbb{S};\mathbb{C}%
;|\mathfrak{\mathfrak{a}}|)$, we define a so-called approximating
(self-adjoint, short-range) interaction by
\begin{equation}
\Phi _{\mathfrak{m}}(c)\doteq \Phi +2\int_{\mathbb{S}}\mathrm{Re}\left\{
\overline{c_{\Psi }}\Psi \right\} \mathfrak{a}\left( \mathrm{d}\Psi \right)
\in \mathcal{W}^{\mathbb{R}}\ .  \label{approx interaction}
\end{equation}%
The integral in the last definition, which refers to a self-adjoint
interaction, i.e., an element of the space $\mathcal{W}^{\mathbb{R}}$, has
to be understood as follows:
\begin{equation}
\left( \int_{\mathbb{S}}\mathrm{Re}\left\{ \overline{c_{\Psi }}\Psi \right\}
\mathfrak{a}\left( \mathrm{d}\Psi \right) \right) _{\Lambda }\doteq \int_{%
\mathbb{S}}\mathrm{Re}\left\{ \overline{c_{\Psi }}\Psi _{_{\Lambda
}}\right\} \mathfrak{a}\left( \mathrm{d}\Psi \right) \ ,\qquad \Lambda \in
\mathcal{P}_{f}\text{ }.  \label{def integral approx interaction}
\end{equation}%
Note that the integral in the definiens is well-defined because, for each $%
\Lambda \in \mathcal{P}_{f}$, the integrand is an absolutely integrable
(measurable) function taking values in a finite-dimensional normed space,
which is $\mathcal{U}_{\Lambda }$.

Then, by \cite[Theorem 6.5]{BruPedra-MFII}, if $\mathfrak{m=}(\Phi ,%
\mathfrak{\mathfrak{a}})\in \mathcal{M}_{0}$ (see (\ref{M0})--(\ref{S00bis})
for the definition of the dense subspace $\mathcal{M}_{0}\subseteq \mathcal{M%
}$) there is a unique continuous\footnote{%
We endow the set $C\left( E;E\right) $ of continuous functions from $E$ to
itself with the topology of uniform convergence. See \cite[Equation (100)]%
{BruPedra-MFII} for more details.} mapping $\mathbf{\varpi }^{\mathfrak{m}}$
from $\mathbb{R}$ to the space of automorphisms\footnote{%
I.e., elements of $C\left( E;E\right) $ with inverse. Note the inverse (in
the sense of functions) of an element of $C\left( E;E\right) $ is again an
element of this space, i.e., it is continuous, by (weak$^{\ast }$)
compactness of $E$.} (or self-homeomorphisms) of $E$ such that
\begin{equation}
\mathbf{\varpi }^{\mathfrak{m}}\left( t;\rho \right) =\rho \circ \tau
_{t,0}^{\Phi ^{\left( \mathfrak{m},\rho \right) }}\ ,\qquad t\in {\mathbb{R}}%
\text{ },\ \rho \in E\text{ },  \label{self-consistency equation}
\end{equation}%
where $\Phi ^{\left( \mathfrak{m},\rho \right) }\in C(\mathbb{R};\mathcal{W}%
^{\mathbb{R}})$ is defined for any $\mathfrak{m}\in \mathcal{M}_{0}$ and $%
\rho \in E$ by
\begin{equation}
\Phi ^{\left( \mathfrak{m},\rho \right) }(t)\doteq \Phi _{\mathfrak{m}}(%
\mathbf{\varpi }^{\mathfrak{m}}\left( t;\rho \right) (\mathfrak{e}_{(\cdot
)}))\ ,\qquad t\in {\mathbb{R}}\text{ },  \label{self-consistency equation2}
\end{equation}%
the mapping $\mathfrak{e}_{(\cdot )}:\mathbb{S}\rightarrow \mathcal{U}$
being defined by (\ref{eq:enpersite}), while the strongly continuous two-para%
\-%
meter family $(\tau _{t,s}^{\Phi ^{\left( \mathfrak{m},\rho \right)
}})_{s,t\in {\mathbb{R}}}$ is the unique solution to (\ref{cauchy trivial1})
for $\Psi =\Phi ^{\left( \mathfrak{m},\rho \right) }$. For any fixed $t\in {%
\mathbb{R}}$ and $\rho \in E$, note that $\mathbf{\varpi }^{\mathfrak{m}%
}\left( t;\rho \right) (\mathfrak{e}_{(\cdot )})$ is a continuous bounded
function on $\mathbb{S}$. In particular, it belongs to $L^{2}(\mathbb{S};%
\mathbb{C};|\mathfrak{\mathfrak{a}}|)$ and thus, at fixed $t\in {\mathbb{R}}$
and $\rho \in E$, the right-hand side of (\ref{self-consistency equation2})
is an approximating (short-range) interaction, as defined by Equation (\ref%
{approx interaction}). The continuity of $\Phi ^{\left( \mathfrak{m},\rho
\right) }$ is a consequence of the continuity of the mappings
\begin{equation*}
\mathbf{\varpi }^{\mathfrak{m}}:{\mathbb{R}}\rightarrow C\left( E;E\right)
\qquad \text{and}\qquad \Phi _{\mathfrak{m}}(\cdot ):L^{2}(\mathbb{S};%
\mathbb{C};|\mathfrak{\mathfrak{a}}|)\rightarrow \mathcal{W}^{\mathbb{R}}.
\end{equation*}%
Equation (\ref{self-consistency equation}) is named here the (dynamical)
self-consistency equation.

\subsubsection{Limit Long-Range Dynamics in the Schr\"{o}dinger Picture\label%
{Section Limit Long-Range Dynamics}}

Any long-range model $\mathfrak{m}\in \mathcal{M}$ leads to a sequence of
finite volume dynamics $(\tau _{t}^{(L,\mathfrak{m})})_{t\in \mathbb{R}}$, $%
L\in \mathbb{N}$, defined by (\ref{finite vol long range dyna}). At length $%
L\in \mathbb{N}$, the time-evolution $(\rho _{t}^{(L)})_{t\in \mathbb{R}}$
of any state $\rho \in E$ is given by
\begin{equation}
\rho _{t}^{(L)}\doteq \rho \circ \tau _{t}^{(L,\mathfrak{m})}\ .
\label{long-range dyn0}
\end{equation}%
Equation (\ref{long-range dyn0}) refers to the Schr\"{o}dinger picture of
quantum mechanics.

At fixed $A\in \mathcal{U}$ and $t\in \mathbb{R}$, the thermodynamic limit $%
L\rightarrow \infty $ of $\tau _{t}^{(L,\mathfrak{m})}(A)$ does not
necessarily exist in $\mathcal{U}$, but the limit $L\rightarrow \infty $ of $%
\rho _{t}^{(L)}$ can still make sense: Fix once and for all a
translation-invariant long-range model $\mathfrak{m}\in \mathcal{M}_{1}\cap
\mathcal{M}_{0}$. Recall that $E_{1}$ denotes the set (\ref{periodic
invariant states}) of translation-invariant states, with set $\mathcal{E}%
(E_{1})$ of extreme points, and that, for any $\rho \in E_{1}$, there is an
orthogonal (unique) probability measure $\mu _{\rho }$ on $E_{1}$ with
support in $\mathcal{E}(E_{1})$ such that%
\begin{equation*}
\rho \left( A\right) =\int_{\mathcal{E}\left( E_{1}\right) }\hat{\rho}\left(
A\right) \ \mathrm{d}\mu _{\rho }\left( \hat{\rho}\right) \ ,\qquad A\in
\mathcal{U}\text{ },
\end{equation*}%
by Theorem \ref{theorem choquet}. By the ergodicity property of extreme
translation-invariant states (see (\ref{Ergodicity2})), one can prove that,
for any time $t\in \mathbb{R}$ and every element $A\in \mathcal{U}$,
\begin{equation}
\lim_{L\rightarrow \infty }\rho _{t}^{(L)}\left( A\right) =\int_{\mathcal{E}%
\left( E_{1}\right) }\mathbf{\varpi }^{\mathfrak{m}}\left( t;\hat{\rho}%
\right) \left( A\right) \ \mathrm{d}\mu _{\rho }\left( \hat{\rho}\right)
=\int_{\mathcal{E}\left( E_{1}\right) }\hat{\rho}\circ \tau _{t,0}^{\Psi
^{\left( \mathfrak{m},\hat{\rho}\right) }}\left( A\right) \ \mathrm{d}\mu
_{\rho }\left( \hat{\rho}\right) \ ,  \label{long-range dyn}
\end{equation}%
$\mathbf{\varpi }^{\mathfrak{m}}$ being the solution to the self-consistency
equation\emph{\ }(\ref{self-consistency equation}). See \cite[Theorem 5.8]%
{BruPedra-MFIII}. This result is not restricted to translation-invariant
states but it can be extended to all periodic states, which form a weak$%
^{\ast }$-dense subset of the set of all even states, the physically
relevant ones. See Remark \ref{periodic states} and \cite[Proposition 2.3]%
{BruPedra-MFII}.

\subsection{Equilibrium States of Long-Range Models\label{sect
Lieb--Robinson copy(4)}}

\subsubsection{The Space-Averaging Functional on Translation-Invariant
States \label{Section space averaging}}

In addition to the energy density and entropy density functionals,
respectively defined in Sections \ref{energy density1}--\ref{energy density2}%
, we need the so-called space-averaging functional in order to study the
thermodynamic properties of long-range models. This new density functional
is defined on the set $E_{1}$ of translation-invariant states as follows:
For any $A\in \mathcal{U}$, the mapping $\Delta _{A}:E_{1}\rightarrow
\mathbb{R}$ is (well-)defined by%
\begin{equation*}
\rho \mapsto \Delta _{A}\left( \rho \right) \doteq \lim\limits_{L\rightarrow
\infty }\rho \left( \left\vert A_{L}\right\vert ^{2}\right) \in \left[ |\rho
(A)|^{2},\Vert A\Vert _{\mathcal{U}}^{2}\right] \text{ },
\end{equation*}%
where $|A_{L}|^{2}\doteq A_{L}^{\ast }A_{L}$ (see (\ref{carre de A})) and $%
A_{L}$ is defined by (\ref{Limit of Space-Averages}) for any $L\in \mathbb{N}
$. Compare with Equation (\ref{Ergodicity}). See also \cite[Section 1.3]%
{BruPedra2}. By \cite[Theorem 1.18]{BruPedra2}, the functional $\Delta _{A}$
is affine and weak$^{\ast }$-upper semi-continuous. Thanks again to \cite[%
Theorem 1.18]{BruPedra2}, note additionally that, at any fixed
(translation-invariant state) $\rho \in E_{1}$,%
\begin{equation*}
|\Delta _{A}\left( \rho \right) -\Delta _{B}\left( \rho \right) |\leq (\Vert
A\Vert _{\mathcal{U}}+\Vert B\Vert _{\mathcal{U}})\Vert A-B\Vert _{\mathcal{U%
}}\ ,\qquad A,B\in \mathcal{U}\text{ }.
\end{equation*}

For any signed Borel measure $\mathfrak{a}$ of bounded variation on $\mathbb{%
S}$, we define the space-averaging functional $\Delta _{\mathfrak{a}%
}:E_{1}\rightarrow \mathbb{R}$ on translation-invariant states by
\begin{equation}
\rho \mapsto \Delta _{\mathfrak{a}}\left( \rho \right) \doteq \int_{\mathbb{S%
}}\Delta _{\mathfrak{e}_{\Psi }}\left( \rho \right) \mathfrak{a}\left(
\mathrm{d}\Psi \right) \text{ },  \label{Free-energy density long range0}
\end{equation}%
the continuous mapping $\mathfrak{e}_{(\cdot )}:\mathcal{W}\rightarrow
\mathcal{U}$ being\ defined by Equation (\ref{eq:enpersite}). By \cite[%
Theorem 1.18]{BruPedra2}, $\Delta _{\mathfrak{a}}$ is a well-defined, affine
and weak$^{\ast }$-upper semi-continuous functional on the set $E_{1}$ of
translation-invariant states.

\subsubsection{Generalized Equilibrium States\label{Translation Invariant
Equilibrium States copy(1)}}

We give here the extension of the notion of equilibrium states of Section %
\ref{Translation Invariant Equilibrium States}, to general long-range
models, by using again the variational principle associated with the
infinite volume pressure. An important issue appears in this more general
situation, because of the lack of weak$^{\ast }$-continuity of the free
energy density functional in presence of long-range repulsions (Section \ref%
{Section purely attrac}), as explained below in more detail.

We start by giving the (grand-canonical) pressure in the thermodynamic
limit: At any given inverse temperature $\beta \in \mathbb{R}^{+}$, the
infinite volume pressure $\mathrm{P}$ for translation-invariant long-range
models is, by definition, the real-valued function on the Banach space $%
\mathcal{M}_{1}$ (\ref{translatino invariatn long range models}) of
translation-invariant long-range models, defined by%
\begin{equation*}
\mathfrak{m}\mapsto \mathrm{P}_{\mathfrak{m}}\doteq \underset{L\rightarrow
\infty }{\lim }\frac{1}{\beta |\Lambda _{L}|}\ln \mathrm{Trace}(\mathrm{e}%
^{-\beta U_{L}^{\mathfrak{m}}})\ .
\end{equation*}%
By \cite[Theorem 2.12]{BruPedra2}, this mapping is well-defined and, for any
$\mathfrak{m}=(\Phi ,\mathfrak{\mathfrak{a}})\in \mathcal{M}_{1}$,
\begin{equation}
\mathrm{P}_{\mathfrak{m}}=-\inf f_{\mathfrak{m}}\left( E_{1}\right) \in
\mathbb{R}\text{ },  \label{BCS main theorem 1eq}
\end{equation}%
where $f_{\mathfrak{m}}:E_{1}\rightarrow \mathbb{R}$ is the free energy
density functional defined by%
\begin{equation}
f_{\mathfrak{m}}\doteq \Delta _{\mathfrak{a}}+f_{\Phi }=\Delta _{\mathfrak{a}%
}+e_{\Phi }-\beta ^{-1}s\ .  \label{Free-energy density long range}
\end{equation}%
See Equation (\ref{map free energy}), defining the free energy density
functional $f_{\Phi }$ for any self-adjoint translation-invariant and
short-range interaction $\Phi \in \mathcal{W}_{1}^{\mathbb{R}}$. Observe
that Equation (\ref{BCS main theorem 1eq}) is an extension of (\ref{pressure
free energy}) -- which refers the space $\mathcal{W}_{1}^{\mathbb{R}}$ of
short-range models only -- to the space $\mathcal{M}_{1}\supseteq \mathcal{W}%
_{1}^{\mathbb{R}}$ of long-range models.

Similar to (\ref{minimizer short range}), for any translation-invariant
long-range model $\mathfrak{m}\in \mathcal{M}_{1}$, one might define the set
of equilibrium states by%
\begin{equation}
\mathit{M}_{\mathfrak{m}}\doteq \left\{ \omega \in E_{1}:f_{\mathfrak{m}%
}\left( \omega \right) =\inf \,f_{\mathfrak{m}}(E_{1})=-\mathrm{P}_{%
\mathfrak{m}}\right\} \ .  \label{minimizer long range}
\end{equation}%
Note however that the free energy density functional $f_{\mathfrak{m}}$ is
in general not weak$^{\ast }$-lower semi-continuous on $E_{1}$ and it is
thus a priori not clear whether $\mathit{M}_{\mathfrak{m}}$ is empty or not.
In fact, by Equation (\ref{Hahn decomposition}), for any
translation-invariant long-range model $\mathfrak{m}=(\Phi ,\mathfrak{%
\mathfrak{a}})\in \mathcal{M}_{1}$,
\begin{equation*}
f_{\mathfrak{m}}=\underset{\text{weak}^{\ast }\text{-upper semi-cont.}}{%
\underbrace{\Delta _{\mathfrak{a}_{+}}}}+\underset{\text{weak}^{\ast }\text{%
-lower semi-cont.}}{\underbrace{\left( -\Delta _{\mathfrak{a}_{-}}+f_{\Phi
}\right) }}\ .
\end{equation*}%
Therefore, instead of considering $\mathit{M}_{\mathfrak{m}}$, we define
\begin{equation}
\mathit{\Omega }_{\mathfrak{m}}\doteq \left\{ \omega \in E_{1}:\exists
\{\rho _{n}\}_{n=1}^{\infty }\subseteq E_{1}\mathrm{\ }\text{weak}^{\ast }%
\text{ converging to}\ \omega \text{ such\ that\ }\underset{n\rightarrow
\infty }{\lim }f_{\mathfrak{m}}(\rho _{n})=\inf \,f_{\mathfrak{m}%
}(E_{1})\right\}  \label{definition equilibirum state}
\end{equation}%
as being the set of \emph{generalized} equilibrium states of any fixed
translation-invariant long-range model $\mathfrak{m}\in \mathcal{M}_{1}$ (at
inverse temperature $\beta \in \mathbb{R}^{+}$). Observe for instance that,
under periodic boundary conditions, the accumulation points of
(finite-volume) Gibbs states associated with any long-range model $\mathfrak{%
m}\in \mathcal{M}_{1}$ and $\beta \in \mathbb{R}^{+}$\ always belong to $%
\mathit{\Omega }_{\mathfrak{m}}$, but not necessarily to $\mathit{M}_{%
\mathfrak{m}}$, by \cite[Theorem 3.13]{BruPedra2}.

Obviously, by weak$^{\ast }$-compactness of $E_{1}$, the set $\mathit{\Omega
}_{\mathfrak{m}}$ is non-empty and $\mathit{\Omega }_{\mathfrak{m}}\supseteq
\mathit{M}_{\mathfrak{m}}$. This definition can be expressed in terms of the
graph of $f_{\mathfrak{m}}$:
\begin{equation*}
\mathit{\Omega }_{\mathfrak{m}}\times \{\inf \,f_{\mathfrak{m}%
}(E_{1})\}=\left( E_{1}\times \{\inf \,f_{\mathfrak{m}}(E_{1})\}\right) \cap
\overline{\mathrm{Graph}(f_{\mathfrak{m}})}\text{ },
\end{equation*}%
where the closure of the graph of $f_{\mathfrak{m}}$ refers to the product
topology of the weak$^{{\ast }}$ topology on $E_{1}$ and the usual topology
on $\mathbb{R}$. It follows that $\mathit{\Omega }_{\mathfrak{m}}$ is weak$^{%
{\ast }}$-closed and convex, by affineness of $f_{\mathfrak{m}}$. Thus, $%
\mathit{\Omega }_{\mathfrak{m}}$\ is a weak$^{\ast }$-compact convex subset
of $E_{1}$. See \cite[Lemma 2.16]{BruPedra2}. If $\mathfrak{a}_{+}=0$ then $%
\mathit{\Omega }_{\mathfrak{m}}=\mathit{M}_{\mathfrak{m}}$ is a (non-empty)\
weak$^{\ast }$-closed face of the Poulsen simplex $E_{1}$. By contrast, as
already mentioned above, a long-range repulsion $\mathfrak{a}_{+}$ has
generally a \emph{geometrical} effect on the set $\mathit{\Omega }_{%
\mathfrak{m}}$, by possibly breaking its face structure in $E_{1}$. This
effect can lead to long-range order of generalized equilibrium states. See
\cite[Section 2.9]{BruPedra2}.

\subsubsection{Thermodynamic Game\label{Section thermo game}}

Through a version of the approximating Hamiltonian method \cite[Section 2.10]%
{BruPedra2}, \cite[Theorem 2.36]{BruPedra2} shows that, for any long-range
model $\mathfrak{m}=(\Phi ,\mathfrak{\mathfrak{a}})\in \mathcal{M}_{1}$, the
pressure $\mathrm{P}_{\mathfrak{m}}$ is given by a (Bogoliubov) min-max
variational problem on the Hilbert space $L^{2}(\mathbb{S};\mathbb{C};|%
\mathfrak{\mathfrak{a}}|)$ of square integrable functions on the sphere $%
\mathbb{S}$, which is interpreted as the result of a two-person zero-sum
game, as it is explained in this section.

For any translation-invariant long-range model $\mathfrak{m}=(\Phi ,%
\mathfrak{\mathfrak{a}})\in \mathcal{M}_{1}$, recall that functions $%
c=(c_{\Psi })_{\Psi \in \mathbb{S}}\in L^{2}(\mathbb{S};\mathbb{C};|%
\mathfrak{\mathfrak{a}}|)$ are parameters of approximating interactions $%
\Phi _{\mathfrak{m}}(c)\in \mathcal{W}_{1}^{\mathbb{R}}$, which are defined
by (\ref{approx interaction}). By Equation (\ref{equation fininte vol dynam0}%
), the energy observables associated with $\Phi _{\mathfrak{m}}(c)$ equal%
\begin{equation*}
U_{L}^{\Phi _{\mathfrak{m}}(c)}=U_{L}^{\Phi }+\int_{\mathbb{S}}2\mathrm{Re}%
\left\{ \overline{c_{\Psi }}U_{L}^{\Psi }\right\} \mathfrak{a}\left( \mathrm{%
d}\Psi \right) \ ,\qquad L\in \mathbb{N}\ .
\end{equation*}%
One then deduces from Equations (\ref{pressure free energy})--(\ref{map free
energy}) that%
\begin{equation}
\mathrm{P}_{\Phi _{\mathfrak{m}}(c)}=-\inf f_{\Phi _{\mathfrak{m}}(c)}\left(
E_{1}\right) \text{ },\qquad c\in L^{2}(\mathbb{S};\mathbb{C};|\mathfrak{%
\mathfrak{a}}|)\text{ },  \label{variational problem approx}
\end{equation}%
where, for any translation-invariant state $\rho \in E_{1}$,%
\begin{equation*}
f_{\Phi _{\mathfrak{m}}(c)}(\rho )=\int_{\mathbb{S}}2\mathrm{Re}\left\{
\overline{c_{\Psi }}e_{\Psi }(\rho )\right\} \mathfrak{a}\left( \mathrm{d}%
\Psi \right) +e_{\Phi }(\rho )-\beta ^{-1}s(\rho )\ .
\end{equation*}%
As compared to the pressure $\mathrm{P}_{\mathfrak{m}}$ for
translation-invariant long-range models $\mathfrak{m}\in \mathcal{M}_{1}$, $%
\mathrm{P}_{\Phi _{\mathfrak{m}}(c)}$ is, in principle, easier to analyze,
because it comes from a purely short-range interaction $\Phi _{\mathfrak{m}%
}(c)\in \mathcal{W}_{1}^{\mathbb{R}}$.

Recall Equation (\ref{Hahn decomposition}): $\mathfrak{a=a}_{+}-\mathfrak{a}%
_{-}$ with $\mathfrak{a}_{\pm }$ being two positive finite measures
vanishing on any subset of $\mathbb{S}_{\mp }$, respectively, where $\mathbb{%
S}_{\pm }$ are Borel sets referring to the Jordan decomposition of $%
\mathfrak{a}$. Then, we define two Hilbert spaces corresponding respectively
to the long-range repulsive and attractive components, $\mathfrak{a}_{+}$
and $\mathfrak{a}_{-}$, of any translation-invariant long-range model $%
\mathfrak{m}\in \mathcal{M}_{1}$:
\begin{equation}
L_{\pm }^{2}(\mathbb{S};\mathbb{C})\doteq L^{2}(\mathbb{S};\mathbb{C};%
\mathfrak{a}_{\pm })\ .  \label{definition of positive-negative L2 space}
\end{equation}%
Note that we canonically have the equality%
\begin{equation*}
L^{2}(\mathbb{S};\mathbb{C};|\mathfrak{a}|)=L_{+}^{2}(\mathbb{S};\mathbb{C}%
)\oplus L_{-}^{2}(\mathbb{S};\mathbb{C})\ .
\end{equation*}%
The approximating free energy density functional%
\begin{equation*}
\mathfrak{f}_{\mathfrak{m}}:L_{-}^{2}(\mathbb{S};\mathbb{C})\times L_{+}^{2}(%
\mathbb{S};\mathbb{C})\rightarrow \mathbb{R}
\end{equation*}%
is defined by%
\begin{equation*}
\mathfrak{f}_{\mathfrak{m}}\left( c_{-},c_{+}\right) \doteq -\left\Vert
c_{+}\right\Vert _{2}^{2}+\left\Vert c_{-}\right\Vert _{2}^{2}-\mathrm{P}%
_{\Phi _{\mathfrak{m}}\left( c_{-}+c_{+}\right) }\ ,\qquad c_{\pm }\in
L_{\pm }^{2}(\mathbb{S};\mathbb{C})\text{ }.
\end{equation*}%
The \emph{thermodynamic game}\ is the two-person zero-sum game defined from $%
\mathfrak{f}_{\mathfrak{m}}$, with one of its conservative values being
equal (up to a minus sign) to the pressure $\mathrm{P}_{\mathfrak{m}}$ (see
\cite[Theorem 2.36 ($\sharp $)]{BruPedra2}):
\begin{equation*}
\mathrm{P}_{\mathfrak{m}}=-\inf_{c_{-}\in L_{-}^{2}(\mathbb{S};\mathbb{C}%
)}\sup_{c_{+}\in L_{+}^{2}(\mathbb{S};\mathbb{C})}\mathfrak{f}_{\mathfrak{m}%
}\left( c_{-},c_{+}\right) \ ,\qquad \mathfrak{m}\in \mathcal{M}_{1}\ .
\end{equation*}%
Compare this equality with Equations (\ref{BCS main theorem 1eq})-(\ref%
{Free-energy density long range}). The $\sup $ and $\inf $ in the above
optimization problem are attained, i.e., they are respectively a $\max $ and
a $\min $ and the set%
\begin{equation}
\mathcal{C}_{\mathfrak{m}}\doteq \left\{ d_{-}\in L_{-}^{2}(\mathbb{S};%
\mathbb{C}):\max_{c_{+}\in L_{+}^{2}(\mathbb{S};\mathbb{C})}\mathfrak{f}_{%
\mathfrak{m}}\left( d_{-},c_{+}\right) =-\mathrm{P}_{\mathfrak{m}}\right\}
\label{eq conserve strategy}
\end{equation}%
(of conservative strategies of the \textquotedblleft attractive
player\textquotedblright ) is non-empty, norm-bounded and weakly compact, by
\cite[Lemma 8.4 ($\sharp $)]{BruPedra2}. In the particular case of purely
repulsive long-range models, i.e., when $\mathfrak{a}_{-}=0$, $\mathcal{C}_{%
\mathfrak{m}}=\{0\}=L_{-}^{2}(\mathbb{S};\mathbb{C})$, which is in this case
the unique equivalent class of all complex-valued functions on $\mathbb{S}$,
as $\mathfrak{f}_{\mathfrak{m}}$ is independent of $c_{-}$.

Note that, in general, there is no saddle point, since the thermodynamic
game and the $\sup $ and $\inf $ do generally not commute. See \cite[p. 42]%
{BruPedra2}. In \cite[Lemma 8.3 ($\sharp $)]{BruPedra2} it is proven that,
when $\mathfrak{a}_{+}\neq 0$, for all functions $c_{-}\in L_{-}^{2}(\mathbb{%
S};\mathbb{C})$, the set
\begin{equation}
\left\{ d_{+}\in L_{+}^{2}(\mathbb{S};\mathbb{C}):\max_{c_{+}\in L_{+}^{2}(%
\mathbb{S};\mathbb{C})}\mathfrak{f}_{\mathfrak{m}}\left( c_{-},c_{+}\right) =%
\mathfrak{f}_{\mathfrak{m}}\left( c_{-},d_{+}\right) \right\}
\label{eq conserve strategybis}
\end{equation}%
has exactly one element, which we denote by $\mathrm{r}_{+}(c_{-})$. By \cite%
[Lemma 8.8]{BruPedra2}, if $\mathfrak{a}_{+}\neq 0$ then the mapping%
\begin{equation}
\mathrm{r}_{+}:c_{-}\mapsto \mathrm{r}_{+}\left( c_{-}\right)
\label{thermodyn decision rule}
\end{equation}%
defines a continuous functional from $L_{-}^{2}(\mathbb{S};\mathbb{C})$ to $%
L_{+}^{2}(\mathbb{S};\mathbb{C})$, where $L_{-}^{2}(\mathbb{S};\mathbb{C})$
and $L_{+}^{2}(\mathbb{S};\mathbb{C})$ are endowed with the weak and norm
topologies, respectively. This mapping is called the thermodynamic decision
rule\ of the translation-invariant long-range model $\mathfrak{m}\in
\mathcal{M}_{1}$. In the particular case of purely attractive long-range
models, i.e., when $\mathfrak{a}_{+}=0$, $\mathfrak{f}_{\mathfrak{m}}$ is
independent of $c_{+}$ and one trivially has $\mathrm{r}_{+}=0$, since $%
L_{+}^{2}(\mathbb{S};\mathbb{C})=\{0\}$ in this case.

\subsubsection{Self-Consistency of Generalized Equilibrium States \label%
{Section effective theories}}

The structure of the set $\mathit{\Omega }_{\mathfrak{m}}$ (\ref{definition
equilibirum state}) of generalized (translation-invariant) equilibrium
states can be now discussed in detail, with respect to the thermodynamic
game.

For any translation-invariant long-range model $\mathfrak{m}=(\Phi ,%
\mathfrak{\mathfrak{a}})\in \mathcal{M}_{1}$ and every function $c\in L^{2}(%
\mathbb{S};\mathbb{C};|\mathfrak{\mathfrak{a}}|)$, we define the (possibly
empty) set
\begin{equation}
\mathit{\Omega }_{\mathfrak{m}}\left( c\right) \doteq \left\{ \omega \in
\mathit{M}_{\Phi _{\mathfrak{m}}(c)}:e_{(\cdot )}\left( \omega \right)
=c\right\} \subseteq E_{1}\text{ },  \label{subset of a face}
\end{equation}%
where, for any fixed translation-invariant state $\rho \in E_{1}$, the
continuous and bounded mapping $e_{(\cdot )}\left( \rho \right) :\mathbb{%
S\rightarrow C}$ is defined from (\ref{ssssssssss})--(\ref{eq:enpersite}) by%
\begin{equation}
e_{\Psi }\left( \rho \right) \doteq \rho \left( \mathfrak{e}_{\Psi }\right)
\ ,\qquad \Psi \in \mathbb{S}\ ,  \label{ssssssssssssssssssss}
\end{equation}%
while $\mathit{M}_{\Phi _{\mathfrak{m}}(c)}$ is the set (\ref{minimizer
short range}) of equilibrium states associated with the approximating
interaction $\Phi =\Phi _{\mathfrak{m}}(c)\in \mathcal{W}_{1}^{\mathbb{R}}$\
defined by (\ref{approx interaction}). Recall that $\mathit{M}_{\Phi _{%
\mathfrak{m}}(c)}$ is a weak$^{\ast }$-closed face of $E_{1}$. Then, we
obtain a (static) self-consistency condition\ for generalized equilibrium
states, which says that any extreme point of $\mathit{\Omega }_{\mathfrak{m}%
} $ must belong to the set
\begin{equation}
\mathit{\Omega }_{\mathfrak{m}}\left( d_{-}+\mathrm{r}_{+}(d_{-})\right)
\label{set gap}
\end{equation}%
for some $d_{-}\in \mathcal{C}_{\mathfrak{m}}$, where $\mathrm{r}_{+}$ is
defined by (\ref{thermodyn decision rule}), and $\mathcal{C}_{\mathfrak{m}}$
is the non-empty, norm-bounded, weakly compact set defined by (\ref{eq
conserve strategy}). This self-consistency condition refers, in a sense, to
Euler-Lagrange equations for the variational problem defining the
thermodynamic game. More precisely, we have the following statements:

\begin{theorem}[Self-consistency of generalized equilibrium states --
\protect\cite{BruPedra2}]
\label{theorem structure of omega}\mbox{ }\newline
Let $\mathfrak{m}\in \mathcal{M}_{1}$ be any translation-invariant
long-range model. \newline
\emph{(i)}
\begin{equation*}
\mathit{\Omega }_{\mathfrak{m}}=\overline{\mathrm{co}}\left( \underset{%
d_{-}\in \mathcal{C}_{\mathfrak{m}}}{\cup }\mathit{\Omega }_{\mathfrak{m}%
}\left( d_{-}+\mathrm{r}_{+}(d_{-})\right) \right) \ .
\end{equation*}%
\emph{(ii)} The set $\mathcal{E}(\mathit{\Omega }_{\mathfrak{m}})$ of
extreme points of the weak$^{\ast }$-compact convex set $\mathit{\Omega }_{%
\mathfrak{m}}$ is included in the union of the sets%
\begin{equation*}
\mathcal{E}\left( \mathit{\Omega }_{\mathfrak{m}}\left( d_{-}+\mathrm{r}%
_{+}(d_{-})\right) \right) \text{ },\qquad d_{-}\in \mathcal{C}_{\mathfrak{m}%
}\text{ },
\end{equation*}%
of all extreme points of $\mathit{\Omega }_{\mathfrak{m}}\left( d_{-}+%
\mathrm{r}_{+}(d_{-})\right) $, $d_{-}\in \mathcal{C}_{\mathfrak{m}}$, which
are non-empty, convex, mutually disjoint, weak$^{\ast }$-closed subsets of $%
E_{1}$.
\end{theorem}

\noindent Assertion (i) results from \cite[Theorem 2.21 (i)]{BruPedra2} and
\cite[Theorem 2.39 (i)]{BruPedra2}, while (ii) corresponds to \cite[Theorem
2.39 (ii)]{BruPedra2}.

Theorem \ref{theorem structure of omega} implies in particular that, for any
extreme state $\hat{\omega}\in \mathcal{E}(\mathit{\Omega }_{\mathfrak{m}})$
of $\mathit{\Omega }_{\mathfrak{m}}$, there is a unique $d_{-}\in \mathcal{C}%
_{\mathfrak{m}}$ such that
\begin{equation}
d\doteq d_{-}+\mathrm{r}_{+}(d_{-})=e_{(\cdot )}(\hat{\omega})\ .
\label{gap equations}
\end{equation}%
In the Physics literature on superconductors, the above equality refers to
the so-called gap equations. Conversely, for any $d_{-}\in \mathcal{C}_{%
\mathfrak{m}}$, there is some generalized equilibrium state $\omega $
satisfying the condition above, but $\omega $ is not necessarily an extreme
point of $\mathit{\Omega }_{\mathfrak{m}}$. In the case of purely attractive
long-range models $\mathfrak{m}=(\Phi ,\mathfrak{\mathfrak{a}})\in \mathcal{M%
}_{1}$, i.e., if $\mathfrak{a}_{+}=0$ (see Section \ref{Section purely
attrac}, in particular (\ref{Hahn decomposition})), we get a stronger
version of Theorem \ref{theorem structure of omega} (ii) as a direct
consequence of (its previous version and) the following proposition:

\begin{proposition}[Self-consistency of generalized equilibrium states --
\protect\cite{BruPedra2}]
\label{theorem structure of omega copy(1)}\mbox{ }\newline
If $\mathfrak{m}\in \mathcal{M}_{1}$ is purely attractive then, for all $%
d_{-}\in \mathcal{C}_{\mathfrak{m}}$, one has
\begin{equation*}
\mathit{\Omega }_{\mathfrak{m}}\left( d_{-}+\mathrm{r}_{+}(d_{-})\right) =%
\mathit{M}_{\Phi _{\mathfrak{m}}(d_{-}+\mathrm{r}_{+}(d_{-}))}\text{ }.
\end{equation*}%
In particular, the sets $\mathit{\Omega }_{\mathfrak{m}}\left( d_{-}+\mathrm{%
r}_{+}(d_{-})\right) $ are weak$^{\ast }$-closed faces of $E_{1}$.
\end{proposition}

\noindent See \cite[Proposition 7.4]{BruPedra2}. By \cite[Remark 2.40]%
{BruPedra2} and Theorem \ref{theorem choquet}, if $\mathfrak{m}\in \mathcal{M%
}_{1}$ is purely attractive then $\mathit{\Omega }_{\mathfrak{m}}$ is a
Choquet simplex with%
\begin{equation}
\mathcal{E}(\mathit{\Omega }_{\mathfrak{m}})=\underset{d_{-}\in \mathcal{C}_{%
\mathfrak{m}}}{\cup }\mathcal{E}\left( \mathit{\Omega }_{\mathfrak{m}}\left(
d_{-}+\mathrm{r}_{+}(d_{-})\right) \right) \subseteq \mathcal{E}(E_{1})\ .
\label{ddfdsfdf}
\end{equation}%
Compare with Theorem \ref{theorem structure of omega} (ii) and see Lemma \ref%
{Thm cool (2)} below. This result also holds true for long-range models
which are not necessarily purely attractive, but have instead the following
property:

\begin{definition}[Simple long-range models]
\label{def simple}\mbox{ }\newline
We say that the long-range model $\mathfrak{m}\in \mathcal{M}_{1}$ is simple
iff, for all $d_{-}\in \mathcal{C}_{\mathfrak{m}}$, the set $\mathit{M}%
_{\Phi _{\mathfrak{m}}(d_{-}+\mathrm{r}_{+}(d_{-}))}$ consists of one single
point.
\end{definition}

This definition means that the effective interactions describing a simple
long-range model, via the so-called Bogoliubov approximation, refer to
fermion systems without (first-order) phase transitions, i.e., with a unique
equilibrium state. Remark that this property is always true for long-range
models leading to approximating (short-range) interactions that are
quasi-free, like in the BCS theory. Such a property is relevant here because
it prevents the long-range repulsion from breaking the face structure of the
set of generalized equilibrium states (see also \cite[Lemma 9.8]{BruPedra2}%
). This is a consequence of the following assertion, which is similar to
Proposition \ref{theorem structure of omega copy(1)}\ for purely attractive
models:

\begin{proposition}[Self-consistency of equilibrium states of simple models]

\label{theorem structure of omega copy(2)}\mbox{ }\newline
If $\mathfrak{m}\in \mathcal{M}_{1}$ is simple then, for all $d_{-}\in
\mathcal{C}_{\mathfrak{m}}$, one has
\begin{equation*}
\mathit{\Omega }_{\mathfrak{m}}\left( d_{-}+\mathrm{r}_{+}(d_{-})\right) =%
\mathit{M}_{\Phi _{\mathfrak{m}}(d_{-}+\mathrm{r}_{+}(d_{-}))}\ .
\end{equation*}%
In particular, the sets $\mathit{\Omega }_{\mathfrak{m}}\left( d_{-}+\mathrm{%
r}_{+}(d_{-})\right) $ are (trivially) weak$^{\ast }$-closed faces of $E_{1}$%
.
\end{proposition}

\begin{proof}
For any model $\mathfrak{m}\in \mathcal{M}_{1}$,%
\begin{equation*}
\emptyset \neq \mathit{\Omega }_{\mathfrak{m}}\left( d_{-}+\mathrm{r}%
_{+}(d_{-})\right) \subseteq \mathit{M}_{\Phi _{\mathfrak{m}}(d_{-}+\mathrm{r%
}_{+}(d_{-}))}\text{ },\qquad d_{-}\in \mathcal{C}_{\mathfrak{m}}\text{ },
\end{equation*}%
by Theorem \ref{theorem structure of omega} (ii). Hence, if $\mathfrak{m}$
is simple, i.e., $\mathit{M}_{\Phi _{\mathfrak{m}}(d_{-}+\mathrm{r}%
_{+}(d_{-}))}$ consists of one single point for every $d_{-}\in \mathcal{C}_{%
\mathfrak{m}}$, then the equality stated in the proposition must be
satisfied.
\end{proof}

By Proposition \ref{theorem structure of omega copy(2)}, similar to purely
attractive long-range models, $\mathit{\Omega }_{\mathfrak{m}}$ is a Choquet
simplex and Equation (\ref{ddfdsfdf}) also holds true for all simple models $%
\mathfrak{m}\in \mathcal{M}_{1}$. See also Lemma \ref{Thm cool (2)} below.
This is another improvement of Theorem \ref{theorem structure of omega} (ii)
in the case of simple long-range models.

\subsubsection{Extreme Decompositions of Generalized Equilibrium States}

By \cite[Lemma 2.16]{BruPedra2}, for any long-range model $\mathfrak{m}%
=(\Phi ,\mathfrak{\mathfrak{a}})\in \mathcal{M}_{1}$, the non-empty set $%
\mathit{\Omega }_{\mathfrak{m}}\subseteq E_{1}$ is weak$^{\ast }$-compact
and convex. If the model is purely attractive, i.e., $\mathfrak{a}_{+}=0$,
or simple then $\mathit{\Omega }_{\mathfrak{m}}$ is a face of $E_{1}$, by
Equation (\ref{ddfdsfdf}). Nevertheless, in general, $\mathit{\Omega }_{%
\mathfrak{m}}$ may not be a face (see \cite[Lemma 9.8]{BruPedra2}) and we
would like to know whether, despite of this fact, the Choquet measures
representing elements of $\mathit{\Omega }_{\mathfrak{m}}$ are \emph{%
orthogonal} measures. This is important in order to be able to use the
theory of direct integrals of measurable families of Hilbert spaces,
operators, von Neumann algebras, and $C^{\ast }$-algebra representations, as
described in \cite[Sections 5-6]{BruPedra-MFIII}, together with the Effros
Theorem \cite[Theorem 4.4.9]{BrattelliRobinsonI}. Unfortunately, as soon as
we have long-range repulsions, the orthogonality property can be lost:

\begin{theorem}[Non-orthogonality of extremal decompositions in $\mathit{%
\Omega }_{\mathfrak{m}}$]
\label{lemma explosion l du mec copacabana2}\mbox{ }\newline
Assume that $\left\vert \mathrm{S}\right\vert \geq 4$. Then, there are
uncountably many models $\mathfrak{m}=(\Phi ,\mathfrak{\mathfrak{a}})\in
\mathcal{M}_{1}$, with $\mathfrak{a}=\mathfrak{a}_{+}$ (i.e., the long-range
model is purely repulsive), having a generalized equilibrium state $\omega
\in \mathit{\Omega }_{\mathfrak{m}}$, whose (Choquet) decomposition on the
set $\mathcal{E}(\mathit{\Omega }_{\mathfrak{m}})$ of extreme points of $%
\mathit{\Omega }_{\mathfrak{m}}$ is non-orthogonal.
\end{theorem}

\begin{proof}
Given any fixed $\theta \in \mathbb{R}/(2\pi \mathbb{Z)}$, recall that $%
\mathrm{g}_{\theta }$ is the unique $\ast $-automorphism of the $C^{\ast }$%
-algebra $\mathcal{U}$ defined by Equation (\ref{automorphism gauge
invariance}). Pick a (non self-adjoint) local element $A\in \mathcal{U}_{0}$
satisfying $\left\Vert A\right\Vert _{\mathcal{U}}=1$ and%
\begin{equation}
A=-\mathrm{g}_{\theta _{1}}(A)=-\mathrm{g}_{\theta _{2}}(A)
\label{ssssssfffggbbb}
\end{equation}%
for some $\theta _{1},\theta _{2}\in \mathbb{R}/(2\pi \mathbb{Z})$, $\theta
_{1}\neq \theta _{2}$. Assume also that, for some translation-invariant
state $\hat{\rho}_{0}\in E_{1}$, one has $\hat{\rho}_{0}(A)\neq 0$ as well
as
\begin{equation}
\hat{\rho}_{0}\circ \mathrm{g}_{\theta _{1}}\neq \hat{\rho}_{0}\circ \mathrm{%
g}_{\theta _{2}}\text{ }.  \label{dfdfdf}
\end{equation}%
For instance, for all $\lambda \in \mathbb{C}$ and any $\left( x_{1},\mathrm{%
s}_{1}\right) ,\ldots ,\left( x_{4},\mathrm{s}_{4}\right) \in \mathbb{Z}%
^{d}\times \mathrm{S}$,%
\begin{equation*}
\mathrm{g}_{-\pi /4}\left( \lambda a_{x_{1},\mathrm{s}_{1}}a_{x_{2},\mathrm{s%
}_{2}}a_{x_{3},\mathrm{s}_{3}}a_{x_{4},\mathrm{s}_{4}}\right) =-\lambda
a_{x_{1},\mathrm{s}_{1}}a_{x_{2},\mathrm{s}_{2}}a_{x_{3},\mathrm{s}%
_{3}}a_{x_{4},\mathrm{s}_{4}}=\mathrm{g}_{\pi /4}\left( \lambda a_{x_{1},%
\mathrm{s}_{1}}a_{x_{2},\mathrm{s}_{2}}a_{x_{3},\mathrm{s}_{3}}a_{x_{4},%
\mathrm{s}_{4}}\right) \ .
\end{equation*}%
Note that
\begin{equation*}
A\doteq \lambda a_{x_{1},\mathrm{s}_{1}}a_{x_{2},\mathrm{s}_{2}}a_{x_{3},%
\mathrm{s}_{3}}a_{x_{4},\mathrm{s}_{4}}\neq 0
\end{equation*}%
if $\lambda \neq 0$ and $\left( x_{1},\mathrm{s}_{1}\right) ,\left( x_{2},%
\mathrm{s}_{2}\right) ,\left( x_{3},\mathrm{s}_{3}\right) ,\left( x_{4},%
\mathrm{s}_{4}\right) $ are different from each other. If $\left\vert
\mathrm{S}\right\vert \geq 4$ and $\mathrm{s}_{1},\mathrm{s}_{2},\mathrm{s}%
_{3},\mathrm{s}_{4}\in \mathrm{S}$ are different from each other, then there
is a product state $\hat{\rho}_{0}\in E_{1}$ such that, for all $x\in
\mathbb{Z}^{d}$,%
\begin{equation*}
\hat{\rho}_{0}(a_{x,\mathrm{s}_{1}}a_{x,\mathrm{s}_{2}}a_{x,\mathrm{s}%
_{3}}a_{x,\mathrm{s}_{4}})=\hat{\rho}_{0}(a_{x,\mathrm{s}_{1}}a_{x,\mathrm{s}%
_{2}})\hat{\rho}_{0}(a_{x,\mathrm{s}_{3}}a_{x,\mathrm{s}_{4}})\neq 0\ ,
\end{equation*}%
because on-site states separate the elements of the on-site $C^{\ast }$%
-algebra $\mathcal{U}_{\{0\}}$ and $a_{0,\mathrm{s}_{1}}a_{0,\mathrm{s}%
_{2}}a_{0,\mathrm{s}_{3}}a_{0,\mathrm{s}_{4}}\in \mathcal{U}_{\{0\}}$ is a
non-vanishing even\footnote{%
On the one hand, a product state is constructed from an even on-site state,
by \cite[Theorem 11.2]{Araki-Moriya}. On the other hand, there is a on-site
state separating $0$ and $a_{0,\mathrm{s}_{1}}a_{0,\mathrm{s}_{2}}a_{0,%
\mathrm{s}_{3}}a_{0,\mathrm{s}_{4}}$ and, since $a_{0,\mathrm{s}_{1}}a_{0,%
\mathrm{s}_{2}}a_{0,\mathrm{s}_{3}}a_{0,\mathrm{s}_{4}}$ is even, one can
assume that this on-site state is even.} element. Observe that, in this
case, Equation (\ref{dfdfdf}) holds true for $\theta _{1}=-\pi /4$ and $%
\theta _{2}=\pi /4$, i.e., $\hat{\rho}_{0}\circ \mathrm{g}_{-\pi /4}\neq
\hat{\rho}_{0}\circ \mathrm{g}_{\pi /4}$, because
\begin{equation*}
\hat{\rho}_{0}\circ \mathrm{g}_{-\pi /4}(a_{0,\mathrm{s}_{1}}a_{0,\mathrm{s}%
_{2}})=\mathrm{e}^{-i\pi /2}\hat{\rho}_{0}(a_{0,\mathrm{s}_{1}}a_{0,\mathrm{s%
}_{2}})\neq \mathrm{e}^{i\pi /2}\hat{\rho}_{0}(a_{0,\mathrm{s}_{1}}a_{0,%
\mathrm{s}_{2}})=\hat{\rho}_{0}\circ \mathrm{g}_{\pi /4}(a_{0,\mathrm{s}%
_{1}}a_{0,\mathrm{s}_{2}})\ ,
\end{equation*}%
since $\hat{\rho}_{0}(a_{0,\mathrm{s}_{1}}a_{0,\mathrm{s}_{2}})\neq 0$. Note
also that
\begin{equation*}
\{\lambda a_{0,\mathrm{s}_{1}}a_{0,\mathrm{s}_{2}}a_{0,\mathrm{s}_{3}}a_{0,%
\mathrm{s}_{4}}:\lambda \in \mathbb{C},\left\Vert \lambda a_{0,\mathrm{s}%
_{1}}\cdots a_{0,\mathrm{s}_{4}}\right\Vert _{\mathcal{U}}=1\}\subseteq
\mathcal{U}
\end{equation*}%
is an uncountable set.

We can also assume that $\hat{\rho}_{0}\in \mathcal{E}(E_{1})$, i.e., $\hat{%
\rho}_{0}$ is ergodic. In fact, note that the above example already
corresponds to this special case, for product states are always ergodic. By
Equation (\ref{ssssssfffggbbb}),
\begin{equation}
\hat{\rho}_{0}\left( A\right) =-\hat{\rho}_{1}\left( A\right) =-\hat{\rho}%
_{2}\left( A\right) \neq 0\text{ },  \label{mec copa eq1}
\end{equation}%
where $\hat{\rho}_{1}\doteq \hat{\rho}_{0}\circ \mathrm{g}_{\theta _{1}}$
and $\hat{\rho}_{2}\doteq \hat{\rho}_{0}\circ \mathrm{g}_{\theta _{2}}$.
Since $\mathrm{g}_{\theta }$ is a $\ast $-auto%
\-%
morphism of $\mathcal{U}$ and $\mathrm{g}_{\theta _{1}}\neq \mathrm{g}%
_{\theta _{2}}$, $\hat{\rho}_{1}\neq \hat{\rho}_{0}$ and $\hat{\rho}_{2}\neq
\hat{\rho}_{0}$ are two different states, see (\ref{dfdfdf}). As $\hat{\rho}%
_{0}\in \mathcal{E}(E_{1})$, by using the relations $\alpha _{x}\circ
\mathrm{g}_{\theta }=\mathrm{g}_{\theta }\circ \alpha _{x}$ for all $\theta
\in \mathbb{R}/(2\pi \mathbb{Z)}$ and $x\in \mathbb{Z}^{d}$, we infer from
Equations (\ref{Limit of Space-Averages})--(\ref{Ergodicity2}) that the
states $\hat{\rho}_{1}$ and $\hat{\rho}_{2}$ are also extreme states of $%
E_{1}$, i.e., $\hat{\rho}_{1}\in \mathcal{E}(E_{1})$ and $\hat{\rho}_{2}\in
\mathcal{E}(E_{1})$.

As explained in the proof of \cite[Lemma 4.18]{BruPedra2}, for any local
element $A\in \mathcal{U}_{0}$, there exists a finite-range
translation-invariant interaction $\Phi ^{A}\in \mathcal{W}_{1}$ such that
\begin{equation}
\Vert \Phi ^{A}\Vert _{\mathcal{W}_{1}}=\Vert A\Vert _{\mathcal{U}}\qquad
\text{and}\qquad \rho (A)=e_{\Phi ^{A}}(\rho )\ ,\qquad \rho \in E_{1},
\label{mec copa eq2}
\end{equation}%
with $e_{\Phi ^{A}}:E_{1}\rightarrow \mathbb{C}$ being defined by Equation (%
\ref{ssssssssss}) for $\Phi =\Phi ^{A}$. For instance, assuming that $%
\left\vert \mathrm{S}\right\vert \geq 4$, if $A=\lambda a_{0,\mathrm{s}%
_{1}}a_{0,\mathrm{s}_{2}}a_{0,\mathrm{s}_{3}}a_{0,\mathrm{s}_{4}}$, where $%
\mathrm{s}_{1},\mathrm{s}_{2},\mathrm{s}_{3},\mathrm{s}_{4}\in \mathrm{S}$
are different from each other and $\lambda \in \mathbb{C}$ is such that $%
\Vert A\Vert _{\mathcal{U}}=1$, then $\Phi ^{A}\in \mathcal{W}_{1}$ is the
interaction defined by
\begin{equation*}
\Phi _{\{x\}}^{A}=\lambda a_{x,\mathrm{s}_{1}}a_{x,\mathrm{s}_{2}}a_{x,%
\mathrm{s}_{3}}a_{x,\mathrm{s}_{4}},\quad x\in \mathbb{Z}^{d},\qquad \text{%
and}\qquad \Phi _{\Lambda }^{A}=0\quad \text{otherwise.}
\end{equation*}%
Observe that the constant $\lambda \in \mathbb{C}$ is fixed in such a way $%
\Phi ^{A}\in \mathbb{S}\subseteq \mathcal{W}_{1}$, i.e., $\Vert \Phi
^{A}\Vert _{\mathcal{W}_{1}}=\Vert A\Vert _{\mathcal{U}}=1$. Let $\mathfrak{a%
}_{+}$ be defined, for all Borel subset $\mathfrak{B}\subseteq \mathbb{S}$,
by
\begin{equation*}
\mathfrak{a}_{+}\left( \mathfrak{B}\right) =\mathbf{1}\left[ \Phi ^{A}\in
\mathfrak{B}\right] \ .
\end{equation*}%
Since $\hat{\rho}_{0},\hat{\rho}_{1},\hat{\rho}_{2}\in \mathcal{E}(E_{1})$
are all extreme states, thanks to \cite[Lemma 9.7]{BruPedra2}, there is $%
\Phi \in \mathcal{W}_{1}^{\mathbb{R}}$ such that $\hat{\rho}_{0},\hat{\rho}%
_{1},\hat{\rho}_{2}\in \mathit{M}_{\Phi }$, see (\ref{minimizer short range}%
). Let $\mathfrak{m}_{A}\doteq \left( \Phi ,\mathfrak{a}_{+}\right) \in
\mathcal{M}_{1}$, which is a purely repulsive translation-invariant
long-range model. By convexity of $\mathit{M}_{\Phi }$,%
\begin{equation}
\omega _{1}\doteq \frac{1}{2}\hat{\rho}_{0}+\frac{1}{2}\hat{\rho}_{1}\in
\mathit{M}_{\Phi }\qquad \text{and}\qquad \omega _{2}\doteq \frac{1}{2}\hat{%
\rho}_{0}+\frac{1}{2}\hat{\rho}_{2}\in \mathit{M}_{\Phi }\ .
\label{mec copa eq3}
\end{equation}%
By assumption, $\hat{\rho}_{1}\neq \hat{\rho}_{2}$ and thus, $\omega
_{1}\neq \omega _{2}$.

Consider the convex and weak$^{\ast }$-lower semi-continuous functional $g_{%
\mathfrak{m}_{A}}:E_{1}\rightarrow \mathbb{R}$ defined by%
\begin{equation*}
g_{\mathfrak{m}_{A}}(\rho )=|\rho (A)|^{2}+f_{\Phi }(\rho )\text{ },\qquad
\rho \in E_{1}.
\end{equation*}%
It turns out that the generalized equilibrium states of $\mathfrak{m}_{A}$
are exactly the minimizers of this functional, by \cite[Theorem 2.25 (+)]%
{BruPedra2}. From Equations (\ref{Limit of Space-Averages})--(\ref%
{Ergodicity2}), (\ref{Free-energy density long range0})--(\ref{Free-energy
density long range}), (\ref{definition equilibirum state}) and (\ref{mec
copa eq1})--(\ref{mec copa eq3}), it follows that $\omega _{1},\omega
_{2}\in \mathit{\Omega }_{\mathfrak{m}_{A}}$ and any minimizer $\omega \in
E_{1}$ of $g_{\mathfrak{m}_{A}}$ satisfies $e_{\Phi ^{A}}(\omega )=0$. Thus,
generalized equilibrium states $\omega \in \mathit{\Omega }_{\mathfrak{m}%
_{A}}$ have to satisfy the equality $e_{\Phi ^{A}}(\omega )=0$. Observe that
this refers to the (static) self-consistency of generalized equilibrium
states of long-range models. See, for instance, Theorem \ref{theorem
structure of omega} (ii) with $\mathcal{C}_{\mathfrak{m}}=\{0\}$ and $%
\mathrm{r}_{+}(0)=0$.

Now, using (\ref{mec copa eq3}) and Theorem \ref{theorem choquet} (in
particular that $E_{1}$ is a Choquet simplex), we deduce that
\begin{equation*}
\omega _{1},\omega _{2}\in \mathcal{E}(\mathit{\Omega }_{\mathfrak{m}%
_{A}})\subseteq E_{1}\ ,
\end{equation*}%
by construction of the long-range model $\mathfrak{m}_{A}$. In fact, if $%
\omega _{1},\omega _{2}$ were not both extreme in $\mathit{\Omega }_{%
\mathfrak{m}_{A}}$ then we would have states $\omega _{1}^{\prime },\omega
_{1}^{\prime \prime },\omega _{2}^{\prime },\omega _{2}^{\prime \prime }\in
\mathit{\Omega }_{\mathfrak{m}_{A}}$, $\omega _{1}^{\prime }\neq \omega
_{1}^{\prime \prime }$, $\omega _{2}^{\prime }\neq \omega _{2}^{\prime
\prime }$, such that
\begin{equation*}
\omega _{1}=\frac{1}{2}\omega _{1}^{\prime }+\frac{1}{2}\omega _{1}^{\prime
\prime }\qquad \text{and}\qquad \omega _{2}=\frac{1}{2}\omega _{2}^{\prime }+%
\frac{1}{2}\omega _{2}^{\prime \prime }\text{ }.
\end{equation*}%
In this case, as $\hat{\rho}_{0},\hat{\rho}_{1}\in \mathcal{E}(E_{1})$, the
supports of the (Choquet) measures $\mu _{\omega _{1}^{\prime }}$ and $\mu
_{\omega _{1}^{\prime \prime }}$ decomposing $\omega _{1}^{\prime }$ and $%
\omega _{1}^{\prime \prime }$ in $E_{1}$ are contained in $\{\hat{\rho}_{0},%
\hat{\rho}_{1}\}$. This means that $\omega _{1}^{\prime }$ is a convex
combination of $\hat{\rho}_{0}$ and $\hat{\rho}_{1}$. But, because of the
above (static) self-consistency condition for generalized equilibrium
states, the unique convex combination of $\hat{\rho}_{0}$ and $\hat{\rho}%
_{1} $ which is an element of $\mathit{\Omega }_{\mathfrak{m}_{A}}$, is $%
\omega _{1}$ itself. From this we would conclude that $\omega _{1}^{\prime
}=\omega _{1}$ and, hence, $\omega _{1}^{\prime }=\omega _{1}^{\prime \prime
}$. Using exactly the same argument for the state $\omega _{2}$, we would
arrive at $\omega _{2}^{\prime }=\omega _{2}^{\prime \prime }$.

Finally, by \cite[Lemma 4.1.19 and Definition 4.1.20]{BrattelliRobinsonI}, $%
\omega _{1}$ and $\omega _{2}$, which are two \emph{different} elements of $%
\mathcal{E}(\mathit{\Omega }_{\mathfrak{m}_{A}})$, are \emph{not} orthogonal
because $\hat{\rho}_{0}/2\leq \omega _{1}$, $\hat{\rho}_{0}/2\leq \omega
_{2} $ and $\hat{\rho}_{0}/2$ is a non-zero positive functional (as $\hat{%
\rho}_{0}$ is a state). In particular, the state
\begin{equation*}
\omega _{0}\doteq \frac{1}{2}\omega _{1}+\frac{1}{2}\omega _{2}=\frac{1}{2}%
\hat{\rho}_{0}+\frac{1}{4}\hat{\rho}_{1}+\frac{1}{4}\hat{\rho}_{2}\in
\mathit{\Omega }_{\mathfrak{m}_{A}}\backslash \mathcal{E}(\mathit{\Omega }_{%
\mathfrak{m}_{A}})
\end{equation*}%
has a non-orthogonal (Choquet) decomposition on the set $\mathcal{E}(\mathit{%
\Omega }_{\mathfrak{m}})$ of extreme points of $\mathit{\Omega }_{\mathfrak{m%
}}$.
\end{proof}

By Theorem \ref{theorem structure of omega} (i), note meanwhile that, for
all translation-invariant long-range models $\mathfrak{m}\in \mathcal{M}_{1}$%
,
\begin{equation}
\mathit{\Omega }_{\mathfrak{m}}\subseteq \overline{\mathrm{co}}\left(
\mathbf{M}_{\mathfrak{m}}\right) \subseteq E_{1}\qquad \text{with}\qquad
\mathbf{M}_{\mathfrak{m}}\doteq \underset{d_{-}\in \mathcal{C}_{\mathfrak{m}}%
}{\cup }\mathit{M}_{\Phi _{\mathfrak{m}}(d_{-}+\mathrm{r}_{+}(d_{-}))}%
\subseteq E_{1}\ .  \label{def M}
\end{equation}%
Therefore, we can alternatively decompose generalized equilibrium states
within the weak$^{\ast }$-compact convex set $\overline{\mathrm{co}}\left(
\mathbf{M}_{\mathfrak{m}}\right) $. In contrast to the (Choquet)
decomposition within $\mathit{\Omega }_{\mathfrak{m}}$, in this situation
the decomposition of an arbitrary generalized equilibrium state is always
orthogonal, for it coincides with its ergodic decomposition:

\begin{theorem}[Ergodic decomposition of generalized equilibrium states]
\label{Thm cool}\mbox{ }\newline
For any $\mathfrak{m}\in \mathcal{M}_{1}$, $\overline{\mathrm{co}}\left(
\mathbf{M}_{\mathfrak{m}}\right) $ is a weak$^{\ast }$-closed face of $E_{1}$
and, for any $\omega \in \mathit{\Omega }_{\mathfrak{m}}\subseteq \overline{%
\mathrm{co}}\left( \mathbf{M}_{\mathfrak{m}}\right) \subseteq E_{1}$, the
unique Choquet probability measure $\mu _{\omega }$ on $\overline{\mathrm{co}%
}\left( \mathbf{M}_{\mathfrak{m}}\right) $ (or $E_{1}$) representing $\omega
$ (see Theorem \ref{theorem choquet}) satisfies%
\begin{equation*}
\mu _{\omega }\left( \mathcal{E}(E_{1})\cap \mathbf{M}_{\mathfrak{m}}\right)
=1\ .
\end{equation*}
\end{theorem}

\begin{proof}
Fix without loss of generality any long-range model $\mathfrak{m}=(\Phi ,%
\mathfrak{\mathfrak{a}})\in \mathcal{M}_{1}$ such that $\mathfrak{\mathfrak{a%
}}_{\pm }\neq 0$. (The cases $\mathfrak{\mathfrak{a}}_{-}=0$ or $\mathfrak{%
\mathfrak{a}}_{+}=0$ are clearly simpler.) Assume that the set $\mathbf{M}_{%
\mathfrak{m}}$ is weak$^{\ast }$-closed. By the Milman theorem \cite[Theorem
10.13 (ii)]{BruPedra2}, $\mathcal{E}(\overline{\mathrm{co}}\left( \mathbf{M}%
_{\mathfrak{m}}\right) )\subseteq \mathbf{M}_{\mathfrak{m}}$. Thus, since,
for any $\Psi \in \mathcal{W}_{1}^{\mathbb{R}}$, the non-empty convex set $%
\mathit{M}_{\Psi }$ is a (weak$^{\ast }$-closed) face of $E_{1}$, one has
\begin{equation}
\mathcal{E}(\overline{\mathrm{co}}\left( \mathbf{M}_{\mathfrak{m}}\right)
)\subseteq \underset{d_{-}\in \mathcal{C}_{\mathfrak{m}}}{\cup }\mathcal{E}%
\left( \mathit{M}_{\Phi _{\mathfrak{m}}(d_{-}+\mathrm{r}_{+}(d_{-}))}\right)
\subseteq \mathcal{E}\left( E_{1}\right) \ .  \label{extremal state utiles}
\end{equation}%
Hence, $\overline{\mathrm{co}}\left( \mathbf{M}_{\mathfrak{m}}\right) $ is a
weak$^{\ast }$-closed face of the simplex $E_{1}$. Therefore, by Theorem \ref%
{theorem choquet}, for any $\omega \in \mathit{\Omega }_{\mathfrak{m}%
}\subseteq \overline{\mathrm{co}}\left( \mathbf{M}_{\mathfrak{m}}\right) $,
there is a unique probability measure $\mu _{\omega }$ on $\mathbf{M}_{%
\mathfrak{m}}$ such that%
\begin{equation}
\mu _{\omega }(\mathcal{E}(\overline{\mathrm{co}}\left( \mathbf{M}_{%
\mathfrak{m}}\right) ))=1\mathrm{\quad }\text{and}\mathrm{\quad }\omega
\left( A\right) =\int_{\mathcal{E}(\overline{\mathrm{co}}\left( \mathbf{M}_{%
\mathfrak{m}}\right) )}\hat{\omega}\left( A\right) \ \mu _{\omega }(\mathrm{d%
}\hat{\omega})\text{ },\qquad A\in \mathcal{U}\text{ }.
\label{extremal decomposition eq states}
\end{equation}%
By Equation (\ref{extremal state utiles}), $\mu _{\omega }$ is supported on $%
\mathcal{E}(E_{1})$. Therefore, the theorem follows from (\ref{extremal
decomposition eq states}), provided one proves the weak$^{\ast }$-closedness
of the set $\mathbf{M}_{\mathfrak{m}}$.

In order to prove that $\mathbf{M}_{\mathfrak{m}}$ is indeed weak$^{\ast }$%
-closed, take any sequence $(\omega _{n})_{n=1}^{\infty }\subseteq \mathbf{M}%
_{\mathfrak{m}}$ converging in the weak$^{\ast }$-topology to $\omega
_{\infty }\in E_{1}$. (Note that $E_{1}$ is weak$^{\ast }$-closed, being
even weak$^{\ast }$-compact, and the use of nets is here not necessary, as
the weak$^{\ast }$ topology of $E_{1}$ is metrizable.) Then, by Equation (%
\ref{def M}), for every $n\in \mathbb{N}$, there is an element $%
d_{-}^{(n)}\in \mathcal{C}_{\mathfrak{m}}$ such that
\begin{equation*}
\omega _{n}\in \mathit{M}_{\Phi _{\mathfrak{m}}(d_{-}^{(n)}+\mathrm{r}%
_{+}(d_{-}^{(n)}))}\ .
\end{equation*}%
Since $\mathcal{C}_{\mathfrak{m}}\subseteq L_{-}^{2}(\mathbb{S};\mathbb{C};|%
\mathfrak{\mathfrak{a}}|)$ is a (non-empty) norm-bounded and weakly compact
set \cite[Lemma 8.4 ($\sharp $)]{BruPedra2}], the sequence $%
(d_{-}^{(n)})_{n=1}^{\infty }$ converges (along a subsequence again denoted
by $(d_{-}^{(n)})_{n=1}^{\infty }$) in the weak topology to an element $%
d_{-}^{(\infty )}\in \mathcal{C}_{\mathfrak{m}}$ within a ball $\mathcal{B}%
_{R}\left( 0\right) \subseteq L^{2}(\mathbb{S};\mathbb{C};|\mathfrak{%
\mathfrak{a}}|)$ of sufficiently large radius $R>0$. Since, by \cite[Lemma
8.8]{BruPedra2}, the thermodynamic decision rule (\ref{thermodyn decision
rule}) defines a continuous mapping from $L_{-}^{2}(\mathbb{S};\mathbb{C})$
to $L_{+}^{2}(\mathbb{S};\mathbb{C})$ with $L_{-}^{2}(\mathbb{S};\mathbb{C})$
and $L_{+}^{2}(\mathbb{S};\mathbb{C})$ endowed with the weak and norm
topologies, respectively, we then infer from \cite[Proposition 7.1 (ii)]%
{BruPedra2} that
\begin{equation}
\lim_{n\rightarrow \infty }\mathrm{P}_{\Phi _{\mathfrak{m}}(d_{-}^{(n)}+%
\mathrm{r}_{+}(d_{-}^{(n)}))+\Psi }=\mathrm{P}_{\Phi _{\mathfrak{m}%
}(d_{-}^{(\infty )}+\mathrm{r}_{+}(d_{-}^{(\infty )}))+\Psi }\ ,\qquad \Psi
\in \mathcal{W}_{1}^{\mathbb{R}}\text{ }.  \label{limit}
\end{equation}%
(Recall the notation given by (\ref{variational problem approx}).) Now, by
\cite[Theorem 2.28]{BruPedra2} applied to models with vanishing long-range
components, for any $n\in \mathbb{N}$,
\begin{equation*}
\mathrm{P}_{\Phi _{\mathfrak{m}}(d_{-}^{(n)}+\mathrm{r}_{+}(d_{-}^{(n)}))+%
\Psi }-\mathrm{P}_{\Phi _{\mathfrak{m}}(d_{-}^{(n)}+\mathrm{r}%
_{+}(d_{-}^{(n)}))}\geq -e_{\Psi }(\omega _{n})\ ,\qquad \Psi \in \mathcal{W}%
_{1}^{\mathbb{R}}\text{ },
\end{equation*}%
which, combined with (\ref{ssssssssss})-(\ref{eq:enpersite}) and (\ref{limit}%
), implies that
\begin{equation*}
\mathrm{P}_{\Phi _{\mathfrak{m}}(d_{-}^{(\infty )}+\mathrm{r}%
_{+}(d_{-}^{(\infty )}))+\Psi }-\mathrm{P}_{\Phi _{\mathfrak{m}%
}(d_{-}^{(\infty )}+\mathrm{r}_{+}(d_{-}^{(\infty )}))}\geq -e_{\Psi
}(\omega _{\infty })\ ,\qquad \Psi \in \mathcal{W}_{1}^{\mathbb{R}}\text{ }.
\end{equation*}%
Finally, again by \cite[Theorem 2.28]{BruPedra2}, we deduce that
\begin{equation*}
\omega _{\infty }\in \mathit{M}_{\Phi (d_{-}^{(\infty )}+\mathrm{r}%
_{+}(d_{-}^{(\infty )}))}\qquad \text{with }\qquad d_{-}^{(\infty )}\in
\mathcal{C}_{\mathfrak{m}}\ .
\end{equation*}%
It follows from Equation (\ref{def M}) that $\omega _{\infty }\in \mathbf{M}%
_{\mathfrak{m}}$. In other words, $\mathbf{M}_{\mathfrak{m}}$ is weak$^{\ast
}$-closed.
\end{proof}

\begin{corollary}[Generalized equilibrium states as faithful modular states]

\label{eq.tang.bcs.type0 copy(4)}\mbox{ }\newline
For all $\mathfrak{m}\in \mathcal{M}_{1}$, generalized equilibrium states $%
\omega \in \mathit{\Omega }_{\mathfrak{m}}$ are faithful and modular.
\end{corollary}

\begin{proof}
Let $\mathfrak{m}\in \mathcal{M}_{1}$. Since $\mu _{\omega }$ is an
orthogonal measure (Theorem \ref{theorem choquet}) for any $\omega \in
\mathit{\Omega }_{\mathfrak{m}}$, one can use the theory of direct integrals
of GNS representations of families of states described in \cite[Section 5.6]%
{BruPedra-MFIII}. Since, in each fiber, the corresponding state $\hat{\omega}%
\in \mathcal{E}(\overline{\mathrm{co}}\left( \mathbf{M}_{\mathfrak{m}%
}\right) )$ is faithful and modular, by Corollary \ref{eq.tang.bcs.type0
copy(2)} and Equation (\ref{extremal state utiles}), the corollary easily
follows.
\end{proof}

\noindent The property of generalized equilibrium states of
translation-invariant long-range models stated in Corollary \ref%
{eq.tang.bcs.type0 copy(4)} is important, because it allows to use of the
Tomita-Takesaki modular theory \cite[Section 2.5]{BrattelliRobinsonI}.

Note that the situation is much simpler for purely attractive or simple
long-range models (see Section \ref{Section purely attrac} and Definition %
\ref{def simple}), because Theorem \ref{theorem structure of omega} combined
with Propositions \ref{theorem structure of omega copy(1)} and \ref{theorem
structure of omega copy(2)} directly yields the following assertion, which
is merely a reformulation of Equation (\ref{ddfdsfdf}).

\begin{lemma}[Choquet decompositions of generalized equilibrium states]
\label{Thm cool (2)}\mbox{ }\newline
If $\mathfrak{m}\in \mathcal{M}_{1}$ is purely attractive or simple then $%
\mathit{\Omega }_{\mathfrak{m}}=\overline{\mathrm{co}}\left( \mathbf{M}_{%
\mathfrak{m}}\right) $, the set $\mathbf{M}_{\mathfrak{m}}$ being defined by
(\ref{def M}). In particular, the Choquet decomposition in $\mathit{\Omega }%
_{\mathfrak{m}}$ of any generalized equilibrium state of $\mathfrak{m}$
coincides with its ergodic decomposition.
\end{lemma}

\begin{proof}
Let $\mathfrak{m}\in \mathcal{M}_{1}$ be a purely attractive or simple
translation-invariant long-range model. From Proposition \ref{theorem
structure of omega copy(1)} or \ref{theorem structure of omega copy(2)}, in
both cases,
\begin{equation*}
\mathit{\Omega }_{\mathfrak{m}}(d_{-}+\mathrm{r}_{+}(d_{-}))=\mathit{M}%
_{\Phi _{\mathfrak{m}}(d_{-}+\mathrm{r}_{+}(d_{-}))}\ ,\qquad d_{-}\in
\mathcal{C}_{\mathfrak{m}}\text{ }.
\end{equation*}%
The assertion then follows from Equation (\ref{def M}) and Theorem \ref%
{theorem structure of omega} (i).
\end{proof}

To sum up, there are two natural ways to perform Choquet decompositions of
generalized equilibrium states $\omega \in \mathit{\Omega }_{\mathfrak{m}%
}\subseteq E_{1}$, for any fixed translation-invariant long-range model $%
\mathfrak{m}\in \mathcal{M}_{1}$:

\begin{itemize}
\item We can decompose $\omega $ in the weak$^{\ast }$-compact convex set $%
\mathit{\Omega }_{\mathfrak{m}}$ of all generalized equilibrium states of
the long-range model $\mathfrak{m}$. The advantage of doing so is that the
extreme states of $\mathit{\Omega }_{\mathfrak{m}}$ always satisfy the
(static) self-consistency condition (\ref{gap equations}). A drawback,
however, is that the decomposition may not be orthogonal (Theorem \ref{lemma
explosion l du mec copacabana2}) and the Effros Theorem \cite[Theorem 4.4.9]%
{BrattelliRobinsonI} cannot be used. See \cite[Section 5.6]{BruPedra-MFIII}
for more details.

\item We can decompose $\omega $ in the weak$^{\ast }$-compact convex set $%
\overline{\mathrm{co}}\left( \mathbf{M}_{\mathfrak{m}}\right) $, which turns
out to be equivalent to decompose it in $E_{1}$ (Theorem \ref{Thm cool}). In
this case, in contrast to the previous strategy, the decomposition is always
orthogonal while the extreme (ergodic) states of $\overline{\mathrm{co}}%
\left( \mathbf{M}_{\mathfrak{m}}\right) $ are still equilibrium states of
approximating (short-range) interactions. The drawback is now that elements
of $\mathcal{E}(\overline{\mathrm{co}}\left( \mathbf{M}_{\mathfrak{m}%
}\right) )$ may not anymore satisfy the (static) self-consistency condition (%
\ref{gap equations}), which turns out to be essential in the present study,
as it is apparent in the next section.
\end{itemize}

\noindent If $\mathfrak{m}\in \mathcal{M}_{1}$ is purely attractive or
simple then the situation becomes simpler because the Choquet decompositions
of generalized equilibrium states in $\mathit{\Omega }_{\mathfrak{m}}$
coincide with those in $\overline{\mathrm{co}}\left( \mathbf{M}_{\mathfrak{m}%
}\right) $. Thus, we have, in this case, all the good properties of both
types of extreme decomposition. In this situation, we obtain in Section \ref%
{mainresult} two extensions of Theorem \ref{eq.tang.bcs.type0 copy(1)} to
long-range models: Corollary \ref{lr - equilibrium and kms states (2)} and
Theorem \ref{gen eq states as KMS states}.

\section{Main Results: Generalized Equilibrium States and KMS Conditions
\label{mainresult}}

Recall that in Section \ref{Section KMS States}, we assert that equilibrium
states of lattice fermion systems with short-range interactions, as
translation-invariant minimizers of the free energy density functional, are
KMS states associated with the (well-defined) infinite volume dynamics on
the CAR $C^{\ast }$-algebra $\mathcal{U}$. See Theorem \ref%
{eq.tang.bcs.type0 copy(1)}, which is a direct consequence of results of
\cite{Araki-Moriya}. Here, we aim at contributing an extension of this
result to generalized equilibrium states of long-range models.

Note first that, for a general long-range model $\mathfrak{m}\in \mathcal{M}$%
, at fixed $A\in \mathcal{U}$ and $t\in \mathbb{R}$, the sequence $\tau
_{t}^{(L,\mathfrak{m})}(A)$, $L\in \mathbb{N}$, where (the finite volume
dynamics) $\tau _{t}^{(L,\mathfrak{m})}$\ is defined by (\ref{finite vol
long range dyna}), does not necessarily converge in the $C^{\ast }$-algebra $%
\mathcal{U}$, in contrast to the short-range case. See Sections \ref{sect
Lieb--Robinson copy(2)} and \ref{Section Limit Long-Range Dynamics}. In
other words, for a general long-range model, we do not have at our disposal
a well-defined infinite volume dynamics on the CAR $C^{\ast }$-algebra $%
\mathcal{U}$. There are two ways to get around this problem:

\begin{itemize}
\item The first one is to relax the (global) KMS property by considering it
only \textquotedblleft fiberwise\textquotedblright , using the ergodic
decomposition of Theorem \ref{theorem choquet}. This is performed in Section %
\ref{SECTION IS NEW}.

\item The second approach is to use a faithful representation of the $%
C^{\ast }$-algebra $\mathcal{U}$ in order to make sense of the infinite
volume long-range dynamics, by considering a different topology from the one
associated with the norm, typically the strong\footnote{\cite[Theorem 4.3
and Corollary 4.5]{BruPedra-MFIII} show that, within the cyclic
representation associated with the initial state, the infinite volume limit
of dynamics is well-defined in the $\sigma $-weak operator topology. Using
stronger norms on interactions together with additional estimates, one could
improve \cite[Theorem 4.3 and Corollary 4.5]{BruPedra-MFIII} to get the
convergence in the strong operator topology.} or the $\sigma $-weak operator
topology. This is performed in Section \ref{Modular Group}.
\end{itemize}

\noindent Note in this context that Theorem \ref{Thm cool} is absolutely
crucial to obtain the main outcomes of the present section, for general
long-range models, while Theorem \ref{lemma explosion l du mec copacabana2}
(and its proof) is pivotal to understand the restriction of our study to
long-range models that are either purely attractive (Section \ref{Section
purely attrac}) or simple (Definition \ref{def simple}). Moreover, we only
consider long-range models within a dense subspace of translation-invariant
models, that is, $\mathcal{M}_{0}\cap \mathcal{M}_{1}$ (see Equations (\ref%
{M0}) and (\ref{translatino invariatn long range models})), but this is a
very mild technical restriction, which is related to the well-posedness of
(dynamical) self-consistency equations (\ref{self-consistency equation}).

\subsection{Invariant Self-consistently KMS states\label{SECTION IS NEW}}

As explained above, in this section we relax the (global) KMS property by
considering it with respect to the ergodic decomposition of
translation-invariant states. This leads to the concept of \emph{%
self-consistently} KMS states of translation-invariant long-range models.

\begin{definition}[Self-consistently KMS states of translation-invariant
long-range models]
\label{def KMS self}\mbox{ }\newline
Take any $\rho \in E_{1}$ and let $\mu _{\rho }$ be the unique (Choquet)
measure (Theorem \ref{theorem choquet}) on ergodic states, whose barycenter
is the translation-invariant state $\rho $. We say that $\rho $ is a
(translation-invariant) self-consistently KMS state of $\mathfrak{m}\in
\mathcal{M}_{1}$ iff $\hat{\rho}\in E_{1}$ is $\mu _{\rho }$-almost surely a
KMS state for the strongly continuous group of automorphisms generated by
the approximating interaction $\Phi _{\mathfrak{m}}(e_{(\cdot )}(\hat{\rho}%
)) $ (see (\ref{approx interaction}) and (\ref{ssssssssssssssssssss})), at
inverse temperature $\beta \in \mathbb{R}^{+}$. We denote by $\mathrm{K}_{%
\mathfrak{m}}\subseteq E_{1}$ the set of self-consistently KMS states of $%
\mathfrak{m}\in \mathcal{M}_{1}$.
\end{definition}

\noindent The above notion is reminiscent of van Hemmen's approach \cite%
{Hemmen} to infinite volume equilibrium states for mean-field models. If the
long-range component of $\mathfrak{m}$ is trivial, i.e., $\mathfrak{m}=(\Phi
,0)$ for some $\Phi \in \mathcal{W}_{1}^{\mathbb{R}}$, then $\Phi _{%
\mathfrak{m}}(e_{(\cdot )}(\rho ))=\Phi $ for any $\rho \in E_{1}$ and one
thus has $\mathrm{K}_{(\Phi ,0)}=\mathit{K}_{\Phi }$ in this case, see
Equation (\ref{set of KMS}). Therefore, the above definition generalizes the
notion of KMS states to long-range models.

Note that, even for purely attractive long-range models, a self-consistently
KMS state is not necessarily a generalized equilibrium state. To see this
explicitly, one may use, for instance, the strong-coupling BCS model
explained in \cite{Bru-pedra-proceeding} and consider the equilibrium states
of the same model without its mean-field component. These states are always
ergodic, gauge-invariant and self-consistently KMS (for the original model
with mean-field interactions), but they violate the gap equations (\ref{gap
equations}) at sufficiently large (inverse temperature) $\beta $. This fact
leads us to introduce the notion of \emph{Bogoliubov} states, which is
crucial in order to establish a relation between the self-consistent KMS
condition and generalized equilibrium states of translation-invariant
long-range models.

\begin{definition}[Bogoliubov states of translation-invariant long-range
models]
\label{Bogo states}\mbox{ }\newline
Take any $\rho \in E_{1}$ and let $\mu _{\rho }$ be the unique (Choquet)
measure (Theorem \ref{theorem choquet}) on ergodic states, whose barycenter
is the translation-invariant state $\rho $. We say that $\rho $ is a
Bogoliubov state of $\mathfrak{m}\in \mathcal{M}_{1}$ iff the gap equation $%
e_{(\cdot )}(\hat{\rho})=d_{-}+\mathrm{r}_{+}(d_{-})$ holds true for some $%
d_{-}\in \mathcal{C}_{\mathfrak{m}}$, $\mu _{\rho }$-almost surely for $\hat{%
\rho}\in E_{1}$. We denote by $\mathrm{B}_{\mathfrak{m}}\subseteq E_{1}$ the
set of all Bogoliubov states of $\mathfrak{m}\in \mathcal{M}_{1}$.
\end{definition}

\noindent Compare with Equation (\ref{gap equations}), recalling that $%
\mathrm{r}_{+}$ is the thermodynamic decision rule (\ref{thermodyn decision
rule})\ of $\mathfrak{m}\in \mathcal{M}_{1}$, while $e_{(\cdot )}\left( \rho
\right) :\mathbb{S\rightarrow C}$ is the continuous and bounded mapping
defined from (\ref{ssssssssss})--(\ref{eq:enpersite}) by Equation (\ref%
{ssssssssssssssssssss}) for any state $\rho \in E_{1}$.

Note that if the long-range component of $\mathfrak{m}$ is trivial, i.e., $%
\mathfrak{m}=(\Phi ,0)$ for some $\Phi \in \mathcal{W}_{1}$, then the
equality $e_{(\cdot )}(\rho )=d_{-}+\mathrm{r}_{+}(d_{-})$ holds trivially
true for any $\rho \in E_{1}$, since $L_{-}^{2}(\mathbb{S};\mathbb{C}%
)=L_{+}^{2}(\mathbb{S};\mathbb{C})=L^{2}(\mathbb{S};\mathbb{C})=\{0\}$, the
unique equivalent class of all complex-valued functions on $\mathbb{S}$.
Thus, $\mathrm{B}_{(\Phi ,0)}=E_{1}$ in this case. Recall that, by Lemma \ref%
{Thm cool (2)}, if $\mathfrak{m}\in \mathcal{M}_{1}$ is purely attractive or
simple then the Choquet decompositions of generalized equilibrium states in $%
\mathit{\Omega }_{\mathfrak{m}}$ coincide with those in $\overline{\mathrm{co%
}}\left( \mathbf{M}_{\mathfrak{m}}\right) $. In this case, all generalized
equilibrium states are Bogoliubov states:

\begin{lemma}[Generalized equilibrium states as Bogoliubov states]
\label{lemma Omega subset B}\mbox{ }\newline
If $\mathfrak{m}\in \mathcal{M}_{1}$ is purely attractive or simple then $%
\mathit{\Omega }_{\mathfrak{m}}\subseteq \mathrm{B}_{\mathfrak{m}}$.
\end{lemma}

\begin{proof}
Let $\mathfrak{m}\in \mathcal{M}_{1}$ be a purely attractive or simple
translation-invariant long-range model. Suppose that $\omega \in \mathit{%
\Omega }_{\mathfrak{m}}$. From Theorems \ref{theorem structure of omega}, %
\ref{Thm cool} and Lemma \ref{Thm cool (2)}, for $\mu _{\omega }$-almost all
$\rho \in E_{1}$, $\rho \in \mathit{\Omega }_{\mathfrak{m}}(d_{-}+\mathrm{r}%
_{+}(d_{-}))$ for some $d_{-}\in \mathcal{C}_{\mathfrak{m}}$, that is, $\rho
\in \mathit{M}_{\Phi _{\mathfrak{m}}(d_{-}+\mathrm{r}_{+}(d_{-}))}$ and $%
e_{(\cdot )}(\rho )=d_{-}+\mathrm{r}_{+}(d_{-})$. See Equation (\ref{subset
of a face}) and Definition \ref{Bogo states}.
\end{proof}

From the proof of Theorem \ref{lemma explosion l du mec copacabana2}, note
that, in general, generalized equilibrium states of long-range models are
not Bogoliubov states of these models. This results from the fact that one
can construct (uncountably many) models $\mathfrak{m}\in \mathcal{M}_{1}$
whose set $\mathit{\Omega }_{\mathfrak{m}}$ of generalized equilibrium
states is not a face of $E_{1}$. See \cite[Lemma 9.8]{BruPedra2}. This
singular case is however still important, since it is also related to
long-range order of equilibrium states, as shown in \cite[Section 2.9]%
{BruPedra2}. In fact, in the general case, we have the following result on
the relation between self-consistently KMS, Bogoliubov and generalized
equilibrium states:

\begin{theorem}[Self-consistently KMS, Bogoliubov and generalized
equilibrium states]
\label{lr - equilibrium and kms states (1)}\mbox{ }\newline
For any translation-invariant long-range model $\mathfrak{m}\in \mathcal{M}%
_{1}$, $\mathit{\Omega }_{\mathfrak{m}}\cap \mathrm{B}_{\mathfrak{m}}=%
\mathrm{K}_{\mathfrak{m}}\cap \mathrm{B}_{\mathfrak{m}}$.
\end{theorem}

\begin{proof}
Let $\mathfrak{m}\in \mathcal{M}_{1}$. Assume that $\mathit{\Omega }_{%
\mathfrak{m}}\cap \mathrm{B}_{\mathfrak{m}}\neq \emptyset $ and take $\omega
\in \mathit{\Omega }_{\mathfrak{m}}\cap \mathrm{B}_{\mathfrak{m}}$. From
Theorem \ref{Thm cool}, for $\mu _{\omega }$-almost all $\rho \in E_{1}$, $%
\rho \in \mathit{M}_{\Phi _{\mathfrak{m}}(d_{-}+\mathrm{r}_{+}(d_{-}))}=%
\mathit{M}_{\Phi _{\mathfrak{m}}(e_{(\cdot )}(\rho ))}$ for some $d_{-}\in
\mathcal{C}_{\mathfrak{m}}$. Then, thanks to Theorem \ref{eq.tang.bcs.type0
copy(1)}, $\rho \in \mathit{K}_{\Phi _{\mathfrak{m}}(e_{(\cdot )}(\rho ))}$ $%
\mu _{\omega }$-almost surely. Therefore, $\omega \in \mathrm{K}_{\mathfrak{m%
}}$.

Assume that $\mathrm{K}_{\mathfrak{m}}\cap \mathrm{B}_{\mathfrak{m}}\neq
\emptyset $ and take $\omega \in \mathrm{K}_{\mathfrak{m}}\cap \mathrm{B}_{%
\mathfrak{m}}$. By Definition \ref{def KMS self}, it means in particular
that, for $\mu _{\omega }$-almost all $\rho \in E_{1}$, $\rho \in \mathit{K}%
_{\Phi _{\mathfrak{m}}(e_{(\cdot )}(\rho ))}$ and $e_{(\cdot )}(\rho )=d_{-}+%
\mathrm{r}_{+}(d_{-})$ for some $d_{-}\in \mathcal{C}_{\mathfrak{m}}$.
Thanks again to Theorem \ref{eq.tang.bcs.type0 copy(1)}, it follows that,
for $\mu _{\omega }$-almost all $\rho \in E_{1}$, $\rho \in \mathit{M}_{\Phi
_{\mathfrak{m}}(d_{-}+\mathrm{r}_{+}(d_{-}))}$ and $\rho \in \mathit{\Omega }%
_{\mathfrak{m}}(d_{-}+\mathrm{r}_{+}(d_{-}))$ for some $d_{-}\in \mathcal{C}%
_{\mathfrak{m}}$, by Equation (\ref{subset of a face}). We then conclude
from Theorem \ref{theorem structure of omega} (i) that $\omega \in \mathit{%
\Omega }_{\mathfrak{m}}$.
\end{proof}

In the case of purely attractive or simple long-range models we have the
following improvement of Theorem \ref{lr - equilibrium and kms states (1)}:

\begin{corollary}[Generalized equilibrium as self-consistently KMS and
Bogoliubov states]
\label{lr - equilibrium and kms states (2)}\mbox{ }\newline
If $\mathfrak{m}\in \mathcal{M}_{1}$ is purely attractive or simple then $%
\mathit{\Omega }_{\mathfrak{m}}=\mathrm{K}_{\mathfrak{m}}\cap \mathrm{B}_{%
\mathfrak{m}}$.
\end{corollary}

\begin{proof}
The assertion directly follows from Lemma \ref{lemma Omega subset B} and
Theorem \ref{lr - equilibrium and kms states (1)}.
\end{proof}

For any self-adjoint translation-invariant and short-range interaction $\Phi
\in \mathcal{W}_{1}^{\mathbb{R}}$, recall that $\mathrm{B}_{(\Phi ,0)}=E_{1}$%
, $\mathrm{K}_{(\Phi ,0)}=\mathit{K}_{\Phi }$ and $\mathit{\Omega }_{(\Phi
,0)}=\mathit{M}_{\Phi }$. It means that Corollary \ref{lr - equilibrium and
kms states (2)} is an extension of Theorem \ref{eq.tang.bcs.type0 copy(1)}
to translation-invariant long-range models. This result is complemented by
Theorem \ref{gen eq states as KMS states} below, which asserts that (under
the mild technical condition that $\mathfrak{m}\in \mathcal{M}_{0}$) the
elements of $\mathit{\Omega }_{\mathfrak{m}}=\mathrm{K}_{\mathfrak{m}}\cap
\mathrm{B}_{\mathfrak{m}}$ are KMS states in the usual (or global) sense,
with respect to the infinite volume dynamics generated by the long-range
model $\mathfrak{m}$ in a given representation of the CAR $C^{\ast }$%
-algebra $\mathcal{U}$.

\subsection{The Modular Group of Generalized Equilibrium States\label%
{Modular Group}}

Recall that $\mathcal{M}_{0}$ is the dense subspace of the Banach space $%
\mathcal{M}$ of long-range models defined by Equation (\ref{M0}), while $%
\mathcal{M}_{1}\subseteq \mathcal{M}$ is the Banach space of all
translation-invariant long-range models defined by Equation (\ref%
{translatino invariatn long range models}). A long-range model $\mathfrak{m}%
=(\Phi ,\mathfrak{a})\in \mathcal{M}$ is purely attractive whenever $%
\mathfrak{a}_{+}=0$, see Section \ref{Section purely attrac}. Simple models
are defined by Definition \ref{def simple}. Both situations are important
here because in these cases, when $\mathfrak{m}\in \mathcal{M}_{1}$, the
Choquet decomposition in $\mathit{\Omega }_{\mathfrak{m}}$ of any
generalized equilibrium state of $\mathfrak{m}$ is the same as its ergodic
decomposition (in $E_{1}$), by Lemma \ref{Thm cool (2)}. In particular, $%
\mathcal{E}(E_{1})\cap \mathit{\Omega }_{\mathfrak{m}}\neq \emptyset $. For
such long-range models, we first prove the stationarity of generalized
equilibrium states, in the Schr\"{o}dinger picture of quantum mechanics (see
Equation (\ref{long-range dyn0})).

\begin{proposition}[Stationarity of generalized equilibrium states]
\label{stationarity of gen equilibrium states}\mbox{ }\newline
If $\mathfrak{m}\in \mathcal{M}_{0}\cap \mathcal{M}_{1}$ is purely
attractive or simple then
\begin{equation*}
\lim_{L\rightarrow \infty }\omega _{t}^{(L)}\left( A\right) \doteq
\lim_{L\rightarrow \infty }\omega \circ \tau _{t}^{(L,\mathfrak{m})}\left(
A\right) =\omega \left( A\right) \text{ },\qquad A\in \mathcal{U},\ \omega
\in \mathit{\Omega }_{\mathfrak{m}}\text{ }.
\end{equation*}
\end{proposition}

\begin{proof}
Let $\mathfrak{m}\in \mathcal{M}_{0}\cap \mathcal{M}_{1}$ be a purely
attractive or simple translation-invariant long-range model. If $\hat{\omega}%
\in \mathcal{E}(E_{1})\cap \mathit{\Omega }_{\mathfrak{m}}$ is an ergodic
generalized equilibrium state, then we infer from Lemma \ref{Thm cool (2)}
the existence of a unique $d_{-}\in \mathcal{C}_{\mathfrak{m}}$ such that
\begin{equation}
d\doteq d_{-}+\mathrm{r}_{+}(d_{-})=e_{(\cdot )}(\hat{\omega})\doteq \hat{%
\omega}(\mathfrak{e}_{(\cdot )})\qquad \text{and}\qquad \hat{\omega}\in
\mathit{M}_{\Phi _{\mathfrak{m}}(d)\text{ }},  \label{gap equations copy}
\end{equation}%
recalling that $e_{(\cdot )}\left( \rho \right) :\mathbb{S\rightarrow C}$ is
the continuous and bounded mapping defined from (\ref{ssssssssss})--(\ref%
{eq:enpersite}) by Equation (\ref{ssssssssssssssssssss}) for any state $\rho
\in E_{1}$. Thanks to Theorem \ref{eq.tang.bcs.type0 copy(1)}, $\hat{\omega}%
\in \mathit{K}_{\Phi _{\mathfrak{m}}(d)}$, i.e., $\hat{\omega}$ is a $(\tau
^{\Phi _{\mathfrak{m}}(d)},\beta )$-KMS state, keeping in mind that $\beta
\in \mathbb{R}^{+}$ is fixed in all the paper. It is well-known that KMS
states are stationary, see, e.g., \cite[Proposition 5.3.3]{BrattelliRobinson}%
. In particular, the KMS state $\hat{\omega}$ is $\tau ^{\Phi _{\mathfrak{m}%
}(d)}$-invariant, i.e.,
\begin{equation}
\hat{\omega}=\hat{\omega}\circ \tau _{t}^{\Phi _{\mathfrak{m}}(d)}\ ,\qquad
t\in \mathbb{R}\text{ }.
\label{self-consistency for ergodic equilibrium states0}
\end{equation}%
It follows from Equation (\ref{gap equations copy}) and \cite[Lemma 7.2]%
{BruPedra-MFII} that, for all $t\in \mathbb{R}$,
\begin{equation}
\mathbf{\varpi }^{\mathfrak{m}}(t;\hat{\omega})=\hat{\omega}\qquad \text{and}%
\qquad \Phi ^{\mathfrak{m},\hat{\omega}}(t)=\Phi _{\mathfrak{m}}(\hat{\omega}%
(\mathfrak{e}_{(\cdot )}))=\Phi _{\mathfrak{m}}(d)\ ,
\label{self-consistency for ergodic equilibrium states}
\end{equation}%
where $\mathbf{\varpi }^{\mathfrak{m}}$ is the unique continuous mapping
from $\mathbb{R}$ to the space of automorphisms of $E$ satisfying (\ref%
{self-consistency equation}) and $\Phi ^{\left( \mathfrak{m},\hat{\omega}%
\right) }(t)$ is defined by Equation (\ref{self-consistency equation2}) with
$\rho =\hat{\omega}$. The proposition then follows from Equations (\ref%
{long-range dyn}) and (\ref{self-consistency for ergodic equilibrium states}%
).
\end{proof}

The stationarity of the state of a given physical system is the minimal
requirement characterizing the thermodynamic equilibrium of that system. In
this section, we contribute a much stronger result, which complements
Corollary \ref{lr - equilibrium and kms states (2)}. In fact, we show below
that generalized equilibrium states of purely attractive or simple
long-range models can also be seen as KMS states, in the usual sense. We
first need to provide a well-defined infinite volume dynamics for long-range
models, in order to be able to study the KMS\ property of their
(generalized) equilibrium states.

\begin{proposition}[Infinite volume long-range dynamics for generalized
equilibrium states]
\label{long-range dyn on gen equilibrium states}\mbox{ }\newline
If $\mathfrak{m}\in \mathcal{M}_{0}\cap \mathcal{M}_{1}$ is purely
attractive or simple then, for any $\omega \in \mathit{\Omega }_{\mathfrak{m}%
}$ with cyclic representation $(\mathcal{H}_{\omega },\pi _{\omega },\Omega
_{\omega })$, there exists a unique $\sigma $-weakly continuous\footnote{%
This means here that, for any fixed $t\in \mathbb{R}$ and $A\in \pi _{\omega
}(\mathcal{U})^{\prime \prime }$, the mappings $\mathbf{\Lambda }%
_{t}^{\omega }(\cdot ):\pi _{\omega }(\mathcal{U})^{\prime \prime
}\rightarrow \pi _{\omega }(\mathcal{U})^{\prime \prime }$ and\ $\mathbf{%
\Lambda }_{(\cdot )}^{\omega }(A):\mathbb{R}\rightarrow \pi _{\omega }(%
\mathcal{U})^{\prime \prime }$ are $\sigma $-weak continuous.} group $(%
\mathbf{\Lambda }_{t}^{\omega })_{t\in \mathbb{R}}$ of $\ast $-automorphisms
of the von Neumann algebra $\pi _{\omega }(\mathcal{U})^{\prime \prime }$
such that, with respect to the $\sigma $-weak topology,%
\begin{equation*}
\lim_{L\rightarrow \infty }\pi _{\omega }\left( \tau _{t}^{(L,\mathfrak{m}%
)}\left( A\right) \right) =\mathbf{\Lambda }_{t}^{\omega }\left( \pi
_{\omega }\left( A\right) \right) \ ,\qquad A\in \mathcal{U},\ t\in \mathbb{R%
}\text{ },
\end{equation*}%
where, in the special case of ergodic generalized equilibrium states,
\begin{equation*}
\mathbf{\Lambda }_{t}^{\hat{\omega}}\left( \pi _{\hat{\omega}}\left(
A\right) \right) =\pi _{\hat{\omega}}\circ \tau _{t}^{\Phi _{\mathfrak{m}}(%
\hat{\omega}(\mathfrak{e}_{(\cdot )}))}\left( A\right) \text{ },\qquad A\in
\mathcal{U},\ t\in \mathbb{R},\ \hat{\omega}\in \mathcal{E}(E_{1})\cap
\mathit{\Omega }_{\mathfrak{m}}\text{ }.
\end{equation*}
\end{proposition}

\begin{proof}
Let $\mathfrak{m}\in \mathcal{M}_{0}\cap \mathcal{M}_{1}$ be a purely
attractive or simple translation-invariant long-range model. In the special
case $\hat{\omega}\in \mathcal{E}(E_{1})\cap \mathit{\Omega }_{\mathfrak{m}}$%
, with associated cyclic representation $(\mathcal{H}_{\hat{\omega}},\pi _{%
\hat{\omega}},\Omega _{\hat{\omega}})$, we infer from \cite[Theorem 5.8]%
{BruPedra-MFIII} and Equation (\ref{self-consistency for ergodic equilibrium
states}) that
\begin{equation*}
\lim_{L\rightarrow \infty }\pi _{\hat{\omega}}\left( \tau _{t}^{(L,\mathfrak{%
m})}(A)\right) =\pi _{\hat{\omega}}\left( \tau _{t}^{\Phi _{\mathfrak{m}}(%
\hat{\omega}(\mathfrak{e}_{(\cdot )}))}(A)\right) \text{ },\qquad A\in
\mathcal{U}\text{ },\ t\in \mathbb{R}\text{ },
\end{equation*}%
with respect to the $\sigma $-weak topology. To study general (possibly
non-ergodic) generalized equilibrium states, one applies the theory of
direct integrals of measurable families of Hilbert spaces, operators, von
Neumann algebras and $C^{\ast }$-algebra representations, as already done in
\cite{BruPedra-MFIII}. In fact, we consider the $C^{\ast }$-algebra $%
\mathfrak{U}\doteq C(E;\mathcal{U})$, whose norm is%
\begin{equation*}
\left\Vert f\right\Vert _{\mathfrak{U}}\doteq \max_{\rho \in E}\left\Vert
f\left( \rho \right) \right\Vert _{\mathcal{U}}\text{ },\qquad f\in
\mathfrak{U}\text{ }.
\end{equation*}%
The CAR $C^{\ast }$-algebra $\mathcal{U}$ is canonically identified with the
subalgebra of constant functions of $\mathfrak{U}$, i.e., $\mathcal{U}%
\subseteq \mathfrak{U}$. Fix now, once and for all in the proof, $\omega \in
\mathit{\Omega }_{\mathfrak{m}}\subseteq E_{1}$. From \cite[Proposition 4.2]%
{BruPedra-MFIII}, there exists a unique representation $\Pi _{\omega }$ of $%
\mathfrak{U}$ on $\mathcal{H}_{\omega }$ such that $\Pi _{\omega }|_{%
\mathcal{U}}=\pi _{\omega }$ and $(\mathcal{H}_{\omega },\Pi _{\omega
},\Omega _{\omega })$ is a cyclic representation associated with the state $%
\omega $, seen as a state of $\mathfrak{U}$ via the definition%
\begin{equation}
\omega \left( f\right) \doteq \int_{\mathcal{E}(E_{1})\cap \mathit{\Omega }_{%
\mathfrak{m}}}\hat{\omega}\left( f\left( \hat{\omega}\right) \right) \mu
_{\omega }\left( \mathrm{d}\hat{\omega}\right) \text{ },\qquad f\in
\mathfrak{U}\text{ }.  \label{ddddd}
\end{equation}%
See Theorem \ref{theorem choquet} and Lemma \ref{Thm cool (2)}. Moreover,
one has%
\begin{equation}
\Pi _{\omega }\left( \mathfrak{U}\right) ^{\prime \prime }=\pi _{\omega }(%
\mathcal{U})^{\prime \prime }\text{ }.  \label{eq kip1}
\end{equation}%
From \cite[Theorem 4.3]{BruPedra-MFIII} and Equation (\ref{self-consistency
for ergodic equilibrium states}), for any time $t\in \mathbb{R}$ and all
elements $A\in \mathcal{U}\subseteq \mathfrak{U}$,
\begin{equation}
\lim_{L\rightarrow \infty }\pi _{\omega }\left( \tau _{t}^{(L,\mathfrak{m}%
)}(A)\right) =\Pi _{\omega }\left( \mathfrak{T}_{t}^{\mathfrak{m}}\left(
A\right) \right) \in \mathcal{B}\left( \mathcal{H}_{\omega }\right)
\label{f3}
\end{equation}%
with respect to the $\sigma $-weak topology, where $\mathfrak{T}^{\mathfrak{m%
}}=(\mathfrak{T}_{t}^{\mathfrak{m}})_{t\in \mathbb{R}}$ is a strongly
continuous group of $\ast $-automorphisms of $\mathfrak{U}$ defined by
\begin{equation*}
\mathfrak{T}_{t}^{\mathfrak{m}}\left( f\right) \left( \rho \right) \doteq
\tau _{t,0}^{\Phi ^{\left( \mathfrak{m},\rho \right) }}\left( f\left( \rho
\right) \right) \text{ },\qquad \rho \in E_{1},\ f\in \mathfrak{U}\text{ },\
t\in \mathbb{R},
\end{equation*}%
the interaction $\Phi ^{\left( \mathfrak{m},\rho \right) }$ being defined by
(\ref{self-consistency equation2}). We deduce from (\ref{ddddd}) combined
with (\ref{self-consistency for ergodic equilibrium states0})--(\ref%
{self-consistency for ergodic equilibrium states}) that, for any $t\in
\mathbb{R}$ and $f\in \mathfrak{U}$,
\begin{align*}
\omega \left( \mathfrak{T}_{t}^{\mathfrak{m}}\left( f\right) \right) &
=\int_{\mathcal{E}(E_{1})\cap \mathit{\Omega }_{\mathfrak{m}}}\hat{\omega}%
\left( \left( \mathfrak{T}_{t}^{\mathfrak{m}}f\right) \left( \hat{\omega}%
\right) \right) \mu _{\omega }\left( \mathrm{d}\hat{\omega}\right) =\int_{%
\mathcal{E}(E_{1})\cap \mathit{\Omega }_{\mathfrak{m}}}\hat{\omega}\left(
\tau _{t,0}^{\Phi ^{\left( \mathfrak{m},\hat{\omega}\right) }}\left( f\left(
\hat{\omega}\right) \right) \right) \mu _{\omega }\left( \mathrm{d}\hat{%
\omega}\right) \\
& =\int_{\mathcal{E}(E_{1})\cap \mathit{\Omega }_{\mathfrak{m}}}\hat{\omega}%
\left( f\left( \hat{\omega}\right) \right) \mu _{\omega }\left( \mathrm{d}%
\hat{\omega}\right) =\omega \left( f\right) .
\end{align*}%
In other words, $\omega \in \mathfrak{U}^{\ast }$ is $\mathfrak{T}^{%
\mathfrak{m}}$-invariant. By \cite[Corollary 2.3.17]{BrattelliRobinsonI},
there exists a unique strongly continuous family $(U_{t})_{t\in \mathbb{R}}$
of unitary operators on $\mathcal{B}(\mathcal{H}_{\omega })$ such that
\begin{equation}
\Pi _{\omega }\left( \mathfrak{T}_{t}^{\mathfrak{m}}\left( A\right) \right)
=U_{t}\pi _{\omega }\left( A\right) U_{t}^{-1},\qquad A\in \mathcal{U},\
t\in \mathbb{R}\text{ }.  \label{f1}
\end{equation}%
For any time $t\in \mathbb{R}$, define
\begin{equation}
\mathbf{\Lambda }_{t}^{\omega }\left( A\right) \doteq
U_{t}AU_{t}^{-1},\qquad A\in \pi _{\omega }(\mathcal{U})^{\prime \prime }.
\label{f2}
\end{equation}%
Observe in particular that, for any fixed $t\in \mathbb{R}$,
\begin{equation*}
\mathbf{\Lambda }_{t}^{\omega }\left( \cdot \right) :\pi _{\omega }(\mathcal{%
U})^{\prime \prime }\rightarrow \mathcal{B}(\mathcal{H}_{\omega })
\end{equation*}%
is $\sigma $-weakly continuous. Note from Equation (\ref{eq kip1}) that
\begin{equation*}
\mathbf{\Lambda }_{t}^{\omega }(\pi _{\omega }(\mathcal{U}))\subseteq \Pi
_{\omega }\left( \mathfrak{U}\right) \subseteq \pi _{\omega }\left( \mathcal{%
U}\right) ^{\prime \prime },\qquad t\in \mathbb{R}\text{ }.
\end{equation*}%
Since the bicommutant $\pi _{\omega }(\mathcal{U})^{\prime \prime }$ is the
closure of $\pi _{\omega }(\mathcal{U})$ in the $\sigma $-weak topology and $%
\mathbf{\Lambda }_{t}^{\omega }(\cdot )$ is continuous in this topology at
any fixed $t\in \mathbb{R}$, we arrive at $\mathbf{\Lambda }_{t}^{\omega
}(\pi _{\omega }(\mathcal{U})^{\prime \prime })\subseteq \pi _{\omega }(%
\mathcal{U})^{\prime \prime }$. Then, from the strong continuity of $%
(U_{t})_{t\in \mathbb{R}}$, $(\mathbf{\Lambda }_{t}^{\omega })_{t\in \mathbb{%
R}}$ is a $\sigma $-weakly continuous group of $\ast $-automorphisms of the
von Neumann algebra $\pi _{\omega }(\mathcal{U})^{\prime \prime }$. Finally,
note from Equations (\ref{f3}) and (\ref{f1}) that any $\sigma $-weakly
continuous group of $\ast $-automorphisms of $\pi _{\omega }(\mathcal{U}%
)^{\prime \prime }$ implementing the infinite volume dynamics should be
equal to $(\mathbf{\Lambda }_{t}^{\omega })_{t\in \mathbb{R}}$ on $\pi
_{\omega }(\mathcal{U})$, and thus on $\pi _{\omega }\left( \mathcal{U}%
\right) ^{\prime \prime }$, by the $\sigma $-weak density of $\pi _{\omega }(%
\mathcal{U})$ in $\pi _{\omega }(\mathcal{U})^{\prime \prime }$.
\end{proof}

Having an appropriate and natural notion of infinite volume long-range
dynamics, as given by Proposition \ref{long-range dyn on gen equilibrium
states}, we can now study the (global) KMS property of generalized
equilibrium states for a fixed translation-invariant long-range model $%
\mathfrak{m}=(\Phi ,\mathfrak{a})\in \mathcal{M}_{1}\cap \mathcal{M}_{0}$
that is either purely attractive or simple. The KMS property is in this case
defined as follows: Given a generalized equilibrium state $\omega \in
\mathit{\Omega }_{\mathfrak{m}}$ with associated cyclic representation $(%
\mathcal{H}_{\omega },\pi _{\omega },\Omega _{\omega })$, we define $\tilde{%
\omega}\doteq \left\langle \Omega _{\omega },(\cdot )\Omega _{\omega
}\right\rangle _{\mathcal{H}_{\omega }}$, which is the unique normal
extension\footnote{%
In fact, we say here that the (generalized equilibrium) state $\omega $,
which is a state on the $C^{\ast }$-algebra $\mathcal{U}$, has an extension
to $\pi _{\omega }(\mathcal{U})^{\prime \prime }$, because, by Corollary \ref%
{eq.tang.bcs.type0 copy(4)}, the representation $\pi _{\omega }$ is faithful
and the $C^{\ast }$-algebras $\mathcal{U}$ and $\pi _{\omega }(\mathcal{U})$
can thus be canonically identified with each other.} of $\omega $ to the von
Neumann algebra $\pi _{\omega }(\mathcal{U})^{\prime \prime }$. Considering
the $\sigma $-weakly continuous group $\mathbf{\Lambda }^{\omega }\equiv (%
\mathbf{\Lambda }_{t}^{\omega })_{t\in \mathbb{R}}$ of $\ast $-automorphisms
of the von Neumann algebra $\pi _{\omega }(\mathcal{U})^{\prime \prime }$ of
Proposition \ref{long-range dyn on gen equilibrium states}, for a fixed
inverse temperature $\beta \in \mathbb{R}^{+}$, we say that $\tilde{\omega}$
is a $(\mathbf{\Lambda }^{\omega },\beta )$-KMS\ state if
\begin{equation}
\int_{\mathbb{R}}f\left( t-i\beta \right) \tilde{\omega}\left( A\mathbf{%
\Lambda }_{t}^{\omega }\left( B\right) \right) \mathrm{d}t=\int_{\mathbb{R}%
}f\left( t\right) \tilde{\omega}\left( \mathbf{\Lambda }_{t}^{\omega }\left(
B\right) A\right) \mathrm{d}t  \label{KMS condition v Neumann}
\end{equation}%
for all $A,B\in \pi _{\omega }(\mathcal{U})^{\prime \prime }$ and any
function $f$ being the (holomorphic) Fourier transform of a smooth function
with compact support. Compare this definition of KMS states with the one
related to Equality (\ref{KMS condition}) on the CAR $C^{\ast }$-algebra $%
\mathcal{U}$. Note that the integrals of Equation (\ref{KMS condition v
Neumann}) are well-defined since, for any $A,B\in \pi _{\omega }(\mathcal{U}%
)^{\prime \prime }$, the functions
\begin{equation*}
\tilde{\omega}(A\mathbf{\Lambda }_{(\cdot )}^{\omega }\left( B\right) ),%
\tilde{\omega}(\mathbf{\Lambda }_{(\cdot )}^{\omega }\left( B\right) A):%
\mathbb{R}\rightarrow \mathbb{C}
\end{equation*}%
are clearly continuous and bounded. We can now see generalized equilibrium
states of long-range models as KMS\ states in the following sense:

\begin{theorem}[Generalized equilibrium states as KMS\ states]
\label{gen eq states as KMS states}\mbox{ }\newline
Fix $\beta \in \mathbb{R}^{+}$ and a translation-invariant long-range model $%
\mathfrak{m}\in \mathcal{M}_{1}\cap \mathcal{M}_{0}$ that is purely
attractive or simple. Given $\omega \in \mathit{\Omega }_{\mathfrak{m}}$
with associated cyclic representation $(\mathcal{H}_{\omega },\pi _{\omega
},\Omega _{\omega })$, let $\tilde{\omega}$ be the normal extension of $%
\omega $ to $\pi _{\omega }(\mathcal{U})^{\prime \prime }$ and $\mathbf{%
\Lambda }^{\omega }\equiv (\mathbf{\Lambda }_{t}^{\omega })_{t\in \mathbb{R}%
} $ be the $\sigma $-weakly continuous group of $\ast $-automorphisms of $%
\pi _{\omega }(\mathcal{U})^{\prime \prime }$, whose existence and
uniqueness are stated in Proposition \ref{long-range dyn on gen equilibrium
states}. Then, $\tilde{\omega}$ is a $(\mathbf{\Lambda }^{\omega },\beta )$%
-KMS state.
\end{theorem}

\begin{proof}
Fix all parameters of the theorem. Like in the proof of Proposition \ref%
{long-range dyn on gen equilibrium states}, consider first an arbitrary
ergodic generalized equilibrium state $\hat{\omega}\in \mathcal{E}%
(E_{1})\cap \mathit{\Omega }_{\mathfrak{m}}$. By Equation (\ref{gap
equations copy}) combined with Theorems \ref{eq.tang.bcs.type0 copy(1)} and %
\ref{Thm cool}, $\hat{\omega}\in \mathit{K}_{\Phi _{\mathfrak{m}}(\hat{\omega%
}(\mathfrak{e}_{(\cdot )}))}$. Recall that the strong convergence of a net
of bounded operators implies the boundedness of this net (by the
Banach-Steinhaus uniform boundedness principle), as well as its $\sigma $%
-weak convergence. Thus, from the strong density of $\pi _{\hat{\omega}}(%
\mathcal{U})$ in $\pi _{\hat{\omega}}(\mathcal{U})^{\prime \prime }$, by
Lebesgue's dominated convergence theorem and the $\sigma $-weak continuity
of $\mathbf{\Lambda }^{\hat{\omega}}$ (Proposition \ref{long-range dyn on
gen equilibrium states}), it follows that the normal extension of $\hat{%
\omega}\in \mathcal{E}(E_{1})\cap \mathit{\Omega }_{\mathfrak{m}}$ to $\pi _{%
\hat{\omega}}(\mathcal{U})^{\prime \prime }$ is a $(\mathbf{\Lambda }^{\hat{%
\omega}},\beta )$-KMS state. To study general (possibly non-ergodic)
generalized equilibrium states, one applies again the theory of direct
integrals of measurable families of Hilbert spaces, operators, von Neumann
algebras and $C^{\ast }$-algebra representations, as in \cite[Sections 5-6]%
{BruPedra-MFIII} and in the proof of Proposition \ref{long-range dyn on gen
equilibrium states}: The ergodic decomposition $\mu _{\rho }$ of any
translation-invariant state $\rho \in E_{1}$ is an orthogonal measure
(Theorem \ref{theorem choquet}) and, thanks to the Effros theorem (see,
e.g., \cite[Corollary 5.14]{BruPedra-MFIII}), for any $\rho \in E_{1}$, the
direct integral
\begin{equation}
\left( \mathcal{H}_{\rho }^{\oplus }\equiv \int_{\mathcal{E}(E_{1})}\mathcal{%
H}_{\hat{\rho}}\mu _{\rho }(\mathrm{d}\hat{\rho}),\ \pi _{\rho }^{\oplus
}\equiv \int_{\mathcal{E}(E_{1})}\pi _{\hat{\rho}}\mu _{\rho }(\mathrm{d}%
\hat{\rho}),\ \Omega _{\rho }^{\oplus }\equiv \int_{\mathcal{E}%
(E_{1})}\Omega _{\hat{\rho}}\mu _{\rho }(\mathrm{d}\hat{\rho})\right)
\label{direct integral representation}
\end{equation}%
of the GNS representations $(\mathcal{H}_{\hat{\rho}},\pi _{\hat{\rho}%
},\Omega _{\hat{\rho}})$ of $\mathcal{U}$ associated with the extreme states
$\hat{\rho}\in \mathcal{E}(E_{1})$ is a cyclic representation of the $%
C^{\ast }$-algebra $\mathcal{U}$, associated with the state $\rho \in E_{1}$%
. Moreover, from \cite[Equation (158)]{BruPedra-MFIII}, one has the
inclusion
\begin{equation*}
\pi _{\rho }^{\oplus }\left( \mathcal{U}\right) ^{\prime \prime }\subseteq
\int_{E_{1}}\pi _{\hat{\rho}}\left( \mathcal{U}\right) ^{\prime \prime }\mu
_{\rho }(\mathrm{d}\hat{\rho})\ ,\text{\qquad }\rho \in E_{1}\text{ }.
\end{equation*}%
In particular, any element $A\in \pi _{\rho }(\mathcal{U})^{\prime \prime }$
can be identified with an element $(A_{\hat{\rho}})_{\hat{\rho}\in E_{1}%
\text{ }}$of
\begin{equation*}
\int_{E_{1}}\pi _{\hat{\rho}}\left( \mathcal{U}\right) ^{\prime \prime }\mu
_{\rho }(\mathrm{d}\hat{\rho})\ .
\end{equation*}%
Fix now $\omega \in \mathit{\Omega }_{\mathfrak{m}}\subseteq E_{1}$. From
Lemma \ref{Thm cool (2)} and Proposition \ref{long-range dyn on gen
equilibrium states},
\begin{equation*}
\mathbf{\Lambda }_{t}^{\omega }(A)=\int_{\mathcal{E}(E_{1})\cap \mathit{%
\Omega }_{\mathfrak{m}}}\mathbf{\Lambda }_{t}^{\hat{\omega}}(A_{\hat{\omega}%
})\mu _{\omega }(\mathrm{d}\hat{\omega})\ ,\text{\qquad }A\in \pi _{\omega }(%
\mathcal{U})^{\prime \prime },\ t\in \mathbb{R}\text{ }.
\end{equation*}%
Note that this identity follows from the uniqueness of the $\sigma $-weakly
continuous group $(\mathbf{\Lambda }_{t}^{\omega })_{t\in \mathbb{R}}$
stated in Proposition \ref{long-range dyn on gen equilibrium states}
together with general properties of direct integrals that can be found, for
instance, in \cite[Section 6]{BruPedra-MFIII}. Since in each fiber the
normal extension of $\hat{\omega}\in \mathcal{E}(E_{1})\cap \mathit{\Omega }%
_{\mathfrak{m}}$ to $\pi _{\hat{\omega}}(\mathcal{U})^{\prime \prime }$ is a
$(\mathbf{\Lambda }^{\hat{\omega}},\beta )$-KMS state, again denoted by $%
\hat{\omega}$ to simplify the notation, it follows from Lemma \ref{Thm cool
(2)} and Fubini's theorem that, for all $A,B\in \pi _{\omega }(\mathcal{U}%
)^{\prime \prime }$ and any function $f$ being the (holomorphic) Fourier
transform of a smooth function with compact support,
\begin{eqnarray*}
\int_{\mathbb{R}}f\left( t-i\beta \right) \tilde{\omega}(A\mathbf{\Lambda }%
_{t}^{\omega }(B))\mathrm{d}t &=&\int_{\mathbb{R}}f\left( t-i\beta \right)
\left( \int_{\mathcal{E}(E_{1})\cap \mathit{\Omega }_{\mathfrak{m}}}\hat{%
\omega}(A_{\hat{\omega}}\mathbf{\Lambda }_{t}^{\hat{\omega}}(B_{\hat{\omega}%
}))\mu _{\omega }(\mathrm{d}\hat{\omega})\right) \mathrm{d}t \\
&=&\int_{\mathcal{E}(E_{1})\cap \mathit{\Omega }_{\mathfrak{m}}}\left( \int_{%
\mathbb{R}}f\left( t-i\beta \right) \hat{\omega}(A_{\hat{\omega}}\mathbf{%
\Lambda }_{t}^{\hat{\omega}}(B_{\hat{\omega}}))\mathrm{d}t\right) \mu
_{\omega }(\mathrm{d}\hat{\omega}) \\
&=&\int_{\mathcal{E}(E_{1})\cap \mathit{\Omega }_{\mathfrak{m}}}\left( \int_{%
\mathbb{R}}f\left( t\right) \hat{\omega}(\mathbf{\Lambda }_{t}^{\hat{\omega}%
}(B_{\hat{\omega}})A_{\hat{\omega}})\mathrm{d}t\right) \mu _{\omega }(%
\mathrm{d}\hat{\omega}) \\
&=&\int_{\mathbb{R}}f\left( t\right) \left( \int_{\mathcal{E}(E_{1})\cap
\mathit{\Omega }_{\mathfrak{m}}}\hat{\omega}(\mathbf{\Lambda }_{t}^{\hat{%
\omega}}(B_{\hat{\omega}})A_{\hat{\omega}})\mu _{\omega }(\mathrm{d}\hat{%
\omega})\right) \mathrm{d}t \\
&=&\int_{\mathbb{R}}f\left( t\right) \tilde{\omega}(\mathbf{\Lambda }%
_{t}^{\omega }(B)A)\mathrm{d}t\text{ }.
\end{eqnarray*}%
In other words, the normal extension $\tilde{\omega}$ of $\omega $ to $\pi
_{\omega }(\mathcal{U})^{\prime \prime }$ is a $(\mathbf{\Lambda }^{\omega
},\beta )$-KMS state.
\end{proof}

\noindent Theorem \ref{gen eq states as KMS states} is an(other) extension
of Theorem \ref{eq.tang.bcs.type0 copy(1)} to translation-invariant
long-range models, which complements Corollary \ref{lr - equilibrium and kms
states (2)}. These results pave the way to the use of the Tomita-Takesaki
modular theory \cite[Section 2.5]{BrattelliRobinsonI} and the KMS theory
\cite[Sections 5.3-5.4]{BrattelliRobinson} in the study of mean-field
models, like the BCS model of (conventional) superconductivity. For
instance, by Corollary \ref{eq.tang.bcs.type0 copy(4)}, recall that, for $%
\mathfrak{m}\in \mathcal{M}_{1}$, any generalized equilibrium state $\omega
\in \mathit{\Omega }_{\mathfrak{m}}$, with associated cyclic representation $%
(\mathcal{H}_{\omega },\pi _{\omega },\Omega _{\omega })$, is a modular
state, i.e., the vector $\Omega _{\omega }\in \mathcal{H}_{\omega }$ is
cyclic and separating for the von Neumann algebra $\pi _{\omega }(\mathcal{U}%
)^{\prime \prime }\subseteq \mathcal{B}(\mathcal{H}_{\omega })$. Therefore,
the modular $\ast $-automorphism group associated with $\Omega _{\omega }$
and the von Neumann algebra $\pi _{\omega }(\mathcal{U})^{\prime \prime }$
is well-defined. See, e.g.,\cite[Section 2.5.2]{BrattelliRobinsonI}. By
Proposition \ref{long-range dyn on gen equilibrium states} and Theorem \ref%
{gen eq states as KMS states}, it is directly related to the limit
long-range dynamics within the cyclic representation $(\mathcal{H}_{\omega
},\pi _{\omega },\Omega _{\omega })$ of the corresponding generalized
equilibrium state:

\begin{corollary}[Limit long-range dynamics as a modular group]
\label{modular group}\mbox{ }\newline
Fix $\beta \in \mathbb{R}^{+}$ and a translation-invariant long-range model $%
\mathfrak{m}\in \mathcal{M}_{1}\cap \mathcal{M}_{0}$ that is purely
attractive or simple. Given $\omega \in \mathit{\Omega }_{\mathfrak{m}}$
with associated cyclic representation $(\mathcal{H}_{\omega },\pi _{\omega
},\Omega _{\omega })$, let $(\mathbf{\sigma }_{t}^{\omega })_{t\in \mathbb{R}%
}$ be the modular $\ast $-automorphism group associated with $\Omega
_{\omega }$ and the von Neumann algebra $\pi _{\omega }(\mathcal{U})^{\prime
\prime }$. Then, with respect to the $\sigma $-weak topology,%
\begin{equation*}
\lim_{L\rightarrow \infty }\pi _{\omega }\left( \tau _{t}^{(L,\mathfrak{m}%
)}\left( A\right) \right) =\mathbf{\sigma }_{-\beta ^{-1}t}^{\omega }\left(
\pi _{\omega }\left( A\right) \right) \ ,\qquad A\in \mathcal{U},\ t\in
\mathbb{R}\text{ }.
\end{equation*}
\end{corollary}

\begin{proof}
Fix all parameters of the corollary. By Theorem \ref{gen eq states as KMS
states}, the unique normal extension $\tilde{\omega}$ of the generalized
equilibrium state $\omega \in \mathit{\Omega }_{\mathfrak{m}}$ is a $(%
\mathbf{\Lambda }^{\omega },\beta )$-KMS state. As a consequence, $\tilde{%
\omega}$ is also a $(\mathbf{\tilde{\Lambda}}^{\omega },-1)$-KMS state with
respect to the rescaled ($\sigma $-weakly continuous) group $\mathbf{\tilde{%
\Lambda}}^{\omega }\equiv (\mathbf{\Lambda }_{-\beta t}^{\omega })_{t\in
\mathbb{R}}$ of $\ast $-automorphisms of $\pi _{\omega }(\mathcal{U}%
)^{\prime \prime }$. From \cite[Theorem 5.3.10]{BrattelliRobinson}, it
follows that
\begin{equation*}
\mathbf{\Lambda }_{-\beta t}^{\omega }=\mathbf{\sigma }_{t}^{\omega }\
,\qquad t\in \mathbb{R}\text{ }.
\end{equation*}%
By combining this equality with Proposition \ref{long-range dyn on gen
equilibrium states} we arrive at the assertion.
\end{proof}

\bigskip

\noindent \textit{Acknowledgments:} This work is supported by CNPq
(309723/2020-5 and 140782/2020-6), FAPESP (2017/22340-9), as well as by the
Basque Government through the grant IT641-13 and the BERC 2018-2021 program,
and by the Spanish Ministry of Science, Innovation and Universities: BCAM
Severo Ochoa accreditation SEV-2017-0718, MTM2017-82160-C2-2-P.

\noindent \textbf{Jean-Bernard Bru} \newline
Departamento de Matem\'{a}ticas\newline
Facultad de Ciencia y Tecnolog\'{\i}a\newline
Universidad del Pa\'{\i}s Vasco\newline
Apartado 644, 48080 Bilbao \medskip \newline
BCAM - Basque Center for Applied Mathematics\newline
Mazarredo, 14. \newline
48009 Bilbao\medskip \newline
IKERBASQUE, Basque Foundation for Science\newline
48011, Bilbao\bigskip

\noindent \textbf{Walter de Siqueira Pedra} \newline
Departamento de F\'{\i}sica Matem\'{a}tica\newline
Instituto de F\'{\i}sica,\newline
Universidade de S\~{a}o Paulo\newline
Rua do Mat\~{a}o 1371\newline
CEP 05508-090 S\~{a}o Paulo, SP Brasil\bigskip

\noindent \textbf{Rafael S. Yamaguti Miada} \newline
Departamento de F\'{\i}sica Matem\'{a}tica\newline
Instituto de F\'{\i}sica,\newline
Universidade de S\~{a}o Paulo\newline
Rua do Mat\~{a}o 1371\newline
CEP 05508-090 S\~{a}o Paulo, SP Brasil


\begin{thebibliography}{99}
\bibitem{BrattelliRobinson} O. Bratteli and D.W. Robinson, \textit{Operator
Algebras and Quantum Statistical Mechanics, Vol. II, 2nd ed.} New York:
Springer-Verlag, 1997.

\bibitem{PW} W. Pusz and S. L. Woronowicz, Passive States and KMS States for
General Quantum Systems, \textit{Commun. math. Phys.} \textbf{58} (1978)
273-290.

\bibitem{brupedraLR} J.-B. Bru and W. de Siqueira Pedra, \textit{%
Lieb--Robinson Bounds for Multi--Commutators and Applications to Response
Theory}, Springer Briefs in Math. Phys., vol. 13, Springer Nature, (2017).

\bibitem{Araki-Moriya} H. Araki and H. Moriya, Equilibrium Statistical
Mechanics of Fermion Lattice Systems. Rev. Math. Phys. \textbf{15} (2003)
93-198.

\bibitem{Israel} R.B. Israel, \textit{Convexity in the Theory of Lattice
Gases}, Princeton Series in Physics, Princeton Univ. Press, 1979.

\bibitem{Araki} H. Araki, On the equivalence of the KMS condition and the
variational principle for quantum lattice systems, \textit{Commun. Math.
Phys.} \textbf{38} (1974) 1-10.

\bibitem{LRB-continuum} M. Gebert, B. Nachtergaele, J. Reschke and R. Sims,
Lieb-Robinson bounds and strongly continuous dynamics for a class of
many-body fermion systems in $\mathbb{R}^{d}$, Ann. Henri Poincar\'{e}
\textbf{21} (2020) 3609-3637.

\bibitem{haag62} R. Haag. The mathematical structure of the
Bardeen-Cooper-Schrieffer model. Nuovo Cimento, \textbf{25} (1962) 287-298.

\bibitem{BruPedra-MFII} J.-B. Bru and W. de Siqueira Pedra, Classical
Dynamics Generated by Long-Range Interactions for Lattice Fermions and
Quantum Spins, J. Math. Anal. Appl. \textbf{493}(1) (2021) 124434.

\bibitem{Hemmen} L. van Hemmen, Linear Fermion Systems, Molecular Field
Models, and the KMS Condition. Fortschritte der Physik \textbf{26} (1978)
397--439.

\bibitem{BruPedra2} J.-B. Bru and W. de Siqueira Pedra, Non-cooperative
Equilibria of Fermi Systems With Long Range Interactions. Memoirs of the AMS
\textbf{224}, no. 1052 (2013).

\bibitem{BruPedra-MFIII} J.-B. Bru and W. de Siqueira Pedra, Quantum
Dynamics Generated by Long-Range Interactions for Lattice-Fermion and
Quantum Spins, J. Math. Anal. Appl. \textbf{493}(1) (2021) 124517.

\bibitem{Bru-pedra-proceeding} J.-B. Bru and W. de Siqueira Pedra,
Macroscopic Dynamics of the Strong-Coupling BCS-Hubbard Model, \textit{%
Physics of Particles and Nuclei} \textbf{51}(4) (2020) 802--806.

\bibitem{Bru-pedra-MF-IV} J.-B. Bru and W. de Siqueira Pedra, Entanglement
of Classical and Quantum Short-Range Dynamics in Mean-Field Systems, in
preparation.

\bibitem{BrattelliRobinsonI} O. Bratteli and D.W. Robinson, \textit{Operator
Algebras and Quantum Statistical Mechanics, Vol. I, 2nd ed.} New York:
Springer-Verlag, 1987.

\bibitem{Rudin} W. Rudin, \textit{Functional Analysis}. McGraw-Hill Science,
1991

\bibitem{Bru-pedra-MF-I} J.-B. Bru and W. de Siqueira Pedra, Classical
Dynamics From Self-Consistency Equations in Quantum Mechanics -- Extended
Version, \textit{arXiv:2009.04969} (2020).

\bibitem{Sewell} G.L. Sewell, \textit{Quantum Theory of Collective Phenomena}%
, Oxford: Clarendon Press, 1986.
\end{thebibliography}
\end{document}